\documentstyle[12pt,epsf]{article}
\newcommand{\lsim}{\stackrel{\scriptstyle <}{\phantom{}_{\sim}}}
\newcommand{\gsim}{\stackrel{\scriptstyle >}{\phantom{}_{\sim}}}
\begin{document}
\def\huge{\LARGE}
\title{\vspace*{-15mm}Charge Screening at
First Order Phase Transitions and Hadron - Quark Mixed Phase}


\author{D. N. Voskresensky $^{1,2,3}$, M. Yasuhira
$^{1,4}$ and T. Tatsumi $^{4}$}
\date{}
\maketitle

\noindent $^{1}${\it Yukawa Institute of Theoretical Physics,
Kyoto University, Kyoto 606-8502, Japan}
\\
$^{2}${\it Gesellschaft f\"ur Schwerionenforschung (GSI), Planck
Str 1, D-64291 Darmstadt, Germany}
\\
$^{3}${\it Moscow Institute for Physics and Engineering,
Kashirskoe sh. 31, Moscow 115409, Russia}
\\
$^{4}${\it Department of Physics, Kyoto University, Kyoto
606-8502, Japan}

\begin{abstract}
The possibility of structured mixed phases at first order phase
transitions in neutron stars is re-examined by taking into
account charge screening and surface effects. The transition from
the hadron $npe$ phase to the quark phase is  studied. Two
possibilities, the mixed phase and  two separate phases given by
the double-tangent (Maxwell) construction are considered.
Inhomogeneous profiles of the electric potential and their
contribution to the energy are analytically calculated. The
electric field configurations determine the droplet size and the
geometry of structures.

\end{abstract}

\section{Introduction}

It is now commonly accepted that different phase transitions may
occur in neutron star interiors. The possibilities of pion and
kaon condensate states, and quark matter state were studied by many
authors during the last thirty years, see \cite{M78,MSTV90,TPL,T95}
and refs therein. It has been argued that these transitions are of
first order. They were described
in terms of
two spatially separated
phases using the Maxwell construction.
Glendenning raised the question whether
mixed phases in systems composed
of charged particles exist
instead of the configuration described by the Maxwell
construction \cite{G92}. In particular, the possibilities of hadron -
kaon condensate ($npe$ - $npeK_{cond}$) and hadron - quark (H-Q)
mixed phases were discussed. The existence of such kind of mixed phases
in dense neutron star interiors would have important consequences
for the equation of state, also affecting neutrino emissivities
\cite{RBP00}, glitch phenomena and $r$ modes, cf.
\cite{BGP00,G01}.

Basing on the validity of the Gibbs conditions, in particular on the
equality of the electron chemical potentials of the phases, refs
\cite{G92,GS99,CG00,G01} further argued that the Maxwell construction
is {\em{always}} unstable. It is due to the inequality of the electron
chemical potentials of two phases and, thus, due to a possibility
for particles to fall down from the higher energetic levels
characterized by the higher electron chemical potential of the one
phase to the lower levels of the other phase. Thereby, refs
\cite{G92,GS99,CG00,CGS00,G01} argued that, if first order phase
transitions, such as quark, kaon condensate and pion condensate
transitions, indeed, occur in neutron star interiors, the existence of
a wide region of the mixed phase is inevitable. On explicit examples
of $npe$ - $npeK_{cond}$ and H-Q phase transitions authors
demonstrated the energetic preference of the mixed phase.

However, strictly speaking, such an argumentation is in
contradiction to the results of some other works. It has been
recently observed that in some models the Gibbs condition of
equality of electron chemical potentials of two phases can't be
fulfilled at all
\cite{PREPL00,MYTT,ARRW}, whereas
conditions for the Maxwell construction are fulfilled.
The latter construction is stable in these specific cases.
A critics
of the bulk calculations which ignore finite size effects was
given in ref. \cite{NR00}. To include these effects authors used a relaxation
procedure in which they start with an initial guess for the shape of the
electric potential in the Wigner-Seitz cell, solve for the kaon equation of
motion to obtain the charged particle profiles, and then recalculate the
electric potential using the new profiles. This is then repeated until
convergence.
Authors found that  with inclusion of
inhomogeneity effects the region of the kaon condensate mixed
phase is significantly narrowed compared to that obtained using
standard Gibbs conditions, disregarding finite size effects.
Please notice that in order the scheme to be completely self-consistent
the electric field, as the new degree of freedom,
should obey the own equation
of motion, and it
enters equations of motions of other fields (including protons and electrons)
via the gauge shift
of the chemical potentials of all
charged particles. All the charged particle densities are affected by the
inhomogeneous electric field even in the
case, if one artificially suppressed the contribution
of the electric field to
the charged kaon density.

For small size droplets, the problem of the construction of the
mixed phase is analogous to that studied somewhat earlier for
matter at sub-nuclear densities, cf. \cite{RPW83}. The possibility of
the structure, as well as its geometry, are determined by
competition between the Coulomb energy and the surface energy of
droplets. Ref. \cite{HPS93} applied these ideas to the description
of the mixed phase for the H-Q phase transition. The Coulomb plus
surface energy per droplet of the new phase always has a minimum
as a function of the droplet radius. On the one hand, this radius
should be not too small in order for the droplet to have rather large
baryon number ($A\gg 1$)\footnote{This condition, $A\gg 1$,
is assumed to be fulfilled. Otherwise quantum effects, like shell effects,
may affect the consideration.} and, on the other hand, it should be not
too large (less than the Debye screening length) in
order one could ignore  screening effects. The Debye screening
length for the quark matter was evaluated as $\lambda_{\rm D} \simeq
5~$fm. The value of the droplet radius $R_{\rm min}$ depends on a poorly
known surface tension parameter $\sigma$.  With a small value of
the surface tension $\sigma \simeq 10~$ MeV$\cdot$fm$^{-2}$ the
droplet size was  estimated as $R_{\rm min} \geq 3.1~$fm and with $\sigma
\simeq 100~$ MeV$\cdot$fm$^{-2}$, as $R_{\rm min} \geq 6.6~$fm
$>\lambda_{\rm D}$. Thereby, it was argued that the mixed phase is not
permitted in the latter case.
Corrections  to the Coulomb solutions\footnote{By the Coulomb
solutions we call solutions with the charged density profiles in the
form of combination of
step-functions. Screening effects are disregarded in this
case.} due to screening effects were not considered, although one
could intuitively expect that for such a narrow interval of
available values of droplet radii, being of the order of the Debye
screening length, the screening may significantly affect the
results. Also the effect of the inhomogeneity of the field profiles in the
strong interaction part of the energy was disregarded. Therefore,
further detailed study of the effects of inhomogeneous
field profiles, such as screening and surface effects, seems to be
of prime importance.

The problem of the construction of the charged density profiles by
taking into account screening effects is in some sense analogous
to that considered previously in refs \cite{MPV77,VSCh77,VCh78}
for abnormal pion condensate nuclei. Results can be used also for
the description of the charge distributions in strangelets and
kaon condensate droplets.

The value of the surface tension
is poorly known. It has the meaning only if there exists a related
shortest scale in the problem, being
much smaller than the typical scale of the
change of the electric field.
In the case of the  kaon condensate phase
transition there is no necessity to introduce
the surface tension.  One can explicitly solve
the equations of motion for all  the mean meson fields,
cf. refs \cite{CGS00,NR00}. However one should also
include the equation of motion for the electric field. The existence of the
additional typical scale $\lambda_{\rm D}$ appearing in this equation
may affect the
solution. In the case of the H-Q phase transition there is a
natural minimal
scale in the problem related to the
confinement $\lsim 0.5$~fm. Since the latter scale is much shorter than
$\lambda_{\rm D}$
the introduction of the surface tension is well established in the given case.

With this paper we show that, firstly, the Maxwell construction does
not contradict  the Gibbs condition of equality of the electron
chemical potentials, if one properly incorporates the electric
potential. Secondly, inclusion of finite size effects, such as
screening and surface tension, is crucial for understanding
the mixed phase existence and its description.
We will consistently incorporate these effects. Part of the
results of this paper was announced in the Letter \cite{VYT01}.

Our consideration is rather general and main effects do not depend on
what the concrete system is studied. However we need to consider
a concrete example
to demonstrate the quantitative effects.
For that aim we concentrate on the description of the
H-Q phase transition. Thus we avoid the discussion of other
possibilities, such as the charged pion condensation, charged
kaon condensation, see \cite{MSTV90,G01} and refs therein.
We will consider the simplest case of the H-Q phase transition
between normal phases.
Thus we also avoid the consideration of
the diquark condensates and diquark condensates together with
the pion and kaon condensates, see \cite{SS00} and refs therein.
With the inclusion of the diquark superfluid gap, the
corrections to the effective energy functional are expected to be as small as
$(\Delta /\mu_B )^2$, where $\Delta$ is the pairing gap and
$\mu_B$ is the baryon chemical potential, cf. \cite{ARRW,IB02}.
However the response to the electric field can be a more specific. E.g.,
in the case of the phase transition to the
color-flavor-locked (CFL) phase the quark sub-system
is enforced to be electrically neutral without electrons,
cf. \cite{ARRW}. There can be even more involved effects,
which we do not consider, like
the modification of the quark and gluon condensates with the increase of
the baryon density, a modification of masses and widths of all
particles, that needs a special quantum treatment, etc.

In sect. \ref{gcap} we critically review the application of the Gibbs
conditions to the finite size structures and we discuss the stability
of the Maxwell construction. Since the problem of interpreting the
Gibbs conditions and the Maxwell construction is related to the counting
of the electric potential from different levels we discuss
different choices of the gauge in  sect. \ref{gcap}. Starting
from  sect.  \ref{gen} we present our results in an arbitrary gauge.
In sect.  \ref{gen} we develop a general formalism based on the
thermodynamic potential. It serves  as a generating functional, variation of
which in the corresponding independent variables determines
the equations of motion. We demonstrate how one can explicitly treat
electric field effects.  Chemical equilibrium
conditions and the charged particle densities are presented in sect.
\ref{Eq}.
In sect. \ref{sph}  we analytically solve the
equation of motion for the electric potential of the new phase
droplets placed in the Wigner-Seitz lattice
and assuming spherical geometry we recover
corresponding contributions to the energy and thermodynamic
potential. Slab structures are studied in sect. \ref{plane}.
Sect. \ref{Comparison} discusses which mixed phase structures,
spherical droplets or slabs, are realized in
dependence on the value of the quark fraction volume.
The discussion on the specifics of the description of
the H-Q system is deferred
to the Appendices. In Appendix \ref{en} we introduce the model
to calculate the volume part of the energy of quark and hadron phases.
In Appendix
\ref{surf} we discuss surface effects. In
Appendix \ref{Free}  we develop  a formalism
based on the Gibbs potential, as generating functional. In
Appendix \ref{example}
we
study the role of the nonlinear correction terms. In Appendix
\ref{Coul-pec} we discuss peculiarities of the Coulomb
limit.

All over the paper we use units $\hbar =c=1$.

\section{
Gibbs conditions, Maxwell construction and electric field
effects}\label{gcap}

Here, we will illustrate some inconsistencies of a naive
treatment of the Gibbs conditions.  These conditions formulated for
spatially homogeneous systems were  further
applied for the description of
the Maxwell construction and  spatially
inhomogeneous structures  of mixed phase leading to inconsistencies.

Let us assume that we deal with a first order phase transition in
a neutron star interior between two bulk phases (I) and (II)
separated by a thin boundary layer. In discussing rather short
distance effects one may disregard spatial changes of the fields
due to gravity.
In order to describe bulk configurations of phases I and II one
usually applies the Gibbs conditions \cite{M78,MSTV90,TY00}
\begin{eqnarray}\label{onegibbs}
P^{\rm I} =P^{\rm II}, \,\,\,\,
\end{eqnarray}
\begin{eqnarray}\label{onegibbs-b}
\mu_{B}^{\rm I}=\mu_{B}^{\rm II},
\end{eqnarray}
supplemented by the local  charge-neutrality relations
\begin{eqnarray}\label{loc-neutr}
\rho_{\rm ch}^{\rm I} =0,\,\,\,\,  \rho_{\rm ch}^{\rm II} =0.
\end{eqnarray}
Here $P^{\alpha}$ is the pressure, $\mu_B^{\alpha}$ is the baryon chemical
potential, $\rho_{\rm ch}^{\alpha}$ is the net charged density of
given phase, $\alpha = \rm{I,II}$.

The configuration corresponding to the coexistence of  two locally
charge-neutral bulk phases is determined by the double tangent
construction (Maxwell construction). The straight line $\epsilon
(\rho_B )= \frac{\partial \epsilon}{\partial \rho_B}|_{\rm I}(\rho_B
- \rho_{B}^{\rm I})+ \epsilon(\rho_{B}^{\rm I} )$ connecting the
points I and II on the curve of the density dependence of the
energy density $\epsilon (\rho_{B} )$ (for temperature $T=0$) is
characterized by equal derivatives
$$\left.\frac{\partial
(\epsilon/\rho_B )}{\partial \rho_B^{-1}} \right|_{\rm I}=\left.
\frac{\partial (\epsilon/\rho_B )}{\partial
\rho_B^{-1}}\right|_{\rm II},$$
and, thus, the pressure equals in both
phases. For phase transitions between two hadron phases these
points relate to the one and the same curve $\epsilon (\rho_B )$
and correspond to the equal area Maxwell construction on the
$P(\rho_B)$ plot.

Bulk configurations I and II are separated by a boundary layer,
which presence is ignored in the formulation of the above conditions;
e.g., the charge-neutrality condition (\ref{loc-neutr}) is obviously
violated in this layer. Thus, the boundary layer is assumed to be
rather thin as compared to the
sizes of regions occupied by both phases.
There are two typical sizes characterizing the boundary layer between
phases. In case of hadron-quark transition the shortest scale ($d_{\rm S}$)
is related to the change of
nuclear and quark fields in a narrow part of the boundary layer (we call it
the surface
layer) with the length $\sim d_{\rm S} \lsim 1$~fm and the longest scale
($\lambda_{\rm D} \gg d_{\rm S}$) is related to the electric field generated
in the boundary electric charged layer of the length $\sim
\lambda_{\rm D} \sim 5-10$~fm. Only in the case  $R \gg \lambda_{\rm D}$
($R$ be the
minimum of $R^{\rm I}$ and $R^{\rm II}$), the spatial change of the
energy density  in this boundary layer can be  treated
as a
surface contribution (neglecting corrections of the
higher order in $\lambda_{\rm D}/R$),
being rather small compared to the volume
one, since the surface  energy is $\lambda_{\rm D}/R$ times
less than the volume energy. Thus,
in case $R \gg \lambda_{\rm D}$, with an accuracy $O(\lambda_{\rm D} /R)$, one
may neglect  contributions to thermodynamic quantities of the
charged boundary layer and the surface layer between two spatially
separated bulk phases, as it is usually done in  the standard
formulation of the Maxwell construction.

The Maxwell construction treatment of first order phase
transitions has no alternatives for the description of systems with
one conserved charge (baryon charge in our case). However,  neutron
stars are composed of charged particles and the electric
charge is also conserved. Imposing the relations (\ref{loc-neutr}) one
assumes  that the electric charge is conserved locally, contrary
to the other
possibility that the electric charge is conserved only
globally, see \cite{G92,GS99}. Thus, one should still check a new
possibility of the formation of a mixed phase constructed of
inhomogeneous  rather small size charged structures
embedded in the charge-neutral Wigner-Seitz cells.

So, the local charge-neutrality conditions might be relaxed, being
replaced by the global charge-neutrality condition
\begin{eqnarray}\label{global}
  f^{\rm I}\cdot \int_{D^{\rm I}}d\vec{r}
\rho_{\rm ch}^{\rm I} - f^{\rm II}\cdot \int_{D^{\rm II}}d\vec{r}
\rho_{\rm ch}^{\rm II} = 0.
\end{eqnarray}
Relation (\ref{global}) introduces the fraction volumes $f^{\rm I}$
of the domain $D^{\rm I}$ occupied by the phase I
and $f^{\rm II}$, $f^{\rm I} =1- f^{\rm II}$, of the domain $D^{\rm II}$
occupied by the
phase II. In the case of spherical geometry one deals with
spherical droplets
of the one phase (be phase I) embedded into the other phase (be phase
II). This is so called the mixed phase. The condition of the  total baryon
number conservation,
\begin{eqnarray}\label{globalbar}
\rho_{B}=v^{-1}\int_{D^{\rm I}+D^{\rm II}}d\vec{r}\rho_{B}(\vec{r})=
\rho_{B}^{\rm I}f^{\rm I} +\rho_{B}^{\rm II}f^{\rm II},
\end{eqnarray}
is also imposed, $v=v^{\rm I}+v^{\rm II}$ is the total volume occupied by
the phases.

Refs \cite{G92,GS99} suggested that the conservation of the global
electric charge requires an extra Gibbs condition for the
electron chemical potentials to be fulfilled
\begin{eqnarray}\label{muel}
\mu_{e}^{\rm I}=\mu_{e}^{\rm II},
\end{eqnarray}
which is formulated in analogy with relation (\ref{onegibbs-b}).
At the first glance, condition (\ref{muel}) contradicts
relations (\ref{loc-neutr}), which define $\mu_{e}^{\rm I}\neq
\mu_{e}^{\rm II}$, as stated in text books. Therefore, refs
\cite{GS99,CG00,G01} argued that the double-tangent (Maxwell)
construction is {\em{always}} unstable since, due to the inequality of
electron chemical potentials, the particles may fall down from the
higher energetic level characterized by the larger electron
chemical potential (for concreteness in phase II) to the lower one
(in phase I). This argumentation implies the use of the same
definitions of chemical potentials in Gibbs and Maxwell
treatments.

Now we will show that the Gibbs condition (\ref{muel})
and the Maxwell condition (\ref{loc-neutr}) do not contradict each
other since they use {\em{different definitions}} of the electron
chemical potentials, the global constant quantities  $\mu_{e,\rm
Gibbs}^{\rm I}$ and $\mu_{e,\rm Gibbs}^{\rm II}$ ($\mu_{e,\rm
Gibbs}^{\rm I} =\mu_{e,\rm Gibbs}^{\rm II}\equiv \mu_{e,\rm
Gibbs}$) according to the Gibbs condition (\ref{muel}) and the
local quantities $\mu_{e,\rm loc}^{\rm I}$ and $\mu_{e,\rm
loc}^{\rm II}$ ($\mu_{e,\rm loc}^{\rm I} \neq \mu_{e,\rm loc}^{\rm
II}$) according to the Maxwell construction,
eq.~(\ref{loc-neutr}). Briefly speaking,   the Gibbs condition
(\ref{muel}), as it was formulated for spatially homogeneous
configurations,  has no meaning in the application to charged systems
of a small size, if one does not  incorporate electric field
effects. It only fixes the level from which one counts the
electric potential. Therefore, the above mentioned argumentation
against the Maxwell construction is invalid. It does not take into
account effects of the electric field arising in any charged
systems, at least, near the boundary.

Thus, first, we need to fix the definition of the electric chemical
potential of the spatially inhomogeneous system, since we need to
recover this quantity for phase I based on the knowledge of
the quantities of phase II. The two ways how one can do it are
illustrated in Fig.~1.

The first way (we will call it {\bf{way I}}) is as follows.
Consider two bulk matters occupied by phases I and II separated by
a boundary layer ($R-\lambda_{D}^{\rm I}<r<R+ \lambda_{D}^{\rm
II}$),  see
Fig.1. The electron chemical potential of the phase I
is determined as $\mu_{e, \rm Gibbs}\equiv \mu_{e, \rm Gibbs}^{\rm I}
=\mu_{e,\rm Gibbs}^{\rm II}=const$ in agreement with the Gibbs condition
(\ref{muel}). The value  $\mu_{e, \rm Gibbs}^{\rm II}$ is fixed by
the corresponding local charge-neutrality condition, see second
eq. (\ref{loc-neutr}), i.e. $\mu_{e, \rm Gibbs}^{\rm II} =
\mu_{e,\rm loc}^{\rm II}=const$.
Then, such a Gibbs condition contains no further information
except an indication that all the electron energy levels in both
phases are occupied up to one and the same the top
(if we assume that $\mu_{e,\rm loc}^{\rm II}> \mu_{e,\rm
loc}^{\rm I}$, where $\mu_{e,\rm loc}^{\rm I}$ is fixed by the first
local charge-neutrality condition, first eq. (\ref{loc-neutr}))
energy level
$\varepsilon_e =\mu_{e,\rm Gibbs}^{\rm I}=\mu_{e,\rm loc}^{\rm
II}$.
In order to avoid  any contradiction with the  local charge-neutrality
condition for the phase I (first eq. (\ref{loc-neutr})), which
would determine the value $\mu_{e,\rm loc}^{\rm I}\neq \mu_{e,\rm
Gibbs}^{\rm I}$, one needs to introduce an external electric potential well
$V$. The gradient of the latter is necessarily produced
in the charged boundary layer separating the bulk phases. The
value which enters this charge-neutrality condition, second eq.
(\ref{loc-neutr}), is  then $V =\mu_{e,\rm loc}^{\rm I}-\mu_{e,\rm
loc}^{\rm II}<0$.
In the opposite case with $\mu_{e,\rm loc}^{\rm II}< \mu_{e,\rm loc}^{\rm I}$,
the Gibbs condition means that levels are counted from the bottom
level. The electric potential well is completely filled by
electrons from the bottom to the top according to the Pauli
principle, see \cite{MPV77,VSCh77,VCh78}. Therefore, no
instability of such kind arises within the Maxwell construction.

In discussion of the possibility and construction of the finite size
structures of the mixed
phase one also needs to incorporate the electric potential,
because the small charged droplets have an inhomogeneous profile of
the electric field. The potential well satisfies the
equation of motion, which is the Poisson equation,
\begin{eqnarray}\label{el}
\Delta V =4\pi e^2 \sum_{i}\rho_{\rm ch}^i ,
\end{eqnarray}
where $V=-e\phi_e$ is the electric potential well, $e$ is electric charge,
$e^2 =1/137$, and $\rho_{\rm ch}^i$ is the  charge density of the
$i$ particle species.
$V$ is the gauge variant quantity allowing the replacement
$V\rightarrow V-V^0$
and, thus, being determined
up to an arbitrary constant $V^0$, see sect. 3. Accordingly, the chemical
potential is also a gauge dependent quantity. Here in the {\bf{way I}}
we fixed the gauge taking $V^0 =0$, $V\equiv V(V^0=0)$, see Fig.
1. Then it is related to the electron density $\rho_e ({\vec r})$
by means of $V (\vec{r})=\mu_{e, {\rm Gibbs}}-(3\pi^2 \rho_e
(\vec{r}))^{1/3}$. For $R_{\rm W}\gg
R\gg \lambda_{\rm D}^{\rm I,II}$, where $R_{\rm W}$ is the radius of the
Wigner-Seitz cell,
with the Debye
screening lengths $\lambda_{\rm D}^{\rm I,II}$ determined by the same
eq. (\ref{el}), there is no difference in solution of (\ref{el})
for the configuration given by the Maxwell construction and for
spherical droplets of the mixed phase. What configuration is realized
is determined by the minimization of the appropriate thermodynamic
potential over the droplet size.

The second way (which we will call the {\bf{way II}}) is a
quasiclassical treatment also illustrated in Fig. 1. The chemical
potential of electrons is introduced as a local quantity, being
unambiguously determined by the value of the electric potential
well, namely $\mu_{e,\rm loc} (\vec{r})\equiv -V(\vec{r}) $. Here we
imposed the boundary condition
$V(\vec{r})\rightarrow -\mu_{e, \rm Gibbs}$ at large distances
from the droplet, if
the concentration of droplets
$f^{\rm I}$ is  very small. Thus, the gauge constant
is taken to be $V^0 =-\mu_{e,\rm Gibbs}$, i.e. $V (\vec{r})\equiv
V(V^0=-\mu_{e,\rm Gibbs})=-\mu_{e, \rm loc}(\vec{r}).$
Here $\mu_{e, \rm loc} (\vec{r}) =(3\pi^2 \rho_e (\vec{r}))^{1/3}$
is the quasiclassical quantity, being expressed via the local
concentration of the particles, cf. \cite{LP80}, p. 321.  It is
precisely how the chemical potential is defined in the usual treatment of
Fermi systems. Only in order to distinguish between the local
quantity and the global constant used in the {\bf{way I}} we
introduced the subscript ``$\rm loc$'' and ``$\rm Gibbs$''. Then,
the local charge-neutrality conditions automatically yield different
values of $\mu_{e, \rm loc}^{\rm I}$ and $\mu_{e,\rm loc}^{\rm II}
$ in both phases, as it follows from eq.
(\ref{loc-neutr}), i.e. in complete agreement with the description
given by the Maxwell construction.

The {\bf{way II}} chosen, the Gibbs condition (\ref{muel})
should be omitted. One does not need to introduce any additional
constants, since the value of the top energy level is already
fixed by the condition $V(\vec{r})\rightarrow -\mu_{e, \rm
Gibbs}$, as $\vec{r}$ approaches the boundary of the
Wigner-Seitz cell for the case of a tiny concentration $f^{\rm I}$ ($R_{\rm
  W}\rightarrow \infty$).
All the necessary information comes from the Poisson equation for the
electric potential well, which is {\em{the equation of motion for the
electric field}} properly derived from the Lagrangian.
Equivalently,
this equation of motion can be obtained by
minimizing the appropriate
thermodynamic potential, being expressed in the corresponding
variables. If $\lambda_{\rm D}^{\rm I,II} \ll R$,  there
exist constant solutions of eq. (\ref{el}), being valid for each
phase outside the boundary layer of the length $\lambda_{\rm D}^{\rm
I}+\lambda_{\rm D}^{\rm II} $. Namely, these constant solutions
$V=const$ unambiguously guarantee the charge-neutrality conditions
(\ref{loc-neutr}) and, thus, determine the values of the
electron chemical potentials of each phase, i.e. $-V^{\rm I}
=\mu_{e,\rm loc}^{\rm I}$ and $-V^{\rm II} =\mu_{e,\rm loc}^{\rm
II}\neq \mu_{e,\rm loc}^{\rm I}$. In the boundary layer the
potential $-V (\vec{r})$ or, in other words, the electron chemical
potential $\mu_{e, \rm loc} (\vec{r})$ varies from the value
$\mu_{e,\rm loc}^{\rm II}$ to the value $\mu_{e,\rm loc}^{\rm I}$.

An argument
that the difference in the values of the electron chemical potentials of
the two phases produces an instability, since particles may fall down
from the higher energy level corresponding to the larger value of
$\mu_{e, \rm loc} =\mu_{e,\rm loc }^{\rm II}$ to the lower level
($\mu_{e, \rm loc} =\mu_{e,\rm loc }^{\rm I}$), again does not
work. The electric field configuration given by the solution of
the equation of motion (\ref{el}) is completely filled by the charged
fermions from the bottom ($V=-\mu_{e,\rm loc}^{\rm II}$) to the
top ($V=0$)  in accordance with the Pauli principle and there is
no one free state anymore, cf. \cite{MPV77}. If a free state would
be formed it would be immediately filled by chemical reactions
turned out from the equilibrium in this case.

In the case $R\lsim \lambda_{\rm D}^{\rm I}$ the first local
charge-neutrality condition (\ref{loc-neutr}) is irrelevant.
The configuration is determined by the corresponding inhomogeneous
solution of the Poisson equation. Thus, in order to incorporate
the possibility of mixed phase structures in the framework of a
consistent scheme one needs to explicitly introduce the electric
field developing according to its equation of motion.

The equality of the electron chemical potentials  of  two phases
written as $V^{\rm I}=-\mu_{e,\rm loc}^{\rm I} =V^{\rm
II}=-\mu_{e,\rm loc}^{\rm II}$ would mean that the value of the
electric potential well $V$ is constant within the volume occupied
by the phases. But {\em{ with constant $V$ charged objects can't
exist!}} A charged droplet can be formed only, if there is a
competition between the electric field energy and the surface
energy, that permits the existence of a minimum in the Gibbs potential
(or the thermodynamic potential in the corresponding variables) per
droplet at some finite droplet size. This should be treated as a
{\em{necessary condition for the existence of a mixed phase.}} On
the other hand, according to eq (\ref{el}) the constancy of the
electric potential within some extended space region
would mean that the r.h.s. of this equation
is zero. Therefore it coincides with the  local
charge-neutrality condition, see (\ref{loc-neutr}), and we come
back to the construction with $V^{\rm I}\neq V^{\rm II}$
describing the case $R\gg \lambda_{\rm D}^{\rm I,II}$, as in the Maxwell
construction, rather than to a construction of droplets of a small
size. {\em{One definitely should  exclude condition (\ref{muel})
in the {\bf{way II}}
in favor of the solution of the equation of motion (\ref{el}), which
then unambiguously determines the charged configurations.}}

Concluding, the Gibbs condition (\ref{muel}) is
automatically satisfied in the {\bf{way I}}, where the electron
chemical potential is constant. This constant then enters the r.h.s.
of (\ref{el}). In the {\bf{way II}} the electron chemical
potential introduced as $\mu_{e.\rm loc}(\vec{r})$ is no longer constant
and the Gibbs condition (\ref{muel}) should be dropped. Both
ways are simply related to each other by the gauge transformation
and lead to the very same equations of motion, thus describing the
same physics (see sect. 3).

Thus, in the description of  mixed phase
configurations inhomogeneous electric field profiles should be
explicitly found. Also surface effects modify the Gibbs condition
(\ref{onegibbs}). We will discuss the latter effect in Appendix
\ref{surf}.

We consciously paid so much attention to these questions in order
to motivate the necessity of a consistent account of the electric field
effects and to remove the so called ``contradiction'' between
descriptions of the mixed phase and configurations determined by the
Maxwell construction being discussed during many years in the
literature.

\section{General formalism.
}\label{gen}

Consider the structured mixed phase consisting of two phases I and
II. We assume  droplets of phase I to be located in a lattice described by
Wigner-Seitz cells. The exterior of the droplets is phase II.
Each droplet in the cell occupies the domain $D^{\rm I}$ of volume
$v^{\rm I}$ separated by a narrow surface layer $D_{\rm S}$
from matter
in phase II ( domain $D^{\rm II}$ of volume $v^{\rm II}$). We
expect particle species included in phases I and II to coexist in $D_{\rm S}$.
The thermodynamic potential (effective energy) per cell
is represented by a density functional \cite{par},
\begin{eqnarray}
\Omega=E[\rho ] - \mu_i^{\rm I}\int_{D^{\rm I}}
d\vec{r}\rho_i^{\rm I}- \mu_i^{\rm II}\int_{D^{\rm II}}d\vec{r}
\rho_i^{\rm II}- \mu_i^{\rm S}\int_{D_S}d\vec{r} \rho_i^{\rm S} ,
\label{omeg}
\end{eqnarray}
$E[\rho ]$ is the energy of the cell\footnote{We consider here the
case of the zero temperature and do not distinguish between the free energy
and the energy thereby.}, $\rho =\{\rho_i^{\rm I},
\rho_i^{\rm II}\}$ are densities of different particle species,
$i=1,...,N^{\rm I}$ in phase I, $i=1,...,N^{\rm II}$ in phase
II, and $i=1,...,(N^{\rm I}+N^{\rm II})$ in $D_{\rm S}$,
$N^{\rm I}$,  $N^{\rm II}$ are  total number of particle
species per cell in phases I and II.
Summation over the repeated Latin indices
is implied. We assumed that each phase is in the ground state and
the matter in phase I or II is in chemical equilibrium by means of
the weak and strong interactions.
The term labeled by ''$\rm S$''  is the corresponding
contribution of a narrow surface layer ($ D_{\rm S}$) separating phases.
Equations  of motion,
$\frac{\delta\Omega}{\delta\rho_i^\alpha}=0$, render
\begin{eqnarray}\label{omeg-chem}
\mu_i^\alpha  = \frac{\delta E [\rho ]}{\delta \rho_i^{\alpha}},
\,\,\,\alpha = \{{\rm I}, {\rm II}, {\rm S}\} .
\end{eqnarray}

The energy of the cell consists of four contributions:
\begin{equation}\label{enden1}
E
[\rho ] =\int_{D^{\rm I}} d\vec{r} \epsilon^{\rm I}_{\rm kin+str}
[\rho_i^{\rm I}] + \int_{D^{\rm II}} d\vec{r} \epsilon^{\rm
II}_{\rm kin+str} [\rho_i^{\rm II}]+\int_{ D_{\rm S}} dS\epsilon_{\rm S} [
\rho_i^{\rm S}
]+E_V.
\end{equation}
The first two contributions are the sums of the kinetic and
strong-interaction energies  and $\epsilon_{\rm S} [\rho_i^{\rm S} ]$ is the
surface energy density, which depends on all the particle
densities in the surface layer $D_{\rm S}$, and $E_V$ is the Coulomb
interaction energy.
The surface energy is given by integration of $\epsilon_{\rm S}$ over
this narrow region around  surface, where  densities $\rho_i^{\rm S}$
of, at least,
some
particle species change sharply. To be specific in this paper we concentrate
on the discussion of the H-Q phase transition, as a concrete example. The
corresponding values $\epsilon^{\rm I,II}_{\rm kin+str}$ are introduced in
Appendix \ref{en}
and the surface energy of the H-Q interface is discussed in Appendix
\ref{surf}.

We will further
use that there is a shortest scale, $d_{\rm S}$, relevant for variation
of above mentioned
particle densities. In case of the  H-Q phase transition
the quark and nucleon fields change sharply and
this scale is $d_{\rm S}\sim$~1~fm, relating to the confinement radius
($r_\Lambda \simeq (0.2\div 0.4)$~fm ) and to the diffuseness of
the nuclear layer ($d_N\simeq 0.6$~fm for an atomic nucleus).
We will use that  $d_{\rm S}$
being
much less than other scales involved in the problem, namely sizes
of regions occupied by the phases and the Debye screening lengths
$\lambda_{\rm D}^{\rm I,II} \sim (5\div 10)$~fm, see estimations below.
Then one may replace the narrow surface layer by the sharp boundary
$\partial D$ and approximate the surface energy, as we shall do, in
terms of the surface tension $\sigma$, taking $\int_{D_{\rm S}}
dS\epsilon_{\rm S} [\rho_i^{\rm S} ]\equiv \int_{\partial D}
dS\bar{\epsilon}_{\rm S} [\rho_i^{\rm S} ] =4\pi R^2 \sigma$ for spherical
droplet. In reality $\sigma$ is a complicated function of particle
densities at the surface. It should disappear when energy
densities of phases are equal, thus according to \cite{BBP71}
producing $\sigma =a_{\rm S} \mid \epsilon^{\rm I}-\epsilon^{\rm
II}\mid$, with coefficient $a_{\rm S}$ slightly depending on
particle densities of phases and $\epsilon^{\rm I,II}$, as volume
parts of energy densities of the phases. Simplifying, we will
further neglect the density dependence  of $a_{\rm S}(\rho_B )$, considering
$a_{\rm S}$ as constant. In this approximation there appears no
contribution from the variation of $\sigma$ with respect to
the pressure and to the
chemical potentials, see (\ref{gibbs1}), (\ref{gibbs2}) below. The
latter approximation is commonly adopted in literature, cf.
\cite{HH00} and refs therein. Since the values of $a_{\rm S}$ and $\sigma$
are rather
poorly known we will allow their variation in the wide range. We
will more closely discuss this question in Appendix \ref{surf}.

The Coulomb interaction energy
\footnote{We will further
call it the ``electric field energy'', in order not to mix it up
with what we relate in this paper to  ``the Coulomb limit''
with the spatially step-function dependence of $\rho_i$.}
$E_V$ in (\ref{enden1})
is
expressed in terms of particle
densities,
\begin{equation}\label{enden2}
E_V=\frac{1}{2}\int d\vec{r}\,d\vec{r}^{\,\prime}
\frac{Q_i \rho_i (\vec{r}) Q_j \rho_j (\vec{r}^{\,\prime}) }{\mid
\vec{r}-\vec{r}^{\,\prime}\mid},
\end{equation}
with $Q_i$ being the particle charge ($Q =-e <0$ for the
electron).

Then the equations of motion (\ref{omeg-chem}) can be  re-written as
\begin{equation}\label{eom}
\mu_i^\alpha =\frac{\partial\epsilon_{\rm kin+str}^\alpha}
{\partial\rho_i^\alpha}- N^{{\rm{ch}},\alpha }_i V^\alpha
(\vec{r})  ,~~~ N^{{\rm{ch}},\alpha}_i =Q_i^\alpha /e,
\end{equation}
with the electric potential well $V^\alpha (\vec r )$:
\begin{eqnarray}\label{other1}
V(\vec r ) =- \int d\vec{r}^{\,\prime}\, \frac{e Q_i \rho_i
(\vec{r}^{\,\prime}) }{\mid \vec{r}-\vec{r}^{\,\,\prime}\mid }
\equiv\left\{
\begin{array}{ll}
V^{\rm I}(\vec{r}), & \vec{r}\in D^{\rm I}\\ V^{\rm II}(\vec{r}),
& \vec{r}\in D^{\rm II}
\end{array}
\right.
\end{eqnarray}
generated by the
particle distributions.

We will keep the gauge invariance, so that  $V$
can be shifted by an arbitrary constant ($V^0$) due to the gauge
transformation, $V(\vec{r})\rightarrow V(\vec{r}) -V^0$. Formally
varying eq.~(\ref{eom}) with respect to $V^\alpha (\vec{r})$ or
$\mu_i^\alpha$ we have the matrix form relation,
\begin{eqnarray}\label{matrix}
A_{ij}^\alpha\frac{\partial\rho_{j}^\alpha}{\partial V^\alpha }=
N^{{\rm ch},\alpha}_i ,\,\,\,\, A_{ij}^\alpha B_{jk}^\alpha
=\delta_{ik},
\end{eqnarray}
where  matrices $A$ and $B$ are defined as
\begin{eqnarray}\label{a}
A_{ij}^\alpha \equiv\frac{\delta^2 E_{\rm kin+str}^\alpha}
{\delta\rho_i^\alpha\delta\rho_j^\alpha},\,\,\,\,
B_{ij}^\alpha\equiv\frac{\partial\rho_i^\alpha}{\partial\mu_j^\alpha}.
\label{matrix1}
\end{eqnarray}
Eqs.~(\ref{matrix}), (\ref{matrix1})  reproduce the gauge-invariance
relation,
\begin{equation}
\frac{\partial\rho_i^\alpha}{\partial V^\alpha}=N^{{\rm
ch},\alpha}_j \frac{\partial\rho_j^\alpha}{\partial\mu_i^\alpha},
\label{gauge}
\end{equation}
clearly showing that a constant shift of the chemical potential is
compensated by a gauge transformation of $V^\alpha (\vec{r})$:
$\mu_i^\alpha\rightarrow \mu_i^\alpha+N_i^{{\rm ch},\alpha}V^0$, as
$V(\vec{r})\rightarrow V(\vec{r}) -V^0$. Hence the chemical potential
$\mu_i^\alpha$ acquires physical meaning only after fixing of the
gauge of $V^\alpha (\vec{r})$; for the choice $V^0=0$ ({\bf{way I}}),
$\mu_i^{\alpha}=\mu_{i,\rm Gibbs}^\alpha$, and for $V^0=-\mu_{e,\rm Gibbs}$
({\bf{way II}}),
$\mu_e^{\alpha} =0$. \footnote{The notation
$\mu_i^\alpha$ used in the paper \cite{VYT01} should be understood
as $\mu_{i, \rm Gibbs}^\alpha$ in this sense.}

Applying Laplacian ($\Delta$) to the l.h.s. of eq.~(\ref{other1})
we recover the Poisson equation ($\vec{r}\in D^\alpha$),
\begin{eqnarray}\label{qV}
\Delta V^\alpha (\vec{r}) =4\pi e^2\rho^{{\rm
ch},\alpha}(\vec{r})\equiv 4\pi e Q_i^\alpha \rho_i^\alpha
(\vec{r}) . \,\,\,
\end{eqnarray}
The charge density $\rho^{{\rm ch},\alpha}(\vec{r})$  as a
function of $V^\alpha (\vec{r})$ is determined by the equations of
motion (\ref{eom}). Thus eq.~(\ref{qV}) is a nonlinear
differential equation for $V^\alpha (\vec{r})$. The boundary
conditions are
\begin{equation}\label{boun}
V^{\rm I}=V^{\rm II} ,~~~\nabla V^{\rm I}=\nabla V^{\rm II}+4\pi
e^2 a^{\rm ch}_{\rm S },~~~ {\vec r}\in \partial D\, ,
\end{equation}
where $\rho^{\rm ch}_{\rm S }=a^{\rm ch}_{\rm S }\delta
(\vec{r}\in S)$ is the surface charge density. Below we will neglect a
small contribution of  the surface charge accumulated at the interface
of the phases. Physically, typical scales for the spatial change of
the charge are $\lambda_{\rm D}^{\rm I,II}$ and $R$ and, thereby, only a
small charge can be accumulated at a much shorter scale $d_{\rm S}$,
$d_{\rm S} \lsim 1/m_{\pi}$, $m_{\pi}=140$~MeV  is the pion mass, the
typical
scale of the strong interaction. Variation of $\sigma$ over $V$
yields  a contribution to the surface charge $Z_{\rm S}\sim Z_{\rm
dr}d_{\rm S} /R $ at most that is
$\ll Z_{\rm dr}$, where $Z_{\rm dr}$ is typical value for the
charge accumulated in droplet, cf. eq.
(\ref{sch}) of Appendix \ref{surf} below.

We also impose the condition $\nabla V^{\rm II}=0$ at the boundary of
the Wigner-Seitz cell, which implies that each cell must be
charge-neutral. Once eqs.~(\ref{qV}) are solved giving $V^\alpha
(\vec{r})$ and the potentials are matched at the boundary, we have
density distributions of particles in the domain $D^\alpha$.

Note that there are two conservation laws relevant in neutron star
matter: baryon number and charge conservation. These quantities
are well defined over the whole space, not restricted to each
domain. Accordingly, the baryon  and charge chemical
potentials ($\mu_B$ and $\mu_Q$), being linear combinations of
$\mu_i^\alpha$,
become constants over the whole space,
\begin{equation}\label{gibbs1}
\mu^{\rm I}_B=\mu^{\rm II}_B\equiv \mu_B,~~~\mu^{\rm I}_Q=\mu^{\rm
II}_Q \equiv \mu_Q .
\end{equation}
 This fact requires two conditions for $\mu_i^\alpha$
at the boundary $\partial D$, which determine the conversion
of particle species of two phases at the interface.
\footnote{Each particle density is not necessarily continuous
across the boundary, since it is only defined in each phase, while
densities of leptons are well defined over the whole space. When
particles of the same species $i$ are allocated in both domains
and the conversion of particle species becomes trivial, we must
further impose the relations, $\mu_i^{\rm I}=\mu_i^{\rm II}$, and
$\rho_i^{\rm I}=\rho_i^{\rm II}$ at the boundary.}

Once eq.~(\ref{eom}) is satisfied, pressure becomes constant in
each domain,
\begin{eqnarray}\label{pressure}
-P^\alpha v^\alpha&=&\int_{D^\alpha} d{\vec r}
\left\{\epsilon^\alpha_{\rm kin+str} [\rho^\alpha_i
(\vec{r})]-
N_i^{{\rm ch},\alpha}\rho_i^\alpha (\vec{r})V^\alpha
(\vec{r})-\mu_i^\alpha\rho^\alpha_i (\vec{r})\right\}.
\end{eqnarray}
Hence, the  condition of the minimum of $\Omega$ with respect to a
modification of the boundary of arbitrary shape (under the total
volume of the Wigner-Seitz cell being fixed) reads
\begin{equation}\label{gibbs2}
P^{\rm I}=P^{\rm II}+\sigma\frac{d S}{dv^{\rm I}},
\end{equation}
where $S$ is the area of the boundary $\partial D$,  $S=4\pi R^2$,
$v^{\rm I} =4\pi R^3 /3$ for spherical droplet, and  $\frac{d
S}{dv^{\rm I}}=0$ in case of slab. Assuming $a_{\rm S} =const$ we
dropped an extra small contribution to the pressure $\propto
\partial a_{\rm S} /\partial \rho_B$, see further discussion in
Appendix \ref{surf}.

The boundary of the cell does not contribute since all the
densities are continuous quantities at this point.
Eq. (\ref{gibbs2}) is the pressure equilibrium condition between the
two phases.
\footnote{
As was already noted, in a more detailed treatment
of the problem with continuous density
distributions, i.e. in the absence of a sharp boundary, the contribution
of the surface energy is absorbed into $P^\alpha$. Hence $P^{\rm
I}=P^{\rm II}$ in such a more detailed treatment.}

The Debye screening parameter is determined by
the Poisson equation, if one expands the charge density in $\delta
V^\alpha (\vec{r})=V^\alpha (\vec{r})- V^{\alpha}_{\rm ref}$
around a reference value $V^{\alpha}_{\rm ref}$, which is also
gauge dependent. Then eq.~(\ref{qV}) renders
\begin{equation}\label{Pois-lin}
\Delta \delta V^\alpha (\vec{r}) =  4\pi e^2
\rho^{{\rm ch},\alpha} (V^{\alpha}(\vec{r})=V^{\alpha}_{\rm ref}
)+ (\kappa^{\alpha} (V^{\alpha}(\vec{r})=V^{\alpha}_{\rm ref} ))^2
\delta V^{\alpha}(\vec{r})+...,
\end{equation}
with the Debye screening parameter,
\begin{equation}\label{debye}
(\kappa^{\alpha}(V^{\alpha}(\vec{r})=V^{\alpha}_{\rm ref} ))^2 =
4\pi e^2\left[ \frac{\partial\rho^{{\rm ch},\alpha}}{\partial
V}\right]_ {V^{\alpha}(\vec{r})=V^{\alpha}_{\rm ref}} =4\left.\pi
Q_i^{\alpha}Q_j^{\alpha}\frac{\partial\rho_j^{\alpha}}{\partial\mu_i^\alpha}
\right|_{V^{\alpha}(\vec{r})=V^{\alpha}_{\rm ref}},
\end{equation}
where we used eq.~(\ref{gauge}). Then we calculate contribution to
the thermodynamic potential (effective energy) of the cell
up to $O(\delta V^{\alpha}(\vec{r}))^{2}$.
The ``electric field energy'' of the cell (\ref{enden2}) can be written
by way of the Poisson equation (\ref{Pois-lin}) as
\begin{eqnarray}\label{eV}
E_V
=\int_{D^{\rm I}} d{\vec r}\epsilon_{V}^{\rm I} +\int_{D^{\rm
II}}d{\vec r}\epsilon_{V}^{\rm II} =\int_{D^{\rm I}}\frac{(\nabla
V^{\rm I}(\vec{r}))^2}{8\pi e^2}d{\vec r} + \int_{D^{\rm
II}}\frac{(\nabla V^{\rm II}(\vec{r}))^2}{8\pi e^2}d{\vec r},
\end{eqnarray}
that is, in the case of  unscreened distributions, usually called
the Coulomb energy.
Besides the terms given by (\ref{eV}), there
are another contributions arising from effects associated with the
inhomogeneity of the electric potential profile, through implicit
dependence of the particle densities on
$V^{\rm I,II}(\vec{r})$. We will call them ``correlation terms'',
$\omega_{\rm cor}^{\alpha}=\epsilon_{\rm kin+str}^{\alpha}-\mu_i^{\alpha}
\rho_i^{\alpha}.$
Taking $\rho_{i}^{\alpha}$ as function of
$V^{\alpha}(\vec{r})$ we expand $\epsilon_{\rm kin+str}^{\alpha}$
in $\delta V^{\alpha}(\vec{r})$:
\begin{eqnarray}\label{exp3}
&&
\epsilon_{\rm kin+str}^{\alpha}[\rho (\vec{r})]= \epsilon_{\rm
kin+str}^{\alpha}(\rho_i^{\alpha}(V^{\alpha}_{\rm ref}))
\\
&&+\left[ \left(\mu_i^\alpha + N^{{\rm ch},{\alpha}}_i V^{\alpha}
(\vec{r}) \right) \frac{\partial\rho_i^{\alpha}}{\partial V^\alpha
}\right]_{V^{\alpha}(\vec{r})= V^{\alpha}_{\rm ref}}\delta
V^{\alpha}(\vec{r}) \nonumber
\\
&& +\frac{1}{2}\left[ \frac{(\kappa^{\alpha})^2}{4\pi e^2}
+\left(\mu_i^\alpha + N^{{\rm ch},{\alpha}}_i V^{\alpha}(\vec{r})
\right)\frac{\partial^2\rho_i^{\alpha}}{(\partial V^{\alpha})^2}
\right]_{V^{\alpha}(\vec{r})=V^{\alpha}_{\rm ref}} (\delta
V^{\alpha}(\vec{r}))^2 + ...\nonumber
\end{eqnarray}
We used eqs.~(\ref{omeg-chem}), (\ref{other1}), (\ref{matrix}) and
(\ref{matrix1}) in this derivation.
Using the expansion
\begin{eqnarray}
&&-\mu_i^\alpha \rho_i^{\alpha}(\vec{r}) =-[\mu_i^\alpha
\rho_i^{\alpha} ]_{V^{\alpha}_{\rm ref}} -
\left[\mu_i^\alpha\frac{\partial \rho_i^{\alpha} }{\partial
V^{\alpha}} \right]_{V^{\alpha}_{\rm ref}} \delta
V^{\alpha}(\vec{r})\nonumber \\ &&-\frac{1}{2}
\left[\mu_i^\alpha\frac{\partial^2 \rho_i^{\alpha} } {(\partial
V^{\alpha})^2}\right]_{V^{\alpha}_{\rm ref}} (\delta
V^{\alpha}(\vec{r}))^2 +...
\end{eqnarray}
we obtain the corresponding
correlation contribution to the thermodynamic potential
$\Omega_{\rm cor} =\int_{D^{\rm I}} d{\vec r}\omega_{\rm cor}^{\rm
I} +\int_{D^{\rm II}} d{\vec r}\omega_{\rm cor}^{\rm II}$:
\begin{eqnarray}\label{om-cor0}
&&\omega_{\rm cor}^{\alpha}= \epsilon_{\rm
kin+str}^{\alpha}(\rho_i^{\alpha}(V^{\alpha}_{\rm
ref}))-\mu_i^\alpha \rho_i^{\alpha} (V^{\alpha}_{\rm ref})-
\rho^{{\rm ch},\alpha} (V^{\alpha}_{\rm ref} )V^{\alpha}_{\rm ref}
\nonumber \\ &&+\frac{V^{\alpha}_{\rm ref}\Delta
V^{\alpha}(\vec{r})}{4\pi e^2} +\frac{
(\kappa^{\alpha}(V^{\alpha}_{\rm ref }) )^2 (\delta
V^{\alpha}(\vec{r}))^2 }{8\pi e^2 }+...,
\end{eqnarray}
where we also used eqs. (\ref{Pois-lin}) and (\ref{debye}). In
general $V^{\rm I}_{\rm ref }\neq V^{\rm II}_{\rm ref }$ and they
may depend on the droplet size.  Their proper choice should
provide appropriate convergence of the above expansion in $\delta
V (\vec{r})$. Taking $V^{\rm I}_{\rm ref }= V^{\rm II}_{\rm ref
}=V_{\rm ref } = const$
we find
\begin{eqnarray}\label{om-cor}
\omega_{\rm cor}^{\alpha}=\frac{ (\kappa^{\alpha}(V_{\rm ref })
)^2 ( V^{\alpha}(\vec{r}) - V_{\rm ref })^2 }{8\pi e^2 } + const ,
\end{eqnarray}
and one may count the potential from the corresponding
constant value. Here we also took into account that the term
$V^{\alpha}_{\rm ref}\Delta V^{\alpha}(\vec{r})/(4\pi e^2 )$
does not contribute to $\Omega$ according to the boundary
conditions in the droplet center, at the droplet boundary (at zero
surface charge), cf. (\ref{boun}), and at the boundary of the
Wigner-Seitz cell.

\section{Conditions of chemical equilibrium.  Charged particle
densities for H-Q system}\label{Eq}

Here we discuss the chemical equilibrium conditions on the example of
the H-Q
phase transition, although the scheme is quite general,
as one will easily see.

The chemical potentials of charged particles ($e$ and
$u$, $d$, $s$ quarks in our example)
are determined by the equations of
motion (\ref{omeg-chem}). Variation of (\ref{enden2}) over
$\rho_i$ gives extra
\begin{eqnarray}\label{other-em}
\frac{\delta E_V}{\delta \rho_j} =-\frac{Q_i}{e} V (\vec{r})
\end{eqnarray}
contributions to the chemical potentials of charged particles.
Using eq. (\ref{other-em})
and eq.  (\ref{endenq1}) of Appendix \ref{en}, which determines the value
$\epsilon_{\rm kin+str}^{\rm I}$,
we find
\begin{eqnarray}\label{chemexpr}
&&\mu_Q = \mu_e = (3\pi^2 \rho_e )^{1/3}+V^{\rm I}(\vec{r}) ,
\,\,\,\, \nonumber \\ &&\mu_d \simeq \left( 1+\frac{2\alpha_c
}{3\pi} \right)\pi^{2/3}\rho_d^{1/3} +\frac{1}{3} V^{\rm
I}(\vec{r}) \nonumber \\ &&= \left( 1+\frac{2\alpha_c }{3\pi}
\right) p_{{\rm F}d}+\frac{1}{3}V^{\rm I}(\vec{r}), \,\,\,\nonumber \\
&&\mu_u \simeq \left( 1+\frac{2\alpha_c }{3\pi} \right)
p_{{\rm F}u}-\frac{2}{3}V^{\rm I}(\vec{r}), \,\,\,\,\nonumber \\ &&\mu_s
\simeq \left( 1+\frac{2\alpha_c }{3\pi} \right)
p_{{\rm F}s}+\frac{m_s^2}{2p_{{\rm F}s}}+\frac{1}{3} V^{\rm I}(\vec{r}).
\end{eqnarray}
The quark Fermi momenta are $p_{{\rm F}u}=(\pi^2\rho_u )^{1/3}$,
$p_{{\rm F}d}=(\pi^2\rho_d )^{1/3}$ and $p_{{\rm F}s}=(\pi^2\rho_s )^{1/3}$,
$\alpha_c$ is the QCD running coupling constant and $m_s$ is the mass of the
strange quark.
Note again that the chemical potentials are gauge variant before gauge
fixing of $V^{\rm I}(\vec{r})$.

With the help of  eqs (\ref{chemexpr}),
we further obtain
\begin{eqnarray}\label{rhoqch}
\rho_u &\simeq&\frac{1}{\pi^2 }\left( 1-\frac{2\alpha_c}{\pi}
\right) \left[ \mu_u +\frac{2}{3}V^{\rm I}(\vec{r})\right]^3
,\,\,\,\nonumber \\ \rho_d &\simeq& \frac{1}{\pi^2 }\left(
1-\frac{2\alpha_c}{\pi}\right) \left[ \mu_d-\frac{1}{3} V^{\rm
I}(\vec{r})\right]^3 ,\,\,\,\,\nonumber \\ \rho_s
&\simeq&\frac{1}{\pi^2 }\left( 1-\frac{2\alpha_c}{\pi} \right)
\left[ \left(\mu_s -\frac{1}{3}V^{\rm I}(\vec{r})\right)^3 -
\frac{3m_s^2\mu_s }{2}\right] ,\,\,\,\,\nonumber \\ \rho_e
&=&(\mu_e -V^{\rm I}(\vec{r}))^3 /3\pi^2 .
\end{eqnarray}

We use chemical equilibrium conditions for the reactions $u+e
\leftrightarrow s$, $d\leftrightarrow s$, and $n\leftrightarrow
p+e$ in each phase,
\begin{eqnarray}\label{q-c}
\mu_u -\mu_s +\mu_e =0, \,\,\,\, \mu_d =\mu_s, \,\,\,\,
\end{eqnarray}
\begin{eqnarray}\label{h-c}
\mu_n =\mu_p +\mu_e ,
\end{eqnarray}
and the conversion relation at the boundary,
\begin{eqnarray}\label{q-h1}
\mu_B\equiv \mu_n =2\mu_d +\mu_u ,
\end{eqnarray}
which yield relations between quark and nucleon chemical
potentials. \footnote{Other conversion relation $\mu_p =2\mu_u
+\mu_d$ is then automatically satisfied.}
With the help of these conditions
we obtain, cf. \cite{HPS93},
\begin{eqnarray}\label{rhoqch1}
\rho_u &\simeq&\frac{1}{\pi^2 }\left( 1-\frac{2\alpha_c}{\pi}
\right) \left[ \frac{\mu_B +2(-\mu_e +V^{\rm
I}(\vec{r}))}{3}\right]^3 ,\,\,\,\nonumber \\ \rho_d &\simeq&
\frac{1}{\pi^2 }\left( 1-\frac{2\alpha_c}{\pi}\right) \left[
\frac{\mu_B- (-\mu_e +V^{\rm I}(\vec{r}))}{3}\right]^3
,\,\,\,\,\nonumber \\ \rho_s &\simeq&\frac{1}{\pi^2 }\left(
1-\frac{2\alpha_c}{\pi} \right) \left[ \left(\frac{\mu_B -(-\mu_e
+V^{\rm I}(\vec{r}))}{3}\right)^3 - \frac{3m_s^2\mu_s }{2}\right]
,\,\,\,\,\nonumber \\ \rho_e &=&(\mu_e -V^{\rm I}(\vec{r}))^3
/3\pi^2 .
\end{eqnarray}
Since these formulae are invariant under the gauge transformation,
$V^{\rm I}({\bf r})\rightarrow V^{\rm I}({\bf r})-V^0$ and
accordingly $\mu_e\rightarrow \mu_e-V^0$ by way of
Eq.~(\ref{gauge}), we must fix the gauge to give a definite
meaning of $\mu_e$. In the {\bf{way I}} discussed above in sect.
\ref{gcap} one puts $V^0 =0$, and accordingly $\mu_e=\mu_{e,\rm
Gibbs}$.
In the {\bf{way II}} one takes $V^0 =-\mu_{e, \rm Gibbs}$, and
$\mu_e =0$ in this gauge.
The expression for the electron density then renders $\rho_e =-(V^{\rm
I}(\vec{r}))^3 /3\pi^2$, cf. \cite{MPV77}. Thus, as we saw in
sect. \ref{gcap},  $V^{\rm I}(\vec{r})=-\mu^{\rm I}_{e,\rm
loc}(\vec{r})$ in this gauge. Note that these different gauge
choices are reflected in the boundary conditions for $V(\vec{r})$:
$V^{\rm II}(\vec{r})\rightarrow V^0$ as $\vec{r}$ approaches the
boundary of the Wigner-Seitz cell, for the case $f^{\rm I}
\rightarrow 0$ (single droplet).

For the quark matter with the number of flavors $N_f =3$ that we study,
we obtain from eqs (\ref{qV}), (\ref{rhoqch}), cf. \cite{HPS93},
\begin{eqnarray}\label{qrhoch}
&&\rho_{\rm ch }^{\rm I}\simeq \left( 1-\frac{2\alpha_c}{\pi}
\right)\nonumber \\ &&\times\left[ \frac{2\mu_B^2 (V^{\rm
I}(\vec{r})-\mu_e)}{9\pi^2 } \left( 1+O\left(\frac{(V^{\rm
I}(\vec{r})-\mu_e)}{\mu_B}
\right) \right)+ \frac{\mu_B  m_s^2 }{6\pi^2 } \right]
\end{eqnarray}
with the help of conditions (\ref{q-h1}).
The  electron contribution ($\propto (V^{\rm I}(\vec{r})
-\mu_e)^3$) is numerically rather small  compared to the quark
contribution (for $-V^{\rm I}(\vec{r})+\mu_e \lsim m_{\pi}$ of our
interest) and can be neglected in the description of the quark
phase. This means that the charge screening occurs mainly owing to
the charged quarks rather than the electrons. Since $-V^{\rm
I}(\vec{r}) +\mu_e \ll \mu_B$ we will also  omit small $(V^{\rm
I}(\vec{r}) -\mu_e)^2$ contributions.
Besides, $m_s\sim m_\pi$ and the second term in squared brackets
is essentially smaller than the first one. Numerically $-V^{\rm
I}(\vec{r}) +\mu_e\sim m_\pi$ and the $m_s^2$ term gives a correction
of the order of $(V^{\rm I}(\vec{r}) -\mu_e)^2$. Nevertheless, it
is technically not difficult to keep it. Therefore, for generality
we will retain below both mentioned terms, the linear in $V^{\rm
I}(\vec{r})-\mu_e$ term and the constant one.

The quark phases with $N_f =2$ can be also realized in neutron
star interiors in some models, cf. \cite{SS00}. Therefore, we
also present the charged quark density
for $N_f =2$:
\begin{eqnarray}\label{qrhoch-2}
&&\rho_{\rm ch }^{\rm I}\simeq \left( 1-\frac{2\alpha_c}{\pi}
\right)\nonumber \\ &&\times\left[ \frac{5\mu_B^2 (V^{\rm
I}(\vec{r})-\mu_e) }{27\pi^2 } \left( 1+
O\left( \frac{(V^{\rm I}(\vec{r}) -\mu_e)}{\mu_n} \right)\right)+
\frac{\mu_B^3}{81\pi^2 }
\right] .
\end{eqnarray}
Here, the constant term is numerically not much smaller than the
linear term in $V^{\rm I}(\vec{r}) -\mu_e $.

For phase II, expanding the charge density $\rho_{\rm ch}^{ \rm
II}(\vec{r})$ around a reference value,
\begin{eqnarray}\label{r2}
\rho_{\rm ch }^{\rm II}
(\vec{r})\simeq\rho_p (V^{\rm II}(\vec{r})=V_{\rm ref}^{\rm II} )
+\delta \rho_p (\vec{r})-\rho_e(V^{\rm II}(\vec{r})=V_{\rm
ref}^{\rm II})-\delta\rho_e (\vec{r}),
\end{eqnarray}
and using
eqs.~(\ref{matrix}), (\ref{matrix1}), (\ref{debye}) and
eq. (\ref{eVh11}) of Appendix \ref{en} we find up to linear order
\begin{eqnarray}\label{co}
&&\delta \rho_p (\vec{r})\simeq C_0^{-1}(V^{\rm II}(\vec{r})
-V_{\rm ref}^{\rm II}), \,\, \delta\rho_e (\vec{r})
=\frac{(\mu_e-V_{\rm ref}^{\rm II})^2}{\pi^2} (V^{\rm
II}(\vec{r})-V_{\rm ref}^{\rm II}),
\\ &&C_0=\frac{A_{22}}{|A|},
~~~ p_{{\rm F}p}=(3\pi^2\rho_p (V^{\rm II}(\vec{r})=V_{\rm
ref}^{\rm II}))^{1/3},\nonumber
\end{eqnarray}
where $A_{22}$, and $\mid A\mid$ are the corresponding matrix
element and the determinant of the matrix (\ref{a}).

Note that, at first glance, non-linear corrections to $\delta
\rho_p$ and $\delta \rho_e$ are not too small, $\delta
\rho_e^{(2)} (\vec{r})= \pi^{-2}(V_{\rm ref}^{\rm II}-\mu_e )
(V^{\rm II}(\vec{r}) -V_{\rm ref}^{\rm II})^2$ and one may expect
$\delta \rho_p^{(2)} (\vec{r})\sim m_{\pi} (V^{\rm II}(\vec{r})
-V_{\rm ref}^{\rm II})^2$. Nevertheless, these terms lead to
rather small corrections of the resulting solutions $V(\vec{r})$.
The boundary conditions put additional limitations on values of
corrections, that is often used in variational procedures and we
have more, approximate solutions of the Poisson equation. Finally, the
correction to the resulting electric field profile due to the dropped
nonlinear terms to $\delta \rho_p (r)$ and $\delta\rho_e (r)$ is
proven to be on a ten percent level, cf. also \cite{MPV77},
where the correction to $\delta\rho_e (r)$  in an analogous screening
problem was analytically evaluated resulting in only 1.5 \% correction
to $V$. We discuss the latter example
in Appendix \ref{example}. Explicit expression for the factor $C_0$
is derived in Appendix \ref{en}.

\section{Spherical geometry}\label{sph}

\subsection{Configurations of the electric field}

It is usually assumed that the thermodynamic potential should be invariant
under the interchange of
all the quantities related to the phases
${\rm I}\leftrightarrow {\rm II}$, e.g.
$f^{\rm I}\leftrightarrow f^{\rm II}=1-f^{\rm I}$.
Then
one would deal with the quark droplets
in the hadron surrounding, whether $f^{\rm I}<1/2$, and with the
hadron droplets in the quark surrounding, if $f^{\rm I}>1/2$.
Even in case of unscreened
configurations of a small size
(in the Coulomb limit) this symmetry can hold
only approximately.
As we shall see, in the general case our equations are not invariant under
these replacements and
in order to conclude which structures are realized
for intermediate values of $f^{\rm I}$ in the vicinity of $f^{\rm I}\simeq 1/2$
one should compare different variants.
We postpone this study to the next paper.
At least, for $f^{\rm
I}\ll 1$ we deal with the spherical
quark droplets embedded into the hadron surrounding and, at least, for
$1-f^{\rm I} \ll 1$, with the spherical
hadron droplets embedded into the quark surrounding.
Therefore, let us start with consideration of the
spherical quark droplets of radius $R$ immersed in the
hadron surrounding. Increasing $f^{\rm
I}$ up to $1/2$
we will check then the possibility of the possible transition from the
structure of the quark droplets
within the hadron phase
to the structure of the quark slabs within the same hadron phase.

The Poisson equation (\ref{Pois-lin}) with $\rho_{\rm ch }^{\rm I}$
from (\ref{qrhoch}) describing the electric potential of the quark
droplet  can be solved analytically. For $r<R$, we find
\begin{equation}\label{1-sol}
V^{\rm I}(r)-\mu_e =\frac{V_{0}^{\rm I}}{\kappa^{\rm I} r}{\rm
sh}(\kappa^{\rm I} r)+ U_{0}^{\rm I},
\end{equation}
with an arbitrary constant $V_{0}^{\rm I}$. For  the Debye
parameter $\kappa^{\rm I}$ and for the constant $U_{0}^{\rm I}$ we
obtain:
\begin{eqnarray}\label{qscrl}
(\kappa^{\rm I})^2 =\frac{8e^2 \mu_{B}^2}{9\pi}\left( 1-
\frac{2\alpha_c}{\pi}
\right),\,\,\,\, U_{0}^{\rm I} \simeq -\frac{3 m_s^2}{4\mu_B}.
\end{eqnarray}
For $N_f =2$ we would have
\begin{eqnarray}\label{qscrl-2}
(\kappa^{\rm I})^2 =\frac{20 e^2 \mu_B^2} {27\pi}\left( 1-
\frac{2\alpha_c}{\pi} \right),\,\,\,\, U_{0}^{\rm I}
=-\frac{\mu_B}{15}.
\end{eqnarray}
Thus, the value $U_{0}^{\rm I}$ is rather small, especially for
$N_f =3$, and the main contribution to $V^{\rm I}(\vec{r})$ comes
from the first term in (\ref{1-sol}). Note that solution
(\ref{1-sol}) is independent of the reference value $V_{\rm
ref}^{\rm I}$ in this case, cf. (\ref{debye}), since $\rho_{\rm
ch}^{\rm I}$ in (\ref{qrhoch}) is  the linear function of $V^{\rm
I}(\vec{r})-\mu_e$ in the approximation used.

Charge in the sphere of a radius $r<R$ is given by
\begin{eqnarray}
Q(r) =V_{0}^{\rm I}~r \mbox{ch}(\kappa^{\rm I}r)\left(
1-\frac{\mbox{th}(\kappa^{\rm I}r)}{\kappa^{\rm I}r}\right)<0 ,
\end{eqnarray}
being, thereby, negative, since
$U_0^{\rm II}>U_0^{\rm I}$ and $V_{0}^{\rm I}<0$. This negative
charge is completely screened by the positive charge induced in the
region $R<r\leq R_{\rm W}$.

For $r>R$, the Poisson equation with the boundary condition
$V^{\prime}|_{R_{\rm W}}=0$  yields
\begin{eqnarray}\label{psi}
&&V^{\rm II}(r)-\mu_e = V_{0}^{\rm II} \frac{R}{r}
\mbox{ch}\left(\kappa^{\rm II}(r-R_{\rm W} )\right) \left(
1-\delta\right)+U_0^{\rm II}, \\ &&\delta
=\mbox{th}\left(\kappa^{\rm II}(R_{\rm W} -r)\right) /
(\kappa^{\rm II}R_{\rm W} ), \nonumber
\end{eqnarray}
with an arbitrary constant $V_0^{\rm II}$, where the constant
$U_0^{\rm II}$ is given by
\begin{equation}
U_0^{\rm II}+\mu_e=-\frac{4\pi e^2\rho_{\rm ch}^{\rm II}(V^{\rm
II}= V_{\rm ref}^{\rm II})} {(\kappa^{\rm II})^2}+V_{\rm ref}^{\rm
II}.
\end{equation}

We take the reference value $V_{\rm ref}^{\rm II}=V^{\rm bulk}$,
where
$V^{\rm bulk}$ is a constant bulk solution of the Poisson equation,
$(V^{\rm bulk}-\mu_e )^3 \equiv -\mu_{e,\rm Gibbs}^3=-3\pi^2
\rho_p (V^{\rm bulk}-\mu_e =-\mu_{e, \rm Gibbs} )$,
that coincides with the local charge-neutrality condition for the
case of the spatially homogeneous matter.
Note that $V^{\rm bulk}$ is
a gauge dependent quantity, $V^{\rm bulk}\rightarrow V^{\rm
bulk}-V^0$ at the transformation $V^{\rm II}(r)\rightarrow V^{\rm
II}(r)-V^0$: $\mu_e=\mu_{e, \rm Gibbs}, V^{\rm bulk}=0$ for
$V^0=0$ and $\mu_e=0, V^{\rm bulk}=-\mu_{e, \rm Gibbs}$ for
$V^0=-\mu_{e,\rm Gibbs}$.

The charge screening in the external region is determined by the
Debye parameter
\begin{eqnarray}\label{hscrl}
(\kappa^{\rm II})^2 =\frac{4e^2 (\mu_{e}-V_{\rm ref}^{\rm
II})^2}{\pi} +\frac{4e^2 \pi}{C_0},
\end{eqnarray}
where the second term is the contribution of the proton screening.
Taking $\rho_{B}^{\rm II}=1.5 \rho_0$,   $\mu_{e,\rm Gibbs} \simeq
170$~MeV, $\mu_{B}=\mu_n \simeq 1020$~MeV, $\alpha_c \simeq 0.4$,
we estimate typical Debye screening lengths as
$\lambda_{\rm D}^{\rm I}\equiv 1/\kappa^{\rm I}\simeq 3.4/m_\pi$, and
$\lambda_{\rm D}^{\rm II}\equiv 1/\kappa^{\rm II} \simeq 4.2 /m_\pi$,
whereas one would have $\lambda_{\rm D}^{\rm II}\simeq 8.5 /m_\pi$, if
the proton contribution to the screening (\ref{hscrl})  was absent
($C_0^{-1}=0$). With the estimate $\lambda_{\rm D}^{\rm II}
\simeq 4.2 /m_\pi$ we get that $\kappa^{\rm II}R_{\rm W}>1$
for the droplets with the radii $R>(2f^{\rm I})^{1/3}\cdot 3.3/m_\pi$.

\subsubsection{Realistic case $\kappa^{\rm II}R_{\rm W}>1$}

In this case  eqs (\ref{1-sol}),  (\ref{psi}) can be represented as
\begin{equation}\label{1-sol-1}
V^{\rm I}(r)-\mu_e =\frac{\widetilde{V}_{0}^{\rm I}}{\kappa^{\rm
I} r}{\rm sh}(\kappa^{\rm I} r)+ U_{0}^{\rm I},
\end{equation}
\begin{eqnarray}\label{psi-cor}
V^{\rm II}(r)-\mu_e = \widetilde{V}_{0}^{\rm II} \frac{R}{r}
\mbox{ch}\left(\kappa^{\rm II}(r-\widetilde{R}_{\rm W} )\right)
+U_0^{\rm II},
\end{eqnarray}
where $\widetilde{V}_{0}^{\rm I, II}$ are arbitrary constants and
$$\widetilde{R}_{\rm W} = R_{\rm W}\left( 1-\frac{1}{\kappa^{\rm
II}R_{\rm W}} \mbox{arcth}\left(\frac{1}{\kappa^{\rm
II}R_{\rm W}}\right)\right) .$$
Matching conditions yield
\begin{eqnarray}\label{const1}
\widetilde{V}_{0}^{\rm I} \simeq \frac{ \left(U_0^{\rm
II}-U_0^{\rm I} \right)\left( 1+
\alpha_0 \xi \mbox{th} (\widetilde{\alpha}_1 \xi) \right)}{
\mbox{ch} \xi \left( \alpha_0 \mbox{th}\xi \cdot
\mbox{th}(\widetilde{\alpha}_1 \xi )+1 \right)
},
\end{eqnarray}
\begin{eqnarray}\label{const2}
\widetilde{V}_{0}^{\rm II}\simeq -\frac{ \left(U_0^{\rm
II}-U_0^{\rm I} \right)
\left( 1-\frac{1}{\xi}\mbox{th}\xi \right)} {\mbox{ch}
(\widetilde{\alpha}_1 \xi ) \left( \alpha_0 \mbox{th}\xi \cdot
\mbox{th}(\widetilde{\alpha}_1 \xi )+1 \right)},
\end{eqnarray}
where we introduced  notations
\begin{eqnarray}\label{param}
\alpha_0 = \frac{\kappa^{\rm II}}{\kappa^{\rm I}} =
\frac{\lambda_{\rm D}^{\rm I}}{ \lambda_{\rm D}^{\rm II}},\,\,\,
 \xi =\kappa^{\rm I}R , \,\,\,\, \alpha_1
=\frac{\alpha_0 (1-(f^{\rm I})^{1/3})}{(f^{\rm I})^{1/3}},
\,\,\,\,(f^{\rm I})^{1/3}=\frac{R}{R_{\rm W}},
\end{eqnarray}
and
\begin{eqnarray}\label{al1t}
\widetilde{\alpha}_1 =\alpha_1
-\frac{1}{\xi}\mbox{arcth}\left(\frac{(f^{\rm I})^{1/3}}{\alpha_0\xi}\right)
.
\end{eqnarray}

\paragraph{Tiny quark fraction volume.}

Let us consider the limiting case of a tiny
quark fraction volume $f^{\rm I}\rightarrow 0$. Then,  a given
quark droplet does not feel the presence of other droplets which are
placed at very large distances from it. The limiting expressions are
recovered, if one takes  $R_{\rm W} \rightarrow \infty$. Then,
solutions (\ref{1-sol}), (\ref{psi}) - (\ref{const2}) acquire
simple explicit forms
\begin{eqnarray}\label{v1}
V^{\rm I} (r) -\mu_e =U_0^{\rm I}+(U_{0}^{\rm II} -U_{0}^{\rm I})
\frac{ \left( 1+\kappa^{\rm II}R \right)}{\alpha_0
\mbox{sh}(\kappa^{\rm I}R)+\mbox{ch}(\kappa^{\rm I}R)
}\frac{\mbox{sh}(\kappa^{\rm I}r)}{\kappa^{\rm I}r},
\end{eqnarray}
\begin{eqnarray}\label{const12}
V^{\rm II} (r)-\mu_e=U_{0}^{\rm II}-(U_{0}^{\rm II} -U_{0}^{\rm
I}) \frac{ \left[1- \frac{\mbox{th}(\kappa^{\rm I}R)}{\kappa^{\rm
I}R}\right]R}{( 1+\alpha_0 \mbox{th}(\kappa^{\rm I}R))r }
\mbox{exp}\left( -\kappa^{\rm II}(r-R)\right) .
\end{eqnarray}

\subsubsection{Less realistic case $\kappa^{\rm II}R_{\rm W}<1$}

In this case  eqs (\ref{1-sol}),  (\ref{psi}) can be represented as
\begin{equation}\label{1-sol-11}
V^{\rm I}(r)-\mu_e =\frac{\widetilde{V}_{0}^{\rm I}}{\kappa^{\rm
I} r}{\rm sh}(\kappa^{\rm I} r)+ U_{0}^{\rm I},
\end{equation}
\begin{eqnarray}\label{psi-cor1}
V^{\rm II}(r)-\mu_e = \widetilde{V}_{0}^{\rm II} \frac{R}{r}
\mbox{sh}\left(\kappa^{\rm II}(r-\widetilde{R}_{\rm W} )\right)
+U_0^{\rm II},
\end{eqnarray}
where $\widetilde{V}_{0}^{\rm I, II}$ are again arbitrary
constants and
$$\widetilde{R}_{\rm W} = R_{\rm W}\left( 1-\frac{1}{\kappa^{\rm
II}R_{\rm W}} \mbox{arcth}\left(\kappa^{\rm
II}R_{\rm W}\right)\right) .$$
Matching
conditions yield
\begin{eqnarray}\label{const1-1cor11}
\widetilde{V}_{0}^{\rm I} \simeq \frac{ \left(U_0^{\rm
II}-U_0^{\rm I} \right)\left( 1+
\alpha_0 \xi \mbox{cth} (\widetilde{\alpha}_1 \xi) \right)}{
\mbox{ch} \xi \left( \alpha_0 \mbox{th}\xi \cdot
\mbox{cth}(\widetilde{\alpha}_1 \xi )+1 \right)
},
\end{eqnarray}
\begin{eqnarray}\label{const2-2cor}
\widetilde{V}_{0}^{\rm II}\simeq -\frac{ \left(U_0^{\rm
II}-U_0^{\rm I} \right)
\left( 1-\frac{1}{\xi}\mbox{th}\xi \right)} {\mbox{sh}
(\widetilde{\alpha}_1 \xi ) \left( \alpha_0 \mbox{th}\xi \cdot
\mbox{cth}(\widetilde{\alpha}_1 \xi )+1 \right)},
\end{eqnarray}
where now
\begin{eqnarray}\label{al1nre}
\widetilde{\alpha}_1 =\alpha_1
-\frac{1}{\xi}\mbox{arcth}\left(\frac{\alpha_0
\xi}{(f^{\rm I})^{1/3}}\right).
\end{eqnarray}
At finite $f^{\rm I}$
the Coulomb limit
is recovered at $\alpha_0 \xi \ll (f^{\rm I})^{1/3}$
with the help
of equations (\ref{1-sol-11}) - (\ref{al1nre}), corresponding to
the case $\kappa^{\rm II}R_{\rm W}<1$.
However this limit is never realized for the
realistic parameter choice, as we argue below.

\subsection{Effects of charge inhomogeneity in energy and
thermodynamic potential }

\subsubsection{Realistic case $\kappa^{\rm II}R_{\rm W}>1$}

\paragraph{``Electric field  energy''.}

We will calculate the contribution to
the thermodynamic potential (effective energy) of the Wigner-Seitz
cell per droplet volume. Let us start with the proper ``electric
field energy'' term
$\widetilde{\epsilon}_{V}=\widetilde{\epsilon}_{V}^{\,\rm I}
+\widetilde{\epsilon}_{V}^{\,\rm II}$:
\begin{eqnarray}\label{emV}
 \widetilde{\epsilon}_{V}^{\,\rm I}
  &=& \frac{3}{4\pi R^3}\int_{0}^R
        \frac{(\nabla V^{\rm I}(\vec{r}))^2}{8\pi e^2}4\pi r^2 dr\nonumber\\
  &\simeq& \beta_0
        \frac{\left(1+\alpha_0 \xi \mbox{th}(\widetilde{\alpha}_1 \xi )
\right)^2
        \left(-\frac{1}{\xi}\mbox{th}^2 \xi
        +\frac{\xi}{2}-\frac{\xi}{2}\mbox{th}^2 \xi
        +\frac{1}{2}\mbox{th} \xi\right)}{\xi^3
        \left( \alpha_0\mbox{th}\xi \cdot
          \mbox{th}(\widetilde{\alpha}_1 \xi ) +1\right)^2},
\end{eqnarray}
being expressed in dimensionless units (\ref{param}) and
\begin{eqnarray}\label{param1}
\beta_0 = \frac{3\left(U_0^{\rm II}-U_0^{\rm I} \right)^2
(\kappa^{\rm I})^2}{8\pi e^2},
\end{eqnarray}
where we used eqs.~(\ref{eV}), (\ref{1-sol}), (\ref{const1}). With
the help of  eqs.~(\ref{psi}), (\ref{const2}),
from eq. (\ref{eV}) we find
\begin{eqnarray}\label{ebV}
&&\widetilde{\epsilon}_{V}^{\,\rm II}= \frac{3}{4\pi
R^3}\int_{R}^{R_{\rm W}}
        \frac{(\nabla V^{\rm II}(\vec{r})
)^2}{8\pi e^2}4\pi r^2 dr \nonumber\\
 &&\simeq \beta_0
\frac{\left(1-\frac{1}{\xi}\mbox{th} \xi \right)^2
\left(1-\frac{1}{2}\alpha_0 \alpha_1 \xi^2(1-\mbox{th}^2
(\widetilde{\alpha}_1 \xi )) +\frac{1}{2}\alpha_0 \xi\mbox{th}
(\widetilde{\alpha}_1 \xi )\right) }{\xi^2 \left( \alpha_0
\mbox{th}\xi \cdot \mbox{th}(\widetilde{\alpha}_1 \xi )
+1\right)^2}\nonumber\\
&&-\beta_0 \frac{ \alpha_0^2 \left(1-\frac{1}{\xi}\mbox{th} \xi
\right)^2 \left(1-\mbox{th}^2 (\widetilde{\alpha}_1 \xi )\right)}
{ 2(f^{\rm I})^{1/3}\left( \alpha_0^2 \xi^2 (f^{\rm I})^{-2/3} -1 \right)
\left(
\alpha_0 \mbox{th}\xi \cdot \mbox{th}(\widetilde{\alpha}_1 \xi )
+1\right)^2 } .
\end{eqnarray}
As above, limiting expressions for the case of a tiny quark fraction
volume $f^{\rm I}\rightarrow 0$ are reproduced with the help of the replacement
$R_{\rm W} \rightarrow \infty$.

\paragraph{``Correlation terms'' to the thermodynamic potential.}

Recalling
that the solution $V^{\rm I}$ is almost independent of the
reference value $V_{\rm ref}^{\rm I}$, we choose $V_{\rm ref}^{\rm
I} =V_{\rm ref}^{\rm II}=
V^{\rm bulk}$ in order  to explicitly calculate correlation
terms; then $\delta V^{\rm I}(r) \equiv V^{\rm I}(r)-V^{\rm bulk}$
for $r<R$. Averaging (\ref{om-cor}) over the droplet volume, with
the help of (\ref{1-sol}), (\ref{const1}), we obtain
\begin{eqnarray}\label{emVc}
&&\widetilde{\omega}_{\rm cor}^{\rm I} = \frac{3}{4\pi
R^3}\int_{0}^R 4\pi r^2 dr\omega^{\rm I}_{\rm cor} \simeq
\frac{\beta_0}{2\xi^3 } \frac{ \left( 1+ \alpha_0 \xi
\mbox{th}(\widetilde{\alpha}_1 \xi ) \right)^2 }{ \left(  \alpha_0
\mbox{th}\xi \cdot \mbox{th}(\widetilde{\alpha}_1 \xi ) +1\right)^2
}\left(\mbox{th}\xi -\frac{\xi}{\mbox{ch}^2 \xi} \right)\nonumber
\\ && +\frac{2\beta_0}{\xi^3 }\frac{ \left( 1+ \alpha_0 \xi
\mbox{th}(\widetilde{\alpha}_1 \xi ) \right) }{\left(  \alpha_0 \mbox{th}\xi
\cdot \mbox{th}(\widetilde{\alpha}_1 \xi ) +1\right)}(\mbox{th}\xi -\xi )
+\frac{\beta_0}{3 } .
\end{eqnarray}
In the hadron phase introducing $\delta V^{\rm II}(r)=V^{\rm
II}(r)-V^{\rm bulk}$, and using eqs.~(\ref{om-cor}),
(\ref{psi}),
(\ref{hscrl}), (\ref{const2}),
we find
\begin{eqnarray}\label{ebVc}
&&\widetilde{\omega}_{\rm cor}^{\rm II} =\frac{3}{4\pi
R^3}\int_{R} ^{R_{\rm W}} 4\pi r^2 dr \omega^{\rm II}_{\rm cor}
\simeq  \\ &&\frac{\beta_0 \alpha_0}{2} \frac{ \left(
1-\frac{1}{\xi}\mbox{th} \xi \right)^2 \left[
\frac{1}{\xi}\mbox{th} (\widetilde{\alpha}_1 \xi )+ \alpha_1
(1-\mbox{th}^2 (\widetilde{\alpha}_1 \xi )) \right] }{ \left(
\alpha_0 \mbox{th}\xi \cdot \mbox{th}(\widetilde{\alpha}_1 \xi )
+1\right)^2 }\nonumber \\
&&+\frac{\beta_0 \alpha_0^2}{2} \frac{ \left(
1-\frac{1}{\xi}\mbox{th} \xi \right)^2  (1-\mbox{th}^2
(\widetilde{\alpha}_1 \xi ))  }{(f^{\rm I})^{1/3} \left( \alpha_0
\mbox{th}\xi \cdot \mbox{th}(\widetilde{\alpha}_1 \xi )
+1\right)^2 (\alpha_0^2 \xi^2 (f^{\rm I})^{-2/3}-1 )} .\nonumber
\end{eqnarray}
Without the $\delta$-correction term in (\ref{psi})
we would come to the very same equations but
with $\widetilde{\alpha}_1$ replaced by $\alpha_1$ and
without the last terms in expressions (\ref{ebV}), (\ref{ebVc}), cf.
\cite{VYT01}.\footnote{Please pay attention on two misprints in \cite{VYT01}.
 In eq. (36) factor 3 is missed and in
eq. (39) the term $4/3$ should be replaced to 1}
The latter terms disappear in the limit $f^{\rm I}
\rightarrow 0$, $\widetilde{\alpha}_1
\rightarrow \alpha_1$.
Thus, account of the
$\delta$-correction results in a small correction, at least, in both
limiting cases $\alpha_0 \xi (f^{\rm I})^{-1/3} \gg 1$  (small quark fraction)
and $\alpha_0 \xi (f^{\rm I})^{-1/3} -1 \ll 1$.

We could also use other values for
$V^{\rm I}_{\rm ref}$ and $V^{\rm II}_{\rm ref}$, e.g. we have
checked that using general eq. (\ref{om-cor0}) with $V^{\rm
I}_{\rm ref}= V^{\rm I}(0)$ and $V^{\rm II}_{\rm ref}= V^{\rm II}(
R_{\rm W})$ leads to the very same result. These two choices are
very natural although one could also select $V^{\rm I, II}_{\rm
ref}$ differently varying their values within interval $(V^{\rm
bulk} ,0)$. Corrections are expected to be of the same order of
magnitude, as those coming from the above dropped non-linear terms in
particle densities.

\subsubsection{Less realistic case, $\kappa^{\rm II}R_{\rm W}<1$}

In this case
eqs (\ref{emV}),  (\ref{ebV}), (\ref{emVc}), (\ref{ebVc}) are rewritten as
\begin{eqnarray}\label{emV-cor1}
 \widetilde{\epsilon}_{V}^{\,\rm I}
  &=& \frac{3}{4\pi R^3}\int_{0}^R
        \frac{(\nabla V^{\rm I}(\vec{r}))^2}{8\pi e^2}4\pi r^2
dr\nonumber\\
  &\simeq& \beta_0
        \frac{\left(1+\alpha_0 \xi \mbox{cth}(\widetilde{\alpha}_1 \xi )
\right)^2
        \left(-\frac{1}{\xi}\mbox{th}^2 \xi
        +\frac{\xi}{2}-\frac{\xi}{2}\mbox{th}^2 \xi
        +\frac{1}{2}\mbox{th} \xi\right)}{\xi^3
        \left( \alpha_0\mbox{th}\xi \cdot
          \mbox{cth}(\widetilde{\alpha}_1 \xi ) +1\right)^2},
\end{eqnarray}
\begin{eqnarray}\label{ebV-cor1}
&&\widetilde{\epsilon}_{V}^{\,\rm II}= \frac{3}{4\pi
R^3}\int_{R}^{R_{\rm W}}
        \frac{(\nabla V^{\rm II}(\vec{r})
)^2}{8\pi e^2}4\pi r^2 dr \nonumber\\
 &&\simeq \beta_0
\frac{\left(1-\frac{1}{\xi}\mbox{th} \xi \right)^2
\left(1-\frac{1}{2}\alpha_0 \alpha_1 \xi^2(1-\mbox{cth}^2
(\widetilde{\alpha}_1 \xi )) +\frac{1}{2}\alpha_0 \xi\mbox{cth}
(\widetilde{\alpha}_1 \xi )\right) }{\xi^2 \left( \alpha_0
\mbox{th}\xi \cdot \mbox{cth}(\widetilde{\alpha}_1 \xi )
+1\right)^2}\nonumber\\
&&+\beta_0 \frac{ \alpha_0^2 \left(1-\frac{1}{\xi}\mbox{th} \xi
\right)^2 \left(1-\mbox{cth}^2 (\widetilde{\alpha}_1 \xi )\right)}
{ 2(f^{\rm I})^{1/3}\left( 1-\alpha_0^2 \xi^2 (f^{\rm I})^{-2/3}
\right) \left(
\alpha_0 \mbox{th}\xi \cdot \mbox{cth}(\widetilde{\alpha}_1 \xi )
+1\right)^2 } ,
\end{eqnarray}
for the ``electric field energy'' terms and
\begin{eqnarray}\label{emVc-cor1}
&&\widetilde{\omega}_{\rm cor}^{\rm I} = \frac{3}{4\pi
R^3}\int_{0}^R 4\pi r^2 dr\omega^{\rm I}_{\rm cor} \simeq
\frac{\beta_0}{2\xi^3 } \frac{ \left( 1+ \alpha_0 \xi
\mbox{cth}(\widetilde{\alpha}_1 \xi ) \right)^2 }{ \left( \alpha_0
\mbox{th}\xi \cdot \mbox{cth}(\widetilde{\alpha}_1 \xi )
+1\right)^2 }\left(\mbox{th}\xi -\frac{\xi}{\mbox{ch}^2 \xi}
\right)\nonumber
\\ && +\frac{2\beta_0}{\xi^3 }\frac{ \left( 1+ \alpha_0 \xi
\mbox{cth}(\widetilde{\alpha}_1 \xi ) \right) }{\left(  \alpha_0
\mbox{th}\xi \cdot \mbox{cth}(\widetilde{\alpha}_1 \xi )
+1\right)}(\mbox{th}\xi -\xi ) +\frac{\beta_0}{3 } ,
\end{eqnarray}
\begin{eqnarray}\label{ebVc-corr1}
&&\widetilde{\omega}_{\rm cor}^{\rm II} =\frac{3}{4\pi
R^3}\int_{R} ^{R_{\rm W}} 4\pi r^2 dr \omega^{\rm II}_{\rm cor}
\simeq  \\ &&\frac{\beta_0 \alpha_0}{2} \frac{ \left(
1-\frac{1}{\xi}\mbox{th} \xi \right)^2 \left[
\frac{1}{\xi}\mbox{cth} (\widetilde{\alpha}_1 \xi )+ \alpha_1
(1-\mbox{cth}^2 (\widetilde{\alpha}_1 \xi )) \right] }{ \left(
\alpha_0 \mbox{th}\xi \cdot \mbox{cth}(\widetilde{\alpha}_1 \xi )
+1\right)^2 }\nonumber \\
&&-\frac{\beta_0 \alpha_0^2}{2} \frac{ \left(
1-\frac{1}{\xi}\mbox{th} \xi \right)^2  (1-\mbox{cth}^2
(\widetilde{\alpha}_1 \xi ))  }{(f^{\rm I})^{1/3} \left( \alpha_0
\mbox{th}\xi \cdot \mbox{cth}(\widetilde{\alpha}_1 \xi )
+1\right)^2 (1-\alpha_0^2 \xi^2 (f^{\rm I})^{-2/3} )} ,\nonumber
\end{eqnarray}
for the correlation terms to the thermodynamic potential.

\subsubsection{Surface energy}

In our dimensionless units the total quark
plus hadron surface contribution to the energy per droplet volume
renders
\begin{eqnarray}\label{esV}
\widetilde{\epsilon}_{\rm S}/\beta_0 =\beta_1 /\xi ,\,\,\,\,
\beta_1 =3\kappa^{\rm I}\sigma /\beta_0 ,
\end{eqnarray}
see  (\ref{param1}), and we used that $\epsilon_{\rm S} =
3\sigma/R$, cf. eq. (\ref{eps-fin}) of Appendix \ref{surf}.
The coefficients $\beta_0$, $\beta_1$  are evaluated with
the help of  eqs.~(\ref{qscrl}), (\ref{param1}) and (\ref{esV}).
For $N_f =3$,
\begin{eqnarray}\label{b0est}
\beta_0 \simeq (U_{0}^{\rm II} -U_{0}^{\rm I})^2
(1-\frac{2\alpha_c}{\pi}
) \frac{ \mu^2_{B}}{3\pi^2},\,\,\,\,
\end{eqnarray}
\begin{eqnarray}\label{b1est}
\beta_1 \simeq \sqrt{\frac{2}{\pi}}(U_{0}^{\rm II} -U_{0}^{\rm
I})^{-2} \left( 1-\frac{2\alpha_c}{\pi}
\right)^{-1/2}\frac{6\pi^2 e \sigma}{\mu_B}\simeq
\sqrt{\frac{8(1-\frac{2\alpha_c}{\pi} )}{\pi}}\frac{e\sigma
\mu_{B}}{\beta_0}.
\end{eqnarray}
For the typical values $-U_0^{\rm II}\simeq \mu_{e,\rm Gibbs}\simeq 170~$ MeV,
$\mu_{B} \simeq 1020~$ MeV, $\alpha_c \simeq 0.4$ and $m_s \simeq
120\div 150$~MeV
we estimate $\beta_0 \simeq 1.6 m_{\pi}^4$. Thus, with the value
$\sigma \simeq 1.3 m_{\pi}^3$ we obtain $\beta_1 \simeq 0.7 $,
whereas with $\sigma \simeq 10~$ MeV/fm$^2 \simeq 0.14m_{\pi}^3$
we would get $\beta_1 \simeq 0.08$.

Note that all the energy terms
at $\xi =\xi_{\rm min}$, i.e. at the energy minimum point,
are expressed in terms of only four
parameters: $\alpha_0 =\lambda_{\rm D}^{\rm I}
/\lambda_{\rm D}^{\rm II}$, $f^{\rm I}$,
$\beta_0$ and $\beta_1$, and the dependence on the properties
of the concrete system under
consideration is hidden only in the values of these parameters.

\paragraph{Correlation term to the surface energy.}

In the general case, there
is still a correlation term to the surface energy due to a
dependence of $\sigma (n_i (V))$. However this term  is rather
small being proportional to the variation of the surface charge
$\delta Z_{\rm  S} /\delta V$. The physical reason of this smallness
is obvious. Electric fields are changed on  large scales
$(\lambda_{\rm D}^{\rm I,II} , R)$ and cannot substantially affect
surface quantities which are changed at shorter distances ($\sim
d_{\rm S}$). We have omitted a contribution of the surface charge to the
charge distributions and by the same reason we omit the
correlation term to the surface energy.

 \subsubsection{Coulomb limit}

\paragraph{Tiny quark fraction volume.}

Peculiarities of the Coulomb limit at finite values of the quark concentration
$f^{\rm I}$ are discussed in Appendix \ref{Coul-pec}.
For the case of a tiny
quark fraction volume (very small $f^{\rm I}$),
$R_{\rm W} $ is very large. Then one puts
$\rho^{\rm I}_{\rm ch} =const$, $\rho^{\rm II}_{\rm ch} =0$ within  the Coulomb
limit of Appendix \ref{Coul-pec} and recovers the
corresponding expressions.

>From our general equations describing inhomogeneous charge profiles
we reproduce
the Coulomb limit  for the case
of a tiny quark fraction volume, if we first put
$\alpha_1 \rightarrow \infty$, and then expand the terms
$\widetilde{\epsilon}_{V}^{\,\rm
I}+\widetilde{\epsilon}_{V}^{\,\rm II} +\widetilde{\epsilon}_{\rm
S}$ in $\xi \ll 1$. Thus, we should put $\xi \ll 1$, $\alpha_1 \xi
\gg 1$. Then we, indeed, recover the Coulomb plus surface energy
per droplet volume
\begin{eqnarray}\label{CS}
\widetilde{\epsilon}_{{\rm C},\rm S}=\widetilde{\epsilon}_{\rm C}+
\widetilde{\epsilon}_{\rm S} = \beta_0 \left( \frac{1}{45}\xi^2
+\frac{1}{9}\xi^2+ \frac{\beta_1}{\xi} \right) ,
\end{eqnarray}
where  partial contributions correspond to the terms
$\widetilde{\epsilon}_{V}^{\,\rm I}$,
$\widetilde{\epsilon}_{V}^{\,\rm II}$ and
$\widetilde{\epsilon}_{\rm S}$. We needed the Taylor expansion of
functions entering $\widetilde{\epsilon}_{V}^{\,\rm II}$, see
(\ref{ebV}), up to two non-vanishing terms in order to recover the
contribution $\propto \xi^2$ and we needed the Taylor
expansion of functions entering
$\widetilde{\epsilon}_{V}^{\,\rm I}$, see (\ref{emV}),
in $\xi \ll 1$ up to  three non-vanishing terms.

Both the
correlation terms $\widetilde{\omega}_{\rm cor}^{\rm I}\propto
\xi^4$ and $\widetilde{\omega}_{\rm cor}^{\rm II}\propto \xi^3$
can be dropped in the Coulomb limit, for droplets of a tiny size
$\xi \ll 1$. For $\widetilde{\omega}_{\rm cor}^{\rm I}$, see
(\ref{emVc}), the Taylor expansion in small $\xi$ should include four
non-vanishing terms  to recover the term $\propto \xi^4$, and
one needs to do the Taylor expansion of the functions entering
(\ref{ebVc}) for $\widetilde{\omega}_{\rm cor}^{\rm II}$ up to two
non-vanishing terms with the result $\propto \xi^3$.

With the help
of eqs (\ref{qrhoch}), (\ref{qscrl}) the value
$\widetilde{\epsilon}_{\rm C}$ in (\ref{CS}) is easily recasted
in the standard form,
\begin{eqnarray}
\widetilde{\epsilon}_{\rm C}=\frac{4\pi e^2}{5}(\rho_{\rm ch
}^{\rm I })^2 R^2 .
\end{eqnarray}
The function $\widetilde{\epsilon}_{\rm C,S}$ has the minimum at $\xi
=\xi_{\rm min}$, corresponding to the optimal size of the unscreened
droplet:
\begin{eqnarray}\label{CSm}
\widetilde{\epsilon}_{\rm C,S}(\xi_{\rm min} )= 3
\widetilde{\epsilon}_{\rm C}(\xi_{\rm min} )=\left(
\frac{9}{10}\right)^{1/3}\beta_0\beta_1^{2/3},\,\,\,\, \xi_{\rm min}
=\left( \frac{15\beta_1}{4}\right)^{1/3}.
\end{eqnarray}
The Coulomb limit is recovered only for $\xi_{\rm min} \ll 1$, i.e.
$\beta_1 \ll 0.1$, whereas with above estimate
$\beta_1 \gsim 0.1$, cf. (\ref{b1est}),
we always get $\xi_{\rm min} \gsim 1$.
{\em{Thus, we conclude that the pure Coulomb limit is never
realized within mixed phase.}} On the other hand, the
criterion of the tiny quark
fraction volume $\alpha_1 \xi_{\rm min} \gg 1$, which we have used, is
applicable only for $f^{\rm I} \ll \alpha_0^3 \beta_1$.

\paragraph{Finite quark fraction volume ($f^{\rm I}$) in the Coulomb limit.}

As above, we assume $\xi_{\rm min} \sim \beta_1^{1/3} \ll 1$, but now
$\alpha_0 \xi_{\rm min} \ll (f^{\rm I})^{1/3}$ ($0.5\geq f^{\rm I} \gg
\alpha_0^3 \beta_1)$. Then, from our general equations
(\ref{emV-cor1}), (\ref{ebV-cor1})
we find
\begin{eqnarray}\label{emV-cor11 }
 \widetilde{\epsilon}_{V}^{\,\rm I}+\widetilde{\epsilon}_{V}^{\,\rm I}=
\frac{2\beta_0\alpha_0^4
(1-\frac{3}{2}(f^{\rm I})^{1/3}+\frac{1}{2}f^{\rm I})\xi^2}{15(f^{\rm I}
+\alpha_0^2 (1-f^{\rm I}))^2},
\end{eqnarray}
that coincides with the asymptotic form
expression (\ref{Peth-d3})
of Appendix \ref{Coul-pec}  for the Coulomb energy
obtained there with the step-function charge densities.
Again this limiting
case
is not achieved  for realistic parameter choices, since then we
always have $\xi_{\rm min} \gsim 1$
and $\alpha_0 \xi_{\rm min} \gsim 1$.

\subsubsection{Limit of  extended quark droplets}

In the limit $\alpha_1 \xi \gg 1$, $\xi
\gg 1$, corresponding to the  large size droplets, from
(\ref{emV}), (\ref{ebV}), (\ref{emVc}) and (\ref{ebVc})
we find that all the terms contribute to the effective surface
energy density (neglecting the curvature terms $\propto 1/\xi^2$),
\begin{eqnarray}\label{surfel}
&&\widetilde{\epsilon}_{V}^{\,\rm I} \rightarrow
\frac{\beta_0\alpha_0^2}{2(1+\alpha_0 )^2}\frac{1}{\xi}, \,\,\,
\widetilde{\epsilon}_{V}^{\,\rm II} \rightarrow
\frac{\beta_0\alpha_0}{2(1+\alpha_0 )^2}
\frac{1}{\xi},\,\,\,\,\nonumber \\ &&\widetilde{\omega}_{\rm
cor}^{\rm I} \rightarrow \left[
-\frac{2\beta_0\alpha_0}{(1+\alpha_0 )}
+\frac{\beta_0\alpha_0^2}{2(1+\alpha_0 )^2}\right]
\frac{1}{\xi}+\frac{\beta_0}{3} ,\,\,\, \widetilde{\omega}_{\rm
cor}^{\rm II}\rightarrow \frac{\beta_0 \alpha_0}{2 (1+\alpha_0
)^2}\frac{1}{\xi}.
\end{eqnarray}
Thus, $\delta \widetilde{\omega}_{\rm
tot}/\beta_0=(\widetilde{\epsilon}_V + \widetilde{\omega}_{\rm
cor}^{\rm I}+ \widetilde{\omega}_{\rm cor}^{\rm
II}+\widetilde{\epsilon}_S )/\beta_0 =\beta_1^{\rm eff}\xi^{-1}
+const$ with $\beta_1^{\rm eff} =3\kappa^{\rm I} \sigma_{\rm
tot}^{\rm spher} /\beta_0$ and therefore, in the given limiting
case of extended droplets
the electric field
effects can be treated with the help of an effective surface
tension. The full surface tension $\sigma_{\rm tot}^{\rm spher}$
then renders
\begin{eqnarray}\label{suftenV}
 \sigma_{\rm tot}^{\rm spher}=\sigma +\sigma_{ V}
 =\sigma -
 \lambda_{D}^{\rm I}\frac{\beta_0 \alpha_0
 [\alpha_0 +1 ] }{3(1+\alpha_0 )^2} .
\end{eqnarray}
The first $\sigma$ term is the contribution (related to the scale $d_{\rm S}$)
of the strong
interaction. The second negative term is the contribution of
electric field effects (related to the scales $\lambda_{\rm D}^{\rm
I,II}$). It depends largely on the values of
parameters. For  $\mu_{e, {\rm Gibbs}}=170$~ MeV, $\mu_{B}=1020$~MeV, $m_s
=150$~MeV, $\alpha_c =0.4$, we estimate the contribution to the
surface tension from the electric effects as $\sigma_V \simeq -
60$~MeV$/$fm$^2$.
Thus, for
$\sigma <\sigma_c \simeq 60$~MeV$/$fm$^2$ we deal with the mixed phase,
whereas for $\sigma >\sigma_c$, with the Maxwell construction.
Note that only due to the correlation term
$\widetilde{\omega}_{\rm cor}^{\rm I}$ we obtained negative
contribution to $\sigma_V$.
Also note that
the spatial variation of the electric field within the surface layer
of thickness $d_{\rm S}$
may
only slightly affect the ``old'' value $\sigma$ at
$d_{\rm S} \ll \lambda_{\rm D}^{\rm
I,II}$.

The dropped $\sim 1/\xi^2$ terms are important to determine the
size ($\xi_{\rm min}$) of the droplet within the mixed phase.
For very large droplet radius these terms can't compete anymore
with  $\sim 1/\xi$ terms. For such large radii of droplets
the mixed phase already can't exist.
With this limiting case we may describe the charge distribution
within the Maxwell construction.

\subsection{Numerical results}

In Fig. 2 we demonstrate the dependence of the contribution of
the inhomogeneous charge distributions to the total thermodynamic
potential  per droplet volume for the case of the lattice of
spherical droplets,  $\delta \widetilde{\omega}_{\rm
tot}/\beta_0=(\widetilde{\epsilon}_V^{\rm I}+
\widetilde{\epsilon}_V^{\rm II}+ \widetilde{\omega}_{\rm
cor}^{\rm I}+ \widetilde{\omega}_{\rm cor}^{\rm
II}+\widetilde{\epsilon}_{\rm S} )/\beta_0$,  given by the sum of
partial contributions (\ref{emV}), (\ref{ebV}), (\ref{emVc}),
(\ref{ebVc}) and (\ref{esV}), if $\kappa^{\rm II}R_{\rm W}>1$,
and by the sum of (\ref{emV-cor1}), (\ref{ebV-cor1}), (\ref{emVc-cor1}),
(\ref{ebVc-corr1}) and (\ref{esV}),  if $\kappa^{\rm II}R_{\rm W}<1$.
It is presented
as the function of the droplet size
($\xi$ in dimensionless units)  for  two values of the
Wigner-Seitz parameter $f^{\rm I}\equiv f=0.01$ (dashed curves)
and $f =0.5$ (solid curves) at fixed value $\beta_1$ (each panel)
for $\alpha_0 =1$, as the representative example. The curves for
$f=0.01$ deviate only little from those for $f \rightarrow 0$
related to the single droplet.

The curves labeled by ``C'', the Coulomb curves, demonstrate the
contribution in the Coulomb limit:
$\widetilde{\epsilon}_{{\rm C},\rm S} /\beta_0$ is determined by eq.
(\ref{CS}) for $f\rightarrow 0$ and by eqs.
(\ref{Peth-d3}), (\ref{rhcul3}) in general case, dashed curves
stand for $f=0.01$ and
solid ones, for $f=0.5$.
Each ``C'' curve has a
pronounced minimum at $\xi =\xi_{\rm min} \propto \beta_1^{1/3}$.
For $\xi >\xi_{\rm min}$  the ``C'' curve shows quadratic growth
deviating drastically for $\xi \gsim 1$ from the other curves.

The dashed and solid ``e.m.'' curves shown in each panel
demonstrate the contribution
$\widetilde{\epsilon}_{\rm e.m.}/\beta_0\equiv(\widetilde{\epsilon}_V +
\widetilde{\epsilon}_{\rm S})/\beta_0$ for $f =0.01$ and $f =0.5$
respectively,
ignoring correlation terms.
The quantity $\widetilde{\epsilon}_{\rm e.m.}$ for $f\rightarrow 0$ is
the counterpart of that in the Coulomb limit of $\widetilde{\epsilon}_{\rm
C,S}$,  when the charge screening effect is taken into account.
We see that the minima at the "e.m." curves disappear already at
$\beta_1 > 0.03$.

For $\beta_1 \lsim 0.01$, minimum points of the dashed ``C'' curves
$\xi_{\rm min}$ deviate little from the minima of the dashed
``e.m.'' curves.
Only for such small values of $\beta_1$ and $\xi_{\rm
min}$ we recover the Coulomb limit! However, one may obtain such
small values of $\beta_1$ only for tiny values of the surface tension
and very large values of the baryon chemical potential. With
increase of the latter, the Debye screening parameter $\kappa^{\rm
I}$ is also increased and the droplet radius $R_{\rm min} =\xi_{\rm min}
/\kappa^{\rm I}$ is proved to be essentially smaller than
$1/m_{\pi}$.
For larger values of $\beta_1$, the difference between  the minima of
the dashed ``e.m.''  and the dashed ``C'' curves is proved to be more pronounced.
The difference between the minima at the dash ``C'' and dashed curves (
$\delta \widetilde{\omega}_{\rm
tot}/\beta_0$) for the case of a tiny quark fraction volume ($f=0.01$)
is minor for  $\beta_1 \lsim 0.05$ but becomes pronounced for larger values
$\beta_1$, whereas for the solid curves ($f=0.5$) this difference is always
pronounced, even for very small $\beta_1$. The latter is due to an essential
contribution of the correlation effective energy in this case
($f=0.5$), whereas $\rho^{\rm I}_{\rm ch}$ is fitted by
coincidence of the
``C'' and its generic partner the ``e.m.'' curve in the limit $\xi \ll 1$.
In
the framework of the Coulomb limit the value $\rho^{\rm I}_{\rm ch}$
is defined in the variational procedure:
the total energy is assumed to be dependent
on $\rho^{\rm I}_{\rm ch}$ via $f$
and $\rho^{\rm I}_{\rm ch}$
is determined by the minimization of the energy. Then mentioned
discrepancy could be partially diminished. However even by
normalizing  the solid ``C'' curve to coincide  with
the solid curve
at $\xi \ll 1$ (for that please simply do the shift of the minimum at
the solid ``C'' curve
up to the value given by the minimum at the
solid curve in the panel $\beta_1 =0.01$ and use the same value of the shift
considering other panels)
we obtain that the difference between the minima
at the curves (which determine droplet radii)
becomes  pronounced for
$\beta_1 > 0.05$. This difference is more pronounced for $\alpha_0 \neq 1$.

The dashed and solid curves converge to $1/3$ with the increase of
$\xi$. The large $\xi$ asymptotic of the curves (counted from $1/3$)
is $\propto 1/\xi$, being interpreted as the surface energy term,
characterized by a significantly smaller value of the surface
tension (\ref{suftenV}) than that determined only by the strong
interaction. We see that for $\sigma \geq \mid \sigma_V \mid$,
corresponding to $\beta_{1}>\beta_{1{\rm c}}\simeq 0.5$ for $\alpha_0 =1$,
see  Fig.3,
the structured mixed
phase is proved to be prohibited, since a {\em{ necessary
condition}} of its existence (the presence of a minimum in the droplet
size) is not satisfied. Contrary to this in  the Coulomb limit the
{\em{necessary condition}} is always fulfilled. Large-size droplets
are  realized within the mixed phase if $\sigma < \mid \sigma_V
\mid$, $\mid \sigma + \sigma_V \mid \ll \mid \sigma_V \mid$ and
within the Maxwell construction, if $\sigma \geq \mid \sigma_V
\mid$.

The difference between the solid and dashed curves
and the corresponding "e.m" curves shows the important
contribution of the correlation energy in the H-Q
structured mixed phase. This correlation contribution contains two
parts, the model independent constant term ($\beta_0 /3$ in our
case), as follows from (\ref{emVc}), and a model dependent part.
The latter however depends only on three parameters $\beta_0 /3$
and $\alpha_0 =
\lambda_{\rm D}^{\rm I}$,
$\lambda_{\rm D}^{\rm II}$, $f^{\rm I}$. The constant term $\beta_0 /3$
shows a difference in the energies of the Maxwell construction and the
mixed phase, if one neglected finite size effects in the calculation
of the latter.

The minima of the solid ``e.m.'' curves are always below the corresponding
minima of the dashed curves, however for the
$\delta
\widetilde{\omega}_{\rm tot}(\xi_{\rm min})/\beta_0$
at the dashed and solid curves the
situation changes to the opposite one, $\delta
\widetilde{\omega}_{\rm tot}(\xi_{\rm min})/\beta_0$ increases with $f$.
Comparison of the solid and the dashed  curves for $\beta_1 \geq 0.5$
demonstrates only a minor  dependence of the value $\delta
\widetilde{\omega}_{\rm tot}/\beta_0$ on the Wigner-Seitz
parameter $f$ in this case.

Fig. 3  demonstrates the dependence of the droplet radius $\xi_{\rm
min}$ (in dimensionless units) on the value $\beta_1$ for
different ratios of the Debye screening lengths $\alpha_0$, again for
two values of the Wigner-Seitz parameter $f=0.01$ (a tiny quark
concentration ) and $f =0.5$. As seen from the figure, the
dependence of $\xi_{\rm min}$ on the parameter $\alpha_0$ is
rather sharp, whereas it was completely absent for the Coulomb
curves in previous works since there $\rho_{\rm ch}^{\rm I}$ was considered as
a function of $f$ rather than of
$\kappa^{\rm I}$ and $\kappa^{\rm II}$.
The dependence on $f$ is more sharp for $\alpha_0 <1$
and less sharp for $\alpha_0 > 1$. All the curves demonstrate
the presence of a critical parameter set ($\beta_{1c}, \xi_{\rm
min}^{c})$. The values $\beta_{1c}$
are shown by arrows in Fig. 3,
cf. with Fig. 2. Corresponding values $\xi_{\rm min}^{c}$
are rather large demonstrating that even the droplets of
the
size $\gsim 10 \lambda_{\rm D}^{\rm I}$ may still exist within the mixed phase.
This statement disagrees with what was expected in previous works
using the Coulomb solutions. It was thought that the mixed phase should
disappear for $\xi >\lambda_{\rm D}$ due to the screening.

For larger values  of $\beta_1$ ($\beta_1 >\beta_{1c}$)
the
Maxwell construction becomes energetically favorable.
The Coulomb curves, ``C''
show the  value $\xi_{\rm min}$ related
to the unscreened case, as the function of $\beta_1$,
($\xi_{\rm min} (\beta_1 )=(15\beta_1/4)^{1/3}$
for a tiny quark fraction volume). The
{\em{necessary
condition}} for the existence of the mixed phase (existence of
$\xi_{\rm min}$) is always fulfilled in the Coulomb limit, opposite to what
we find in our general consideration.
Comparison of the curves with the ``C'' curves again demonstrates
the importance of the screening effects.

\section{Plane geometry}\label{plane}

{\subsection{Electric field configurations in Wigner-Seitz cell.}}
Without incorporation of screening effects previous papers
argued that with
increasing of the quark fraction volume the lattice of spherical
droplets first undergoes the phase transition transforming to the
structure of rods and then to the structure of slabs, which
eventually become energetically more favorable compared to spherical
droplets and rods at $f^{\rm I}<0.5$, cf. \cite{RPW83,HPS93}. Therefore,
we will
describe these configurations but now including the charge screening
effects. We will concentrate on discussion of slabs, since in
absence of screening effects, as has been shown in mentioned
works, these structures appear already at sufficiently small quark
fraction volume, and also, since in this case we may easily
proceed further with our analytical approach.

Let the quark slab occupies a layer $x\in (-R,R)$, $-\infty <y,z
<+\infty$, being placed in the Wigner-Seitz cell $x\in (-R_{\rm W}
, R_{\rm W})$, $R_{\rm W} \geq R$. We suppose that in the interval
$x\in (-R,R)$ the electric potential well $V$ is  an even function of
$x$. Also we assume that the solutions of eq. (\ref{qV})
should not change under  simultaneous replacements $x\rightarrow
-x$, $R\rightarrow -R$, $R_{\rm W} \rightarrow -R_{\rm W}$. For
the charged density (\ref{qrhoch}), dropping the terms $\sim O(V^2
(x))$, we easily find a solution of (\ref{qV})
\begin{eqnarray}\label{qsol-1}
 V^{\rm I}(x)-\mu_e =V_{0,\rm I}^{(1)}\mbox{ch} (\kappa^{\rm I}x)+
U_{0}^{\rm I}  , \,\,\,-R<x<R.
\end{eqnarray}
Here, by superscript $(1)$ we indicate the flat geometry of the
structure, $V_{0,\rm I}^{(1)}$ is an arbitrary constant and
the values $\kappa^{\rm I}$, and $U_{0}^{\rm I}$ are determined by
eq. (\ref{qscrl}), as in the case of the spherical geometry.

The solution of (\ref{qV}), $V^{\rm II}(x)=V_{0}^{\rm II}+\delta V^{\rm
II}(x)$, $V_{0}^{\rm II}=const$, $\mid \delta V^{\rm II}(x) \mid
\ll \mid V_{0}^{\rm II}\mid$ for $R<x<R_{\rm W}$, with the
boundary condition $V^{\prime}_x (x=R_{\rm W})=0$,  reads
\begin{eqnarray}\label{qsol1-1}
 V^{\rm II}(x)-\mu_e
=V_{0,\rm II}^{(1)}\mbox{ch} (\kappa^{\rm II} (x-R_{\rm W} ))+
U_{0}^{\rm II} , \,\,\,R<x<R_{\rm W} ,
\end{eqnarray}
where $V_{0,\rm II}^{(1)}$ is an arbitrary constant and the values
$\kappa^{\rm II}$ and $U_{0}^{\rm II}$ are determined by the same
equations, as in the case of the spherical geometry. The corresponding solution
(\ref{qV}) for $-R_{\rm W} <x<-R$, with the boundary condition
$V^{\prime}_x (x=-R_{\rm W}) =0$ renders
\begin{eqnarray}\label{qsol1-2}
 V^{\rm II}(x)-\mu_e
= V_{0,\rm II}^{(1)} \mbox{ch} (\kappa^{\rm II} (x+R_{\rm W} ))+
U_{0}^{\rm II} , \,\,\,-R_{\rm W} <x<-R .
\end{eqnarray}
Matching of potentials  and their derivatives at $x=\pm R$ yields
\begin{eqnarray}\label{const1-1}
V_{0,\rm I}^{(1)} =\frac{ (U_{0}^{\rm II}-U_{0}^{\rm I})
\alpha_0}{\alpha_0 \mbox{ch}\xi + \mbox{sh}\xi \mbox{cth}(\alpha_2
\xi ) },
\end{eqnarray}
\begin{eqnarray}\label{const2-1}
V_{0,\rm II}^{(1)} =-\frac{ (U_{0}^{\rm II}-U_{0}^{\rm I})
\mbox{th}\xi}{\alpha_0 \mbox{sh}(\alpha_2 \xi)+\mbox{th}\xi
\mbox{ch}(\alpha_2 \xi) },
\end{eqnarray}
where for one dimensional structures the parameter $\alpha_1$ is
replaced by $\alpha_2 =\alpha_0 (1-f^{\rm I} )/f^{\rm I}$, $f^{\rm
I}=R/R_{\rm W}$.

Boundary conditions for $x=-R, x= -R_{\rm W}$ are automatically
fulfilled. The solution describing a tiny quark fraction volume, when one may
deal with a single slab, is obtained from here in the limit
$R_{\rm W} \rightarrow \infty$.

\subsection{Effects of charge inhomogeneity in energy and
  thermodynamic potential}

Using (\ref{qsol-1}), (\ref{const1-1}), we first find contribution
to the energy per slab from the term $\epsilon_{V,\rm I}^{(1)}$:
\begin{eqnarray}\label{eVo}
\widetilde{\epsilon}_{V,\rm I}^{\,(1)} =\frac{1}{2 R}\int_{-R}^R
\frac{(\nabla V(x))^2}{8\pi e^2}dx = \beta_0 \frac{\mbox{th}^2
(\alpha_2 \xi ) \left(\mbox{th}^2 \xi -1 +\frac{1}{\xi}\mbox{th}
\xi\right) }{6 \left( \alpha_0^{-1} \mbox{th}\xi
+\mbox{th}(\alpha_2 \xi ) \right)^2} .
\end{eqnarray}

Using eqs (\ref{qsol1-1}), (\ref{const2-1}), we then obtain the term
$\epsilon_{V,\rm II}^{(1)}$:
\begin{eqnarray}\label{eVo1}
&&\widetilde{\epsilon}_{V,\rm II}^{\,(1)}
=\frac{1}{R}\int_{R}^{R_{\rm W}} \frac{(\nabla V(x))^2}{8\pi
e^2}dx \nonumber \\
&&= \beta_0 \frac{\mbox{th}^2 \xi \left(\frac{1}{\alpha_0
\xi}\mbox{th}(\alpha_2 \xi ) + \frac{\alpha_2 }{\alpha_0}\left(
\mbox{th}^2 (\alpha_2 \xi ) -1\right) \right) }{6 \left(
\alpha_0^{-1} \mbox{th}\xi +\mbox{th}(\alpha_2 \xi ) \right)^2}.
\end{eqnarray}
For $\widetilde{\omega}_{\rm cor, I}^{(1)}$, with the help of eqs
(\ref{om-cor}), (\ref{qsol-1}) and (\ref{qscrl}) taking $V_{\rm
ref}^{\rm I}=V^{\rm bulk}$ we find
\begin{eqnarray}\label{eVoc}
&&\widetilde{\omega}_{\rm cor, I}^{\,(1)} =\frac{1}{2R}\int_{-R}^R
dx \omega_{\rm cor, I}^{(1)} = \beta_0\frac{\mbox{th}^2 (\alpha_2
\xi ) \left(1-\mbox{th}^2 \xi +\frac{1}{\xi}\mbox{th}
\xi\right)}{6 \left( \alpha_0^{-1}
\mbox{th}\xi  +\mbox{th}(\alpha_2 \xi ) \right)^2} \nonumber \\
&&-\beta_0\frac{2\mbox{th}(\alpha_2 \xi )\mbox{th} \xi }{3\xi
\left( \alpha_0^{-1} \mbox{th}\xi  +\mbox{th}(\alpha_2 \xi )
\right)}+\frac{\beta_0}{3}.
\end{eqnarray}
Using eqs (\ref{om-cor}), (\ref{qsol1-1}),
(\ref{hscrl}) and taking $V_{\rm ref}^{\rm II}=V^{\rm bulk}$, we
obtain
\begin{eqnarray}\label{ecor1}
&&\widetilde{\omega}_{\rm cor, II}^{(1)}=\frac{1}{R}\int_{R}^{R_W}
dx \omega_{\rm cor, II}^{(1)}\\ &&=\beta_0 \frac{\mbox{th}^2 \xi
\left[\frac{1}{ \xi}\mbox{th}(\alpha_2 \xi ) + \alpha_2 \left(
1-\mbox{th}^2 (\alpha_2 \xi ) \right) \right] }{6\alpha_0 \left(
\alpha_0^{-1} \mbox{th}\xi +\mbox{th}(\alpha_2 \xi )
\right)^2}.\nonumber
\end{eqnarray}
The surface energy contribution per slab is given by
\begin{eqnarray}\label{esoc}
\widetilde{\epsilon}_{\rm S}^{(1)}=\frac{\beta_0 \beta_1}{3\xi},
\end{eqnarray}
see eq. (\ref{eps-fin}) of Appendix \ref{surf}.

\paragraph{Tiny quark fraction volume. Single slab.}

Expressions for a single
slab are obtained taking the limit $R_{\rm W} \rightarrow \infty$
($\alpha_2\rightarrow \infty$).

\paragraph{Slab of a tiny transverse size.}

The solution for a
single slab of a tiny transverse size is recovered for $\xi \ll
1$, $\alpha_2 \xi \gg 1$. This limit is never realized for an
unscreened slab, the potential of which is linearly growing for
$x\rightarrow \pm\infty$. In the realistic case under
consideration any charged slab is screened already at distances
$\sim 1/\lambda_{\rm D}^{\rm II}$ from it. Then, the main contribution to
the energy per slab comes from the terms
$\widetilde{\epsilon}_{V,\rm II}^{\,(1)}\simeq
\widetilde{\omega}_{\rm cor, II}^{(1)}\propto \xi$. The term
$\widetilde{\epsilon}_{V,\rm I}^{\,(1)}$ is $\propto \xi^2$,
and $\widetilde{\omega}_{\rm cor, I}^{\,(1)}\propto \xi^3$. In this
limit, i.e. for a tiny quark fraction volume, we immediately
obtain that unscreened spherical droplets are energetically more
favorable compared to slabs. Indeed, we have
\begin{eqnarray}\label{Coulsl}
\widetilde{\epsilon}_{\rm C,S}^{(1)}\simeq \frac{\beta_0
\xi}{3\alpha_0 } +\frac{\beta_0 \beta_1}{3\xi},
\end{eqnarray}
from where we find the minimum $\xi_{\rm min} =\sqrt{\alpha_0\beta_1 }$
and $\widetilde{\epsilon}_{\rm C,S}^{(1)}(\xi_{\rm min}
)=2\widetilde{\epsilon}_{\rm C} (\xi_{\rm min} )=\frac{2}{3}\beta_0
\sqrt{\beta_1 /\alpha_0}$, that is  larger  than the term
(\ref{CSm}) for the case of the spherical droplet
($\widetilde{\epsilon}_{\rm C,S}(\xi_{\rm min} ) \propto \beta_1^{2/3}$)
for $\beta_1 <0.11 / \alpha_0^3$. This statement holds  in the
limit $\beta_1 \alpha_0 \ll 1$ and $f^{\rm I}\ll \alpha_0 (\beta_1
\alpha_0 )^{1/2}$ corresponding to $\xi_{\rm min} \ll 1$ and $\alpha_2
\xi_{\rm min} \gg 1$ for slabs, and  $\beta_1 \ll 1$, $\alpha_0^3 \beta_1
\gg f^{\rm I}$
corresponding to $\xi_{\rm min} \ll 1$ and $\alpha_1 \xi_{\rm min} \gg 1$
for droplets, accordingly.

\paragraph{Slab in Wigner-Seitz cell, both of tiny transverse sizes.}

This limit is recovered for $\xi \ll 1$, $\alpha_2 \xi \ll 1$.
Then we arrive at expressions
\begin{eqnarray}\label{limC}
\widetilde{\epsilon}_{V,\rm
I}^{(1)}=\frac{\alpha_0^2\alpha_2^2\beta_0\xi^2}
{9(1+\alpha_0\alpha_2 )^2} ,\,\,\,\
\widetilde{\epsilon}_{V,\rm
II}^{(1)} =\frac{\alpha_0 \alpha_2^3\beta_0\xi^2}
{9(1+\alpha_0\alpha_2 )^2},
\end{eqnarray}
\begin{eqnarray}\label{limCtot}
&&\widetilde{\omega}_{\rm
cor, I}^{(1)}=\frac{\beta_0}{3(1+\alpha_0 \alpha_2 )^2}-
\frac{2\beta_0\alpha_0\alpha_2^2 (\alpha_0
+\alpha_2)}{9(1+\alpha_0 \alpha_2 )^3}\xi^2, \nonumber \\
&&\widetilde{\omega}_{\rm cor, II}^{(1)}=\frac{\beta_0 \alpha_0
\alpha_2 }{3(1+\alpha_0 \alpha_2 )^2}+
\frac{2\beta_0\alpha_0\alpha_2^2 (\alpha_0
+\alpha_2)}{9(1+\alpha_0 \alpha_2 )^3}\xi^2,
\end{eqnarray}
and thus $\widetilde{\omega}_{\rm cor,
I}^{(1)}+\widetilde{\omega}_{\rm cor, II}^{(1)}\simeq
\frac{1}{3}\beta_0 (1+\alpha_0\alpha_2 )^{-1} $.

Summation of all the terms (\ref{limC}), (\ref{limCtot})
together with (\ref{esoc})
yields, after minimization over $\xi$:
\begin{eqnarray}\label{limC1}
 \widetilde{\omega}_{\rm tot}^{(1)}(\xi_{\rm min} ) /\beta_0 =
 \frac{\beta_1^{2/3} \alpha_0
(1-f^{\rm I})^{1/3}}{3^{1/3}2^{2/3} [\alpha_0^2 +f^{\rm I} (1-\alpha_0^2
)]^{2/3}}+ \frac{f^{\rm I}}{3[\alpha_0^2 +f^{\rm I} (1-\alpha_0^2 )]}  ,
\end{eqnarray}
where we used that
\begin{eqnarray}
\xi_{\rm min} =\left( \frac{3 \beta_1 [\alpha_0^2 +f^{\rm I}
(1-\alpha_0^2 )]^2}{2
\alpha_0^3 (1-f^{\rm I})}\right)^{1/3} \ll 1,
\end{eqnarray}
and $\alpha_2 \xi_{\rm min} \ll 1$, that corresponds to $(\beta_1 \alpha_0
)^{1/3} \ll 1$ and $f^{\rm I} \gg \alpha_0 (\beta_1 \alpha_0 )^{1/3}  $.

\paragraph{Single slab of large transverse size.}

As follows from eqs
(\ref{eVo}) -
(\ref{esoc}), in the limit of a single  large slab of a much
larger transverse size  than the typical Debye screening lengths
($\xi \gg 1$, $\alpha_{2}\xi \gg 1$), all the terms contribute to
the surface energy per slab with the corresponding surface tension
terms\footnote{The curvature terms $\propto 1/\xi^2$ are
different in case of droplets and slabs.}
$\sigma^{(1)}_i =\sigma^{\rm spher}_i $, and, thus, for the
total surface tension we get $\sigma^{(1)}_{\rm tot} =\sigma^{\rm
spher}_{\rm tot}$.

The slab of a large transverse size has three times smaller negative
contribution to the surface energy per slab than that for the
corresponding spherical droplet in case  $\sigma^{(1)}_{\rm tot}<0$,
when the mixed phase is energetically favorable. Here, the
factor three comes from
the space dimension.  Thus, droplets are
energetically favorable compared to slabs also in this limit.

For the case $\sigma^{(1)}_{\rm tot}>0$ realized within the Maxwell
construction the plane boundary layer is  energetically
preferable. In spite of that,  in any case the boundary layer,
which arises between two separate phases within the Maxwell
construction, is spherical, following the
geometry of the neutron star as a whole. Above discussed slabs could
exist in the neutron star only, as configurations of the mixed
phase, being rather small in their transverse sizes.

{\subsection{Numerical results}}

In Fig. 4 we show the same quantity, $\delta
\widetilde{\omega}_{tot} /\beta_0$, as in Fig. 2, but now for slabs,
for the values of the Wigner-Seitz parameter $f\equiv f^{\rm I}
=0.01$ and $f
=0.5$, and
for the parameter $\alpha_0 =1$. The ``C'' curves are calculated with
the help of eq. (\ref{Coulsl}) for $f=0.01$ (dashed) \footnote{Thus, this is
actually not the Coulomb curve, being  calculated without inclusion of
screening,
rather it is
the ``C'' curve obtained from our general equations in the corresponding limit
of a tiny quark concentration and $\xi \ll 1$. We call it ``C'' in the sense
that expression (\ref{Coulsl})
is applicable for $\xi \ll 1$
but we apply it
at arbitrary $\xi$, as it is  done with the Coulomb solutions.}
and
eqs (\ref{Peth-d1}), (\ref{rhcul3}) of Appendix \ref{Coul-pec}
for $f=0.5$ (solid). Eq.  (\ref{Peth-d1})
gives a divergent result in the limit $f\rightarrow 0$ and is not applicable to
describe single slab of a tiny transverse size.

Difference between the non-labeled curves ($\delta
\widetilde{\omega}_{tot}^{(1)} / \beta_0 $)
and the "e.m." curves shows how
large are correlation terms.  We see that forms of all the
curves are analogous to those for the spherical droplet case.
Minima disappear at the ``e.m.'' curves already for $\beta_1 \geq 0.03$
at $f=0.01$ and for $\beta_1 \geq 0.05$
at $f=0.5$. Minima of the non-labeled curves
($\delta\widetilde{\omega}_{tot}^{(1)} /\beta_0$) disappear at $\beta_1 >0.5$
for both $f=0.01$ and $f=0.5$ cases, cf. Fig. 5.

The dashed
 "C" curve is the appropriate limit of the dashed curve
($\delta\widetilde{\omega}_{tot}^{(1)} /\beta_0$)
at a tiny $f$ and $\xi \ll 1$.
It grows linearly with $\xi$. At their minima the dashed  "C" curves
begin to essentially
deviate from the minima of the corresponding dashed curves for
$\beta_1 \gsim 0.05\div 0.1$. If they were normalized to coincide with the
``e.m.'' dashed curves at a tiny $\beta_1$,
their minima would be shifted down by factor
$1/\sqrt{2}$ since $\widetilde{\epsilon}_{V,\rm II}^{\,(1)}\simeq
\widetilde{\omega}_{\rm cor, II}^{\,(1)}\propto \xi$,  and
$\widetilde{\epsilon}_{V,\rm I}^{\,(1)}\propto \xi^2$ and
$\widetilde{\omega}_{\rm cor, I}^{\,(1)}\propto \xi^3$
are neglected.
Solid ``C'' curves ($f=0.5$)
are normalized to coincide with the corresponding ``e.m.''
curve for $\xi \ll 1$, see discussion in Appendix \ref{Coul-pec}.

Fig. 5 demonstrates the dependence of the slab size in dimensionless
units, $\xi_{\rm min}$, on the parameter $\beta_1$ for $f =0.01$
and $f =0.5$, for three values of $\alpha_0$, cf. with Fig. 2 for
the spherical droplet case. The $\alpha_0$ dependence is here even
stronger than for the spherical geometry. Also the transverse
size of the slab
is smaller.

\section{Comparison of energies of droplets and slabs}
\label{Comparison}

As has been argued in the  previous studies, cf. \cite{G01} sect. 4.3.1,
the bulk energy densities of the quark and hadron phases are about two orders
of magnitude larger than the sum of the Coulomb and surface energies.
This large factor is partially due to the fact that the latter
sum   contains an extra $\propto e^{2/3}$
pre-factor compared
to the energy densities of the bulk phases given by eqs (\ref{endenq1})
and (\ref{eVh11}) of Appendix \ref{en}. One can see this pre-factor after
minimization of the sum of the  surface and the Coulomb energy densities
with respect to $R$, taking into account that
the values
$\rho_{\rm ch}^{\rm I,II}$ entering eq. (\ref{Peth})
of Appendix \ref{Coul-pec} do not
contain this pre-factor and that the resulting
surface energy density is of the order
of the Coulomb one.
An extra small factor arises from the
smallness of the difference $\rho_{\rm ch}^{\rm I}-\rho_{\rm ch}^{\rm II}$
entering eq. (\ref{Peth}).

With our screened potentials, the smallness of the value
$\delta\widetilde{\omega}_{\rm tot} (\xi_{\rm min})$
is due to the parameter
$\beta_1 \propto e$ (cf. eq. (\ref{b1est}))
for typical values $\xi_{\rm min} \gsim 1$.

>From the Figs 1 - 5 and eqs. (\ref{endenq1})
and (\ref{eVh11}) of Appendix \ref{en} we may see that typically
$\delta \widetilde{\omega}_{\rm tot} (\xi_{\rm min})$ (with
$\xi_{\rm min}$ essentially dependent on the geometry of the structure)
is at least by an order of magnitude
smaller than the bulk values of the energy densities.
This justifies the same ``two part approach''
to the problem, as has been used in
previous papers of Glendenning and his co-authors,
of computing first the bulk properties and then on this
background, the geometrical structure imposed by the effective energy
$\delta\widetilde{\omega}_{\rm tot} (\xi_{\rm min})$.

The bulk structure is defined  following the strategy,
sketched in  \cite{G01}, sect. 4.1.3. Begin with  solving the equations
that define quantities of the phase II
using the local charge neutrality condition and increasing
the baryon density. At each baryon density find the values of the chemical
potentials $\mu_{B,\rm loc}^{\rm II}$ and $\mu_{e, \rm loc}^{\rm II}$.
At each density also find the
solution of the high-density
phase I at the very same value of the chemical potentials
thus putting $\mu_{B,\rm loc}^{\rm II}=\mu_{B,\rm Gibbs}=const$ and
$\mu_{e, \rm loc}^{\rm II}=\mu_{e, \rm Gibbs}=const$ and
using the relations (\ref{q-c}) - (\ref{q-h1}). Locate the value of the
chemical potentials for which the pressures in the two phases are equal.
The beginning point of the mixed phase is fixed in this way.
The complementary procedure will locate the boundary between the
mixed phase and the pure
phase I
(quark phase). Then we may find the properties of the mixed phase in which
both phases are present in equilibrium.
To find the volume fraction  of  phase II at fixed
$\rho_B$,  for given $f$ (starting from
infinitely small $f$)
solve the equations defining both phases
subject to the conditions of equal pressure and global charge neutrality,
where the charged densities are defined through  the values of the
corresponding constant chemical potentials which, as well as the pressure, vary
with $f$. The value $f$ is a function of
$\rho_B$, see (\ref{globalbar}).
The increase of the value $f$ corresponds to
the increase of the
baryon density (\ref{globalbar}). The value $f=1$
corresponds to the end point of the mixed phase.

Then, in order to define, which geometrical structure is energetically
favorable,
we may use the values of $\rho_B$ and $f$ thus calculated, and
compare the quantities
$\delta\widetilde{\omega}_{\rm tot}^{(d_1)} (\xi_{\rm min}^{(d_1)})$
and $\delta\widetilde{\omega}_{\rm tot}^{(d_2)} (\xi_{\rm min}^{(d_2)})$.
The latter values relate to the distinct possible
configurations (quark/hadron) of different geometries $d_1$ and $d_2$.
As we have mentioned, in the present paper we restrict ourselves to the
consideration of the quark structures embedded into hadron
matter for
$f\leq 0.5$ and  by studying of
the two geometries $d_1 =3$ (spherical quark droplets)
and $d_2 =1$ (quark slabs). Each quantity
$\delta\widetilde{\omega}_{\rm tot}^{(3)} (\xi_{\rm min}^{(3)})$
and $\delta\widetilde{\omega}_{\rm tot}^{(1)} (\xi_{\rm min}^{(1)})$
is minimized with respect to the structure size ($\xi_{\rm min}^{(3)}$ and
$\xi_{\rm min}^{(1)}$). The values
$\xi_{\rm min}^{(3)}$ and  $\xi_{\rm min}^{(1)}$
are, thereby, essentially different, see
Figs 3 and 5. This is due to the fact that
the minimization in
$\xi$ is performed for the  quantities
$\delta\widetilde{\omega}_{\rm tot}^{(3)} (\xi)$
and $\delta\widetilde{\omega}_{\rm tot}^{(1)} (\xi)$,
being of the same order of magnitude (having the same $e$ pre-factors).
Thus the argument which allowed us to fix $f$ to be the same
for both structure geometries
does not hold for $\xi$.
As we have demonstrated above, cf. Figs 3, 5, the minima
$\xi_{\rm min}^{(1)}$ and $\xi_{\rm min}^{(3)}$ exist not for all values of
$\beta_1$, but only for those  related to not too large values of the
surface tension parameter. Otherwise the presence of the
mixed phase is not
energetically
favorable
compared to the configuration given by the Maxwell construction.

Above procedure implies that the corrections
to $f$ due to its dependence on the  geometrical structure
and accordingly on the values of $\xi_{\rm min}^{(1)}$ or
$\xi_{\rm min}^{(3)}$ are rather small. Therefore they are neglected.
The values of $\rho_{\rm ch}^{\rm I,II}$ used in determining  the
parameters of the inhomogeneous spatial structures,
should not be  constants in the general case, rather
they are now functions of $\vec{r}$
given by (\ref{qrhoch}) and (\ref{r2}).
The global charge neutrality condition
is now
satisfied since we use the boundary condition $\nabla V =0$
at the boundary of
the Wigner - Seitz cell.

Fig. 6 shows which structures (spheres or slabs) are more
energetically favorable for the given value of the
parameter $f$, in dependence on the value of $\beta_1$. The solid curves,
$\delta\widetilde{\omega}_{\rm tot}^{\rm min}/\beta_0
\equiv \delta
\widetilde{\omega}_{\rm tot}^{(3)} (\xi_{\rm min}^{(3)})/\beta_0 $,
present droplets
and the dashed
curves, $\delta\widetilde{\omega}_{\rm tot}^{\rm min}/\beta_0
\equiv \delta
\widetilde{\omega}_{\rm tot}^{(1)} (\xi_{\rm min}^{(1)})/\beta_0$,
the slabs.  The result, which structure is energetically favorable,
essentially depends on the values of
parameters $\alpha_0$ and $\beta_1$.
For $\alpha_0 =0.5$, $\beta_1 \gsim 0.3$
spherical droplets are energetically favored
for any value of the parameter $f<0.5$.
For $\alpha_0 =2$, $\beta_1 \gsim 0.3$
slabs are energetically favored.
For unrealistically large values of $\alpha_0$, see the cases $\alpha_0 =5; 10$
in Fig. 7,
we find that slabs
become  to be energetically favorable structures already for
$\beta_1 \gsim 0.01$. It occurs then for
any $f$.
Our conclusions here essentially differ from those derived in the Coulomb
limit, if one ignored the screening
effects for any $\beta_1$
and assumed the step-function charge-density profiles. In the
latter case the droplet phase is proven to be energetically favorable for
$f<f_c^{\rm rod}\simeq 0.2$, as has been shown in ref.
\cite{RPW83,GP95}. Then for $f_c^{\rm rod}<f <f_c^{\rm slab}\simeq 0.32$
rods are prefered and for $f>f_c^{\rm slab}$ slabs are prefered.
We recover this limit case result from our general expressions as well,
but only for tiny values of $\beta_1$ and $\alpha_0 =1$,
see discussion in Appendix \ref{Coul-pec} and Fig. 8. In the latter
we present the
effective energies of droplets and slabs
obtained with our general expressions in case $\beta_1 =0.001$ for
several values of $\alpha_0$.

\section{Concluding remarks}\label{Concl}

Summarizing, in this paper we have
discussed  the possibility of the presence of
structured mixed phases at first order phase transitions in
multi-component systems of charged particles.  Using the
thermodynamical potential as the generating
functional we have
developed a formalism,
which allows to explicitly
treat electric field effects including the Debye screening.

We have studied a
``contradiction'' between the Gibbs conditions and the Maxwell construction
extensively discussed in previous works. We have
demonstrated that this
contradiction is resolved
if one takes into account the difference in
the treatment of the chemical potentials used in the two
approaches. The different values of the electron
chemical potentials in the Maxwell construction and the ones
used in the Gibbs conditions do not contradict each
other if one
takes into account the electric field arising
at the boundaries of the finite size structures of the
mixed phase and at the boundary between phases within the Maxwell construction.

As an example, our formalism was applied to the hadron-quark
structured mixed phase. Using a linearization procedure
we analytically solved the Poisson equation
for the electric potential and found the electric field profiles of the
structured mixed phase in cases of spherical and one-dimensional geometries.
We have demonstrated that the charge screening effect greatly
modifies the description of the mixed phase, changing its parameters
and affecting the possibility of its existence.
With screening effects included
it has been proven that the  mixed phase structures may exist even if their
sizes are significantly larger than the Debye screening lengths.
The characteristics of the structures essentially depend on the value of the
ratio of the Debye screening lengths in the interior and exterior of
the structure and on the value of the surface tension on its boundary.
We showed the important
role played by the correlation energy terms omitted in previous works.
For realistic values of the parameters
these correlation
contributions are larger than the electromagnetic contribution to the
energy discussed in previous works.

The result, which structure is energetically favorable,
essentially depends on the values of
parameters $\alpha_0$ and $\beta_1$. This our
conclusion essentially differs from
that was previously done for the Coulomb case, when one ignored the screening
effects for any $\beta_1$
and assumed the step-function charge-density profile.
We recover that limit case result from our general expressions as well,
but only for tiny values of $\beta_1$ and $\alpha_0 =1$.


We support the conclusion of \cite{RPW83,HPS93}
that the mixed phase may arise only if the
surface tension is smaller than a critical value.
We also evaluated the surface tension between the hadron and the quark
phases. However this evaluation remains  rather rough.
An uncertainty in the value of the surface tension does not
allow to conclude whether the
mixed H-Q phase exists or not. Both possibilities,
the mixed phase and the Maxwell construction, can be studied within
different models. An advantage of our study is that the characteristics of
the mixed phase are expressed in terms of only few parameters. Thus knowing
these parameters for a given model one may apply our results for the
discussion of the mixed phase, the Maxwell construction, the geometry of the
structures, the energy contributions of the structures
and the electric field profiles.

In absence of a mixed phase our charged
distributions describe the boundary layer between two separated
phases existing within the double-tangent (Maxwell) construction.
The previously obtained
Coulomb limits (when one omitted the screening effects)
are naturally
recovered from our general expressions but only
in the limiting case of tiny size
droplets. This limiting case, however, is never realized for realistic values
of the parameters.

In this paper we considered only
droplets and slabs and their interplay for $f<0.5$ at zero temperature.
More detailed analysis of a possible interplay between different structures
within the whole interval $0<f<1$ and at finite temperatures
will be performed in a forthcoming paper.

\paragraph{Acknowledgements.}

Research of D.N.V. at Yukawa Institute for
Theoretical Physics was sponsored by the COE program of the
Ministry of Education, Science, Sports and Culture of Japan.
D.N.V. acknowledges this support and kind hospitality of the
Yukawa Institute for Theoretical Physics. He thanks Department of
Physics at Kyoto University for the warm hospitality and also
acknowledges hospitality and support of GSI Darmstadt. The present
research of T.T. is partially supported by the REIMEI Research
Resources of Japan Atomic Energy Research Institute, and by the
Japanese Grant-in-Aid for Scientific Research Fund of the Ministry
of Education, Culture, Sports, Science and Technology (11640272,
13640282).
Authors greatly acknowledge E.E. Kolomeitsev
for the help and discussions and M. Lutz for valuable remarks.

\newpage
\appendix
\begin{center}
\begin{eqnarray}
\nonumber\\\nonumber
\end{eqnarray}
{\bf{APPENDICES}}

\end{center}
\section{
Energy density of quark and hadron sub-systems}\label{en}

In the main text, as a concrete example,
we consider the H-Q phase transition. We assume
that the quark matter (phase I) consists of $u,d,s$ quarks and
electrons. To easier compare our results for the description of
the H-Q system with those obtained previously, we choose
a simple model of equation of state used in \cite{HPS93,TT93} with
slight modifications. We could use any other model. We need it here only
to estimate typical values of the parameters.

The kinetic
plus strong interaction energy density of the quark matter
expressed in terms of particle densities is given by the bag model
\begin{eqnarray}\label{endenq1}
\epsilon_{\rm kin+str}^{\rm I} \simeq \frac{3\pi^{2/3}}{4}\left(
1+\frac{2\alpha_c }{3\pi} \right) \left[
\rho_u^{4/3}+\rho_d^{4/3}+\rho_s^{4/3}
+\frac{m_s^2\rho_s^{2/3}}{\pi^{4/3}} \right] +B+\frac{(3\pi^2
\rho_e)^{4/3}}{4\pi^2},
\end{eqnarray}
where $B$ is the bag constant, $\alpha_c $ is the QCD coupling
constant, $m_s $ is the mass of the strange quark. Last term is
the kinetic energy of electrons. Expansion is presented up to the
terms linear in $\alpha_c$ and $m_s^2$, where the strange quark
mass $m_s$ ($\simeq ~$120 $\div$ 150 MeV) is supposed to be rather
small compared to the quark chemical potentials, and much smaller
values of $u$ and $d$ quark bare masses are omitted.

The hadron matter (phase II) in our model consists of protons,
neutrons and electrons and the kinetic plus strong interaction
energy density is given by
\begin{equation}\label{eVh11}
\epsilon_{\rm kin+str}^{\rm II}\simeq \epsilon^n_{\rm kin}[\rho_n
]+ \epsilon^p_{\rm kin}[\rho_p ]+ \epsilon_{\rm
pot}[\rho_n,\rho_p]+\frac{(3\pi^2 \rho_e)^{4/3}}{4\pi^2},
\end{equation}
where  $\epsilon^i_{\rm kin}[\rho_i ],\,\, i=n,p,$ are standard
relativistic kinetic energies of nucleons,
\begin{eqnarray}
\epsilon^i_{\rm kin}=2\int \sqrt{p^2 +m_i^2}\, n_i (p) \frac{d^3
p}{(2\pi )^3}, \,\,\, i=n,p,
\end{eqnarray}
$m_n \simeq m_p =m_N$ is the nucleon mass, $n_i (p)$ are step-functional Fermi
occupations. The potential energy $\epsilon_{\rm pot}$ we take
here in the form
\begin{eqnarray}\label{eVh11p}
\epsilon_{\rm pot}[\rho_n,\rho_p]&=&S_0 \frac{(\rho_n - \rho_p
)^2}{\rho_0}+ (\rho_n +\rho_p ) \epsilon_{\rm bind} +\frac{K_0
(\rho_n +\rho_p ) }{18}
\left( \frac{\rho_n +\rho_p }{\rho_0}-1 \right)^2 \nonumber\\
 &+& C_{\rm sat} (\rho_n +\rho_p ) \left( \frac{\rho_n +\rho_p }
{\rho_0} -1 \right)  ,
\end{eqnarray}
$\rho_0$ is the nuclear density ($\rho_0\simeq 0.16$~fm$^{-3}$)
and values of constants $\epsilon_{\rm bind}$, $K_0$, $C_{\rm sat}$ are
taken to satisfy the nuclear saturation properties. The value
$\epsilon_{\rm bind} = \widetilde{\epsilon}_{\rm bind} + m_N -
2\rho_0^{-1} \epsilon_p^{kin}(\rho_B =\rho_0, \rho_p /\rho_B
=1/2)$
provides appropriate binding energy $\widetilde{\varepsilon}_{\rm
bind}=-15.6$~MeV at the nuclear saturation density $\rho_0$. The
coefficient $K_0$ is defined by the relation $K_0 =
\widetilde{K}_0 - \frac{3p_{{\rm F}0}^2 }{\sqrt{ m_N^2 + p_{{\rm F}0}^2}}-18
C_{sat}\simeq 285$~MeV, where the value $\widetilde{K}_0 \simeq
240$~MeV provides an appropriate compressibility at saturation.
$S_0 =\widetilde{S}_0 -\pi^2 \rho_0 /(4 p_{{\rm F}0} \sqrt{p_{{\rm F}0}^2
+m_N^2} )$ is the symmetry energy term. The parameter
$\widetilde{S}_0$ is chosen to be $\widetilde{S}_0 \simeq 30
~$MeV, that corresponds to $S_0 \simeq 18 ~$MeV, $p_{{\rm F}0}=p_{{\rm F}p}
(\rho_B =\rho_0 ,\rho_p /\rho_B =1/2 ) =p_{{\rm F}n} (\rho_B =\rho_0
,\rho_p /\rho_B =1/2 )$ is the baryon Fermi momentum for the symmetric
nuclear matter at the saturation density.
The value
\begin{eqnarray}
C_{sat} = 2\rho_0^{-1}\epsilon_p^{kin}(\rho_B =\rho_0 ,\rho_p
/\rho_B =1/2) -\epsilon_{F0}
\end{eqnarray}
is introduced to reproduce the saturation property,
$\epsilon_{{\rm F}0}=\sqrt{p_{{\rm F}0}^2 +m_N^2 }$. Higher order terms
$\sim\left(\rho_B/\rho_0-1\right)^3$ in expansion of the
potential energy are for simplicity dropped in (\ref{eVh11p}),
since their explicit form  is not too important for our study
here.
Contribution of mean fields of the heavy mesons  $\sigma$,
$\omega$ and $\rho$ is hidden in the values of phenomenological
parameters $\widetilde{K}_0$ and $\widetilde{S}_0$.

Chemical potentials of neutrons and protons are determined  with
the help of (\ref{endenq1}) - (\ref{eVh11p}):
\begin{eqnarray}\label{chem-n}
&&\mu_n =\mu_B =\sqrt{p_{Fn}^2 +m_N^2}+\frac{2S_0 (\rho_n -\rho_p
)}{\rho_0} +\epsilon_{bind}+ \frac{K_0}{6}\left( \frac{\rho_n
+\rho_p }{\rho_0}-1 \right)^2\nonumber \\ && +\frac{K_0}{9}\left(
\frac{\rho_n +\rho_p }{\rho_0}-1 \right) +\frac{2C_{sat} (\rho_n
+\rho_p )}{\rho_0}-C_{sat}, \\ &&\mu_p =\mu_n -\sqrt{p_{Fn}^2
+m_N^2}+ \sqrt{p_{Fp}^2 +m_N^2} -\frac{4S_0 (\rho_B -2\rho_p
)}{\rho_0}-V^{\rm II} (\vec{r}),\nonumber\\
&&\mu_e=(3\pi^2\rho_e)^{1/3}+V^{\rm II}(\vec{r}). \nonumber
\end{eqnarray}
As in the phase I, $\mu_e=\mu_{e, \rm Gibbs}$ for $V^0=0$ and
$\mu_e=0$ for $V^0=-\mu_{e, \rm Gibbs}$.

Varying these equations of motion,
after straightforward calculations we get
\begin{eqnarray}\label{deltarop1}
C_0 =\frac{\widetilde{C}_{0n}\widetilde{C}_{0p}-K_1
(\widetilde{C}_{0p}+\widetilde{C}_{0n})}{\widetilde{C}_{0n}-K_1},
\end{eqnarray}
with
\begin{eqnarray}
\widetilde{C}_{0j} \equiv \left[C_{0j}+\frac{4S_0}{\rho_0}\right]
,
\end{eqnarray}
where $C_{0j}^{-1}=p_{{\rm F}j}\sqrt{p_{{\rm F}j}^2 +m^2_j}/\pi^2$ is the
density of states for the given baryon species, $j=\{n,p\}$, and
\begin{eqnarray}\label{k1}
K_1 = \frac{2S_0 }{\rho_0} -\frac{K_0 +18C_{sat}}{9\rho_0}-
\frac{K_0 \rho_B}{3\rho_0^2}\left( 1-\frac{\rho_0}{\rho_B}\right) .
\end{eqnarray}

Dropping
$K_1$ we may use a
simplified expression
\begin{eqnarray}\label{deltarop11}
C_0 \simeq \widetilde{C}_{0p},
\end{eqnarray}
relating to $\delta \rho_n (\vec{r})\simeq 0$, that is not too bad
approximation  for $\rho_B \sim \rho_0$ and for $\rho_B \gg \rho_0$,
$p_{Fn} \gg p_{Fp}$. We
use this simplified expression for rough numerical estimates.

\section{Surface energy of H-Q interface}\label{surf}

The surface layer separating quark and hadron sub-systems is very
difficult to describe microscopically mainly due to absence of the
microscopic treatment of the confinement problem. A simplification
comes from the fact that gradients of all the fields except the
electric field are very sharp and this layer can be then treated
in terms of surface quantities. Thus, in our consideration below
we will use the fact that in absence of electric field effects the
surface layer is characterized by a shortest scale $d_{\rm S}$ ($\lsim
1$~fm as we estimate). With this,  we explicitly include electric
field and calculate corrections to all the meson-nucleon fields
due to electric field effects. These corrections,
relate to larger scales $\lambda_{\rm D}^{\rm I,II} \sim (5\div 10)$~fm
corresponding to the Debye screening of the electric field. Due to that
$\lambda_{\rm D}^{\rm I,II} \gg d_{\rm S}$, the electric field effects do not
significantly affect the quantities related to the shortest scale
$d_{\rm S}$.

Let us first discuss the case of the quark-vacuum interface.  The
quark contribution to the energy density of this layer arises due
to sharp spatial gradients  of the quark-gluon fields within the
confinement area ($\delta r \sim d_{\rm S}^{\,q} \simeq 0.2\div 0.4~$ fm)
near the bag surface. Since $d_{\rm S}^{\,q}$ is substantially smaller than
the corresponding length scales $\lambda_{\rm D}^{\rm I,II}$ involved in
the charge screening problem, the quark contribution to the energy
of the surface layer can be reduced to the surface energy
\begin{eqnarray}\label{epsq}
E_{{\rm S},q}=\int d\vec{r} \, \epsilon_{{\rm S},q}= v^{\rm
I}\sigma_q [\rho ] \, d/R , \,\,\,\, \epsilon_{{\rm S},q} =
(v^{\rm I} \sigma_q \, d/R )\delta (\vec{r} \in \partial D),
\end{eqnarray}
$d$ is the parameter of the space dimension of the structure,
$d=3$ for spherical droplets and $d=1$ for slabs, the volume of
the quark phase is $v^{\rm I} = 4\pi R^3 /3$ for $d=3$, and $v^{\rm
I} = 2R$ for $d=1$, and $\epsilon_{{\rm S},q}$ is proportional to
the delta-function $\delta (\vec{r} \in \partial D)$, being zero
everywhere outside the surface,  $\int d \vec{r}\delta (\vec{r}
\in \partial D)=1$. The surface tension parameter $\sigma_q $ depends
on densities of $u$, $d$, and $s$ quarks in the surface region.

The surface tension $\sigma_q [\rho ]$ is different in dependence
of what exterior region is considered (vacuum or hadron matter).
According to ref. \cite{BJ87}, for $m_s \ll
\epsilon_{{\rm F}s}\simeq\epsilon_{{\rm F}n}/3 \sim m_N/3$, where
$\epsilon_{{\rm F}s}$ is the Fermi energy of the $s$-quark,
$\epsilon_{{\rm F}n}$ is the Fermi energy of the neutron, the value
$\sigma_q$ for the quark-vacuum boundary ($\sigma_q^{vac}$) is
estimated as
\begin{eqnarray}\label{qs}
\sigma_q^{\rm vac} \simeq \frac{3}{4\pi^2} \epsilon_{{\rm F}s}^2 m_s .
\end{eqnarray}
>From eq. (\ref{qs})
we easily evaluate a minimal value of  the surface tension
parameter $\sigma_q $ of the quark-vacuum layer as $\sigma^{\rm
min}= m_N^2 m_s /(12\pi^2 )\simeq 0.3 m_{\pi}^3 \simeq
20$~MeV~$\cdot$~fm$^{-2}$ (for $m_s \simeq 120~$MeV). As one can
see, dependence of $\sigma_q$ on the baryon density is rather
weak.
For realistic values of densities with $\epsilon_{{\rm F}n} \simeq 1\div
1.2~$GeV and $m_s \simeq 150~$MeV we estimate  $\sigma_q \simeq
(0.5\div 0.7)  m_{\pi}^3$.\,\footnote{These values are substantially
less than the value $\sigma_q \sim 300$~MeV~$\cdot$~fm$^{-2}$ used
as a dimensional estimate in \cite{ARRW} and larger than an estimate
$\sigma_q \sim 10$~MeV~$\cdot$~fm$^{-2}$ used in \cite{GP95} in description of
the mixed phase structures neglecting the screening effects.}

In the hadronic region gradients of mean fields of the heavy
mesons  $\sigma$, $\omega$ and $\rho$ are also large compared to
gradients of  the electric field. As known, thickness of nucleus
diffuseness layer is $d_{\rm S}^N \sim 0.6$~fm. This follows both from
experimental data and from the calculation of the nuclear structure
within mean field models, cf. \cite{SW86}.
Therefore, contributions arising due to spatial gradients of all
the hadron mean fields (not related to changes of the electric
potential,
the latter we  treat explicitly) can be included in the hadronic
part of the surface energy density $\epsilon_{{\rm S},h}$.  This
contribution to the surface energy can be  fixed also with  the
help of the corresponding surface tension parameter
\begin{eqnarray}\label{esh}
E_{{\rm S},h} =\int d\vec{r}\epsilon_{{\rm S},h} =v^{\rm I}
\sigma_h [\rho ] \,d/R , \,\,\,\, \epsilon_{{\rm S},h} = (v^{\rm
I} \sigma_h \, d/R )\delta (\vec{r} \in \partial D).
\end{eqnarray}
The value $\sigma_{h}^{vac}$ is estimated as $\simeq
1~$MeV~$\cdot$~fm$^{-2}$ for symmetric atomic nuclei in vacuum.
However  $\sigma_{h}$ sharply increases with the density. For the
interface between two hadronic media $\sigma_h$ depends on
particle densities in both the phases. According to ref.
\cite{BBP71}, the surface tension between two hadronic sub-systems is
given by
\begin{eqnarray}\label{BBP71}
\sigma_{h}=a_{{\rm S},h} \cdot\mid \epsilon_h (\vec{r}\in
\partial D-0)- \epsilon_{h}(\vec{r}\in \partial D+0 ) \mid
.\,\,\,\,
\end{eqnarray}
Here $\epsilon_h [\rho_i ]$ is the volume part of the hadronic
energy density. This expression takes into account that the surface
tension should disappear ($\sigma_{h}= 0$) for the case of equal
energy densities of the phases. Ref. \cite{BBP71} presented an
expression for the coefficient $a_{{\rm S},h}\simeq 0.3
\rho^{-1}_B (\vec{r}\in \partial D-0 )\cdot \mid \rho_{B}
(\vec{r}\in
\partial D-0 )
-\rho_{B} (\vec{r}\in \partial D+0)\mid^{2/3}$. As we see, the
dependence of $a_{{\rm S}, h}$ on the density is rather weak.
Therefore, simplifying we will further use constant value $a_{{\rm
S}, h}$. Numerical estimation yields $a_{{\rm S}, h}\simeq
0.4/m_{\pi}$.

Typical value of the full surface energy per droplet volume
$v^{\rm I}$ separating the hadron and quark phases can be
estimated as a half-sum of partial quark and hadron contributions,
\begin{eqnarray}\label{eps}
E_{\rm S} /v^{\rm I}  = \int d\vec{r}\epsilon_{\rm
S}/v^{\rm I} =\frac{1}{2}\left[ \int d\vec{r}\epsilon_{{\rm
S},q}^{\rm vac}/v^{\rm I} + \int d\vec{r}\epsilon_{{\rm S},h}^{\rm
vac}/v^{\rm I} \right] =\sigma [\rho_i ] \,d/R, \,\,\,\,
\end{eqnarray}
where $\sigma \simeq \frac{1}{2} (\sigma_q^{\rm vac}
+\sigma_{h}^{\rm vac})$, $\sigma_q^{\rm vac}$ and $\sigma_h^{\rm
vac}$ are the corresponding values for the quark-vacuum and
hadron-vacuum boundaries. With above rough estimates we would have
$\sigma \sim (50\div 150)$~MeV$/$fm$^2$ in dependence on the
baryon density. This parameter may in general depend on densities
of all particle species in the surface layer.

For a better estimation of surface effects one needs to use
chemical equilibrium conditions between quark and hadron phases
explicitly solving  equations of motion for the fields in a narrow
layer near the bag surface.
Avoiding this complicated and tedious procedure, we take into
account an observation of ref. \cite{BBP71} that surface tension
should vanish, if energy densities of the phases being equal.
Thus, analogously to (\ref{BBP71}) for the phase boundary between
two quark phases we may assume the validity of the relation
\begin{eqnarray}\label{BBP71q}
\sigma_{q}=a_{{\rm S},q} \cdot \mid \epsilon_q ( \vec{r}\in
\partial D-0 )- \epsilon_{q}(\vec{r}\in \partial D+0)\mid
,\,\,\,\,
\end{eqnarray}
where $a_{{\rm S},q}$ is a positive coefficient, which we may
estimate with the help of (\ref{qs}).

Unifying results (\ref{BBP71}), (\ref{BBP71q}) we suggest an
interpolation equation
\begin{eqnarray}\label{BBP71qh}
\sigma =a_{\rm S} \cdot \mid \epsilon_q (\vec{r}\in \partial D-0)-
\epsilon_h (\vec{r}\in \partial D+0)\mid ,\,\,\,\,
\end{eqnarray}
with a common coefficient $a_{\rm S}\simeq \frac{1}{2}(a_{{\rm
S},q}+a_{{\rm S},h})\simeq (0.3\div 0.4)/m_{\pi}$, and with the
property that $\sigma$ is zero for $\epsilon_h (\vec{r} \in
\partial D+0)=\epsilon_q (\vec{r}\in \partial D-0)$. Here,
evaluating $a_{\rm S}$ we neglected its slight density dependence.

Thus, we use that
\begin{eqnarray}\label{eps-fin}
E_{\rm S}/v^{\rm I}  =\sigma \,d/R, \,\,\,\,
\end{eqnarray}
with $v^{\rm I} =4\pi R^3/3$ for the case of the spherical geometry
and $v^{\rm I} =2R$ for the flat geometry, and we will estimate the
value of the surface tension $\sigma$ according to interpolation
equation (\ref{BBP71qh}).

Due to dependence of surface tension only on the difference of the
bulk quantities $\epsilon_q (\vec{r}\in \partial D-0)- \epsilon_h
(\vec{r}\in \partial D+0)$, there appears no contribution from the
surface region to the chemical potentials ($\mu_{\rm S}^i =0$).
With above expressions pressure acquires an additional contribution
given by eq. (\ref{gibbs2}) but no contribution appears from the
density dependence of $\sigma$.

Contrary, there is a contribution to the surface charge density
\begin{eqnarray}\label{sch}
&&\rho_{\rm S}^{\rm ch}=-\partial \epsilon_{\rm S}/\partial V\nonumber \\
&&=-a_{\rm S} v^{\rm I} d \mid \rho_{\rm ch}^{\rm I}(\partial D-0)
-\rho_{\rm ch}^{\rm II} (\partial D+0)\mid \delta (\vec{r}\in
\partial D) /R .
\end{eqnarray}
To get this equation we used
eqs (\ref{BBP71qh}) and (\ref{eps-fin}). One can see that for
$R\gg 1/m_{\pi}$ the surface charge, $Z_{\rm S} = \int
d\vec{r}\rho_{\rm S}^{\rm ch}$, is  substantially smaller than all
partial contributions to the volume charges given by other terms in
r.h.s. (\ref{qV}). Integrating (\ref{sch}) yields $Z_S \sim Z_{\rm
dr}(Rm_{\pi})^{-1}$ with $Z_{\rm dr}=\mid Z^{\rm I}-Z^{\rm
II}\mid$. More generally, the typical scale for the change of the
electric field is related to the Debye screening. Large values of
$\lambda_{\rm D}^{\rm I,II}$ are due to smallness of the fine structure
coupling constant $e^2$ entering r.h.s. of the Poisson equation,
$\lambda_{\rm D}^{-1} \sim e m_{\pi}$. Thus, electric field effects may
only minor affect quantities, like the surface charge, $\sigma$, etc.,
relating to a much more narrow  surface layer. Due to smallness of
$Z_{\rm S}$ and to avoid extra uncertainties, we drop the surface
charge effects, although generalization is straightforward.

Not entering in more details we permit the variation  of  the
value $\sigma$ in a wide region covering the whole interval of
parameters used in the literature, cf. \cite{HPS93,GS99}.

Concluding, we would like to do a remark. As follows
from analysis of many works and as was argued above, in the absence of
electric field effects typical estimate of $d_{\rm S}$ is $\lsim 1$~fm.
The same estimate is also used in recent work \cite{ARRW} studying
the first order phase transition from the nuclear matter to the
color superconductor within the Maxwell construction. On the other
hand, figures of refs \cite{CG00,NR00} which numerically studied
the possibility of the mixed phase at the first order phase
transition to the kaon condensate phase demonstrate a several
fermi scale for all the fields: nucleon, sigma, omega, rho and
kaon fields. There can be two reasons for that. First, these works
discussing inhomogeneous charged structures within mixed phase do
not introduce the Poisson equation for the electric field, as well as electric
field itself,
and
therefore the region of applicability of their results seems to be
not quite clear. Second, the kaon field and the electric field in the region
I  relate to the same scale $\sim \lambda_{\rm D}^{\rm I}$, as follows
from our analysis of this problem, cf. analogous problem of
supercharged pion-condensate nuclei discussed in ref.
\cite{VCh78}. Therefore, surface effects related to the scale
$d_{\rm S}$, being partially smeared by effects related to the larger
scales $\lambda_{\rm D}^{\rm I,II}$, might be not seen in these
numerical evaluations. Contrary, a partial smearing of the fields is
incorporated explicitly in our analysis. Indeed, corrections to
all the fields due to electric field effects, which we treat
explicitly, relate to the scales $\lambda_{\rm D}^{\rm I,II}$, as we argue,
see eqs.
(\ref{co}). In case of the H-Q phase transition that we discuss
in this paper this smearing can be only partial, since the scale
$d_S$ is fixed by the quark confinement and can't be substantially
changed.

\section{Formulation in terms
of Gibbs potential}\label{Free}
In the main text we performed the gauge invariant treatment of the
problem in terms of the thermodynamic potential and the energy, expressed
in terms of
the $\rho_i $ variables. Sometimes it is more convenient to use
another potential and variables, e.g.  the Gibbs potential $G$
instead of $\Omega$. The ground state is then determined by the
minimization of the Gibbs potential
\begin{eqnarray}
G=E+P v ,
\end{eqnarray}
where $P= const$, $E$ is the energy concentrated in the given
volume $v$, \cite{LP80}. With the conservation of the baryon number
$N_B$, appropriate variables of the functional $G$ are $P$,
$\rho_B^{\alpha}$, $V^{\alpha}$, ${\alpha}={\rm I,II}$, and
concentrations of particles. Thus, we start with the Gibbs
potential functional
\begin{eqnarray}\label{Etot}
&&G[x_u , x_d , V^{\alpha} , \rho_B^{\alpha} , x_p ]
=\int_{\vec{r}\in D^{\rm I} }d \vec{r} \epsilon^{\rm I}  [x_u ,
x_d , V, \rho_B ]\nonumber \\ &&+E_{\rm S} [x_u , x_d , V, \rho_B
, x_p ]+ \int_{\vec{r}\in D^{\rm II} }d \vec{r} \epsilon^{\rm II}
[x_p , V, \rho_B ]+P v ,
\end{eqnarray}
which is expressed in terms of independent variables $x_u , x_d ,
V^{\alpha}, \rho_B^{\alpha} , x_p$, where $x_u =\rho_u /\rho_q$,
$x_d =\rho_d /\rho_q$
are concentrations of $u$, $d$ quarks,
baryon density $\rho_B$ in the quark phase $\rho_B^{\rm I} =\rho_q
/3$, $\rho_q$ is the net quark density, $x_p =\rho_p /\rho_B^{\rm
II}$ is the concentration of protons,
$v^{\rm I}$ is the part of the volume filled by the quark phase,
$v^{\rm II}$ is the part of the volume filled by the hadron phase,
$v =v^{\rm I}+v^{\rm II}$. Densities of strange $s$ quarks and
neutrons are fixed by relations $\rho_s =3\rho_B^{\rm I} -\rho_u
-\rho_d$, and $\rho_n =\rho_B^{\rm II} -\rho_p$ demonstrating the
conservation of the baryon number.

Equations of motion are given by the variation of the Gibbs potential
functional.
Equation
\begin{eqnarray}\label{den-prof}
\frac{\delta G[x_u , x_d , V^{\alpha} , \rho_B^{\alpha} , x_p ]
}{\delta \rho_{B}^{\alpha}} =0
\end{eqnarray}
relates the energy density and its density (volume) derivative to
the constant pressure $P$. Other equations of motion are
\begin{eqnarray}
\frac{\delta G[x_u , x_d , V^{\alpha}, \rho_B^{\alpha} , x_p
]}{\delta x_u} &=& \frac{\delta E[x_u , x_d , V^{\alpha},
\rho_B^{\alpha} , x_p ]}{\delta x_u }
=
0,\,\,\,\,  \label{eqm-u}\\ \frac{ \delta E [x_u , x_d ,
V^{\alpha}, \rho_B^{\alpha} , x_p ]}{\delta x_d} &=&0,
 \label{eqm-d} \\
\frac{\delta E[x_u , x_d , V^{\alpha}, \rho_B^{\alpha} , x_p ]
}{\delta x_p} &=&0,\,\,\,\, \label{eqm-p} \\ \frac{\delta E [x_u ,
x_d , V^{\alpha}, \rho_B^{\alpha} , x_p ]}{ \delta V^{\alpha}}
&=&0.\label{eqm-V}
\end{eqnarray}
They do not depend on $P$ since variations are performed at fixed
baryon density (volume).

In the presence of the electric field the kinetic and potential
contributions to the energy density of the quark phase (the energy per
the quark fraction volume $v^{\rm I} =f^{\rm I}v$) are given by
\begin{eqnarray}\label{eqtot}
\epsilon^{\rm I}  [x_u , x_d , V^{\rm I}, \rho_B^{\rm I} ]
=\epsilon^{\rm I}_{\rm kin+str}+\epsilon^e_{\rm kin, I} +
\epsilon_V^{\rm I} .
\end{eqnarray}

The energy density of the quark bag (\ref{endenq1}) expressed in
new variables renders
\begin{eqnarray}\label{qkin}
&&\epsilon^{\rm I}_{\rm kin+str} [x_u , x_d ,\rho_B^{\rm I}
]\simeq B+ \frac{3\pi^{2/3}}{4}\left( 1+\frac{2\alpha_c }{3\pi}
\right)(3\rho_B^{\rm I})^{4/3}  \nonumber \\ &&\times\left[
x_u^{4/3}+x_d^{4/3}+(1 -x_u -x_d )^{4/3} +\frac{m_s^2 (1 -x_u -x_d
)^{2/3}}{\pi^{4/3}(3\rho_B^{\rm I})^{2/3}} \right].
\end{eqnarray}
The terms
\begin{eqnarray}\label{eVq}
&&\epsilon^e_{\rm kin, I} [V^{\rm I}] + \epsilon_V^{\rm I} [x_u ,
V^{\rm I}, \rho_B^{\rm I} ]\nonumber \\ &&= \frac{(V^{\rm
I})^4}{4\pi^2} +\rho_e^{\rm I} (V^{\rm I}) V^{\rm I}  -\rho_B^{\rm
I} (3x_u -1 )V^{\rm I} - \frac{(\nabla V^{\rm I})^2}{8\pi e^2}
\end{eqnarray}
contain an explicit dependence on the electric potential well. To
avoid complications we selected the {\bf{way II}} fixing the gauge
($V^0 =-\mu_{e,\rm Gibbs}$) by condition $V(r\rightarrow \infty
)\rightarrow 0$.

The form (\ref{eVq}) follows directly from the Lagrangian of
the charged particles in the electric field. First term in r.h.s. is
the kinetic energy density of ultrarelativistic electrons,
and next three terms are related to the energy density
of the electric field in the presence of charged sources.
The quark contribution to the electric charge density is
$\rho_{q}^{\rm ch} =\frac{2}{3}\rho_u -\frac{1}{3}\rho_d
-\frac{1}{3}\rho_s =\rho_B^{\rm I} (3x_u  - 1) $. The electron
density $\rho_e$ is found by a simple counting of all occupied
states $\rho_e (V)=2\int_{0}^{p_m} 4\pi p^2 d p/(2\pi )^3 \simeq
-V^3 /(3\pi^2 )$, cf. \cite{MPV77}. The last filled level ($p=p_m
=\sqrt{(\varepsilon_m -V)^2 -m_e^2}$, $m_e \simeq 0.5$~MeV is the
electron mass) corresponds to the electron energy $\varepsilon_m
=m_e$ since particles should not move to infinity. For typical
values $-V\sim m_\pi$
under consideration a small electron mass term can be safely
omitted and, therefore, for  ultra-relativistic electrons we get
$-\rho_e \simeq V^3 /3\pi^2$. For the case of spatially
homogeneous matter of a constant density, the electric potential
is constant, being determined from  the local charge-neutrality
condition $\rho^{{\rm ch}} =0$. Then $\epsilon_V  =0$, that in
terms of the electron chemical potential $\mu_{e,\rm loc}
=-V(\vec{r}) =const$ reads as $\epsilon_{e, \rm kin} =(\mu_{e,\rm
loc })^4 /4\pi^2$.

The energy density of the hadron matter (energy per fraction
volume $v^{\rm II} =(1-f^{\rm I})v$) includes several terms
\begin{eqnarray}\label{ehtot}
\epsilon^{\rm II} [x_p , \rho_B^{\rm II} ]=\epsilon_{\rm
kin+str}^{\rm II} + \epsilon_{e,\rm kin}^{\rm II} +\epsilon_V^{\rm
II} .
\end{eqnarray}
The energy density of the baryon matter contains kinetic and
potential contributions
\begin{eqnarray}\label{eb0}
\epsilon_{\rm kin+str}^{\rm II} [x_p , \rho_B^{\rm II} ]
=\epsilon^n_{\rm kin}[x_p , \rho_B^{\rm II} ] + \epsilon^p_{\rm
kin}[x_p, \rho_B^{\rm II} ] +\epsilon_{\rm pot} [x_p , \rho_B^{\rm
II} ],
\end{eqnarray}
now expressed in new variables.
The electron kinetic energy density plus the electric field energy
density, as follows right from the Lagrangian, render
\begin{eqnarray}\label{eVh}
&&\epsilon^e_{\rm kin, II} [V^{\rm II}] +\epsilon_V^{\rm II} [x_p
, V^{\rm II}, \rho_B^{\rm II} ]\nonumber \\ &&= \frac{(V^{\rm
II})^4}{4\pi^2}+\rho_e^{\rm II} (V^{\rm II})V^{\rm II}
-\rho_B^{\rm II} x_p V^{\rm II}-\frac{(\nabla V^{\rm II})^2}{8\pi
e^2}.
\end{eqnarray}

Equations of motion for the electric field (\ref{eqm-V}) recover
the Poisson equation (\ref{qV}). Eqs (\ref{eqm-u}), (\ref{eqm-d})
for $\vec r \in D^{\rm I} +D_{\rm S}$ yield
\begin{eqnarray}
&&(1+\frac{2\alpha_c}{ 3\pi})\left( p_{{\rm F}u}-p_{{\rm F}s}\right) -
\frac{m_s^2}{2p_{{\rm F}s}} -V^{\rm I}
+\frac{\delta E_{\rm S}[x_u , x_d, V^{\rm I} , \rho_B^{\rm I}
]}{3\rho_B^{\rm I} \delta x_u} =0, \label{qchemeq1a} \\
&&(1+\frac{2\alpha_c}{ 3\pi})\left( p_{{\rm F}d}-p_{{\rm F}s}\right) -
\frac{m_s^2}{2p_{{\rm F}s}} +\frac{\delta E_{\rm S}[x_u , x_d , V^{\rm
I} , \rho_B^{\rm I} ]}{3\rho_B^{\rm I} \delta x_d} =0
,\label{qchemeq1}
\end{eqnarray}
see eqs (\ref{eqtot}), (\ref{qkin}), (\ref{eVq}).

Integrating   equation of motion (\ref{qV}) times $V(\vec{r})$ and
using continuity of the electric potential and its derivative at
the surface (as above, we neglect a surface charge density), the
energy of the quark phase and the surface energy given by eqs
(\ref{eqtot}), (\ref{qkin}),  and (\ref{eVq}) can be rewritten in
terms of the variables $\rho_u$, $\rho_d$, $\rho_s$ and $\rho_e$.
Thereby, we reproduce eq. (\ref{endenq1}). The $\int d\vec{r}\,
\frac{(\nabla V)^2}{8\pi e^2}$ term is expressed in terms of the
particle densities recovering eq. (\ref{enden2}).

With the help of eqs
(\ref{chemexpr}) we  obtain that equations of motion
(\ref{eqm-d}), (\ref{eqm-p}), or equivalently (\ref{qchemeq1a}),
(\ref{qchemeq1}), specify nothing else than chemical equilibrium
conditions (\ref{q-c}).
Surface terms vanish, if eq. (\ref{BBP71qh}) is assumed to be
fulfilled. Then, one can drop surface contributions in chemical
equilibrium conditions.

Integrating equation of motion (\ref{qV}) times $V(\vec{r})$ and
using the continuity of $V(\vec{r})$ and $\nabla V (\vec{r})$ the
hadronic part of the energy  given by (\ref{ehtot}), (\ref{eb0}),
(\ref{eVh}) is presented in terms of new variables $\rho_n ,
\rho_p , \rho_e $, coinciding with (\ref{eVh11}) -(\ref{eVh11p}).
The chemical equilibrium condition for the reaction $n\leftrightarrow
p+e$
coincides with equation of motion (\ref{eqm-p})
which reads
\begin{eqnarray}\label{hchemeq1}
\sqrt{p_{{\rm F}n}^2 +m_N^2}- \sqrt{p_{{\rm F}p}^2 +m_N^2} +\frac{4S_0
(\rho_B^{\rm II} -2\rho_p )}{\rho_0} +V^{\rm II}
=0,
\end{eqnarray}
where we also
used eqs (\ref{ehtot}) - (\ref{eVh}).

The proton density is not constant in the corresponding eq.
(\ref{qV}), rather it obeys eq. (\ref{hchemeq1}) and, thereby,
depends on $V(\vec{r})$. In linear approximation in deviation of
$V(\vec{r})$ from its expression for the spatially homogeneous
system, by variation of (\ref{hchemeq1}) we obtain
\begin{eqnarray}\label{deltaro}
\left[\widetilde{C}_{0p}+\widetilde{C}_{0n}\right]\delta \rho_p
(\vec{r}) =  \delta V^{\rm II}(\vec{r})
  +\widetilde{C}_{0n}\delta\rho_B^{\rm II}(\vec{r}),
\,\,\,\, \widetilde{C}_{0j} \equiv
\left[C_{0j}+\frac{4S_0}{\rho_0}\right] .
\end{eqnarray}

Equation of motion (\ref{den-prof})
acquires the following explicit form
\begin{eqnarray}\label{graveq}
&&\rho_n \sqrt{p_{{\rm F}n}^2 +m_N^2}+\rho_p \sqrt{p_{{\rm F}p}^2
+m_N^2}+\frac{S_0 (\rho_B^{\rm II} -2\rho_p
  )^2}{\rho_0}-\epsilon^n_{\rm kin}-\epsilon^p_{\rm kin}
\nonumber \\ &&+\frac{K_0 (\rho_B^{\rm II})^2 }
{9\rho_0}\left(\frac{\rho_B^{\rm II}}{\rho_0}-1\right) +
C_{\rm sat}\frac{(\rho_B^{\rm II})^2}{\rho_0} +\frac{(V^{\rm
II})^4}{12\pi^2}+ \frac{(\nabla V^{\rm II})^2} {8\pi e^2}
=P .
\end{eqnarray}
This equation
reproduces nothing else than the standard relation $ \rho_B
\frac{\partial \epsilon}{\partial \rho_B}-\epsilon =P$. The same
equation of motion is obtained if one takes constant chemical
potentials from (\ref{omeg-chem}) and puts them into equation
\begin{eqnarray}\label{P-mu}
\mu_i \rho_i - \epsilon =P.
\end{eqnarray}
This can be  explicitly seen with the help of relations
(\ref{chem-n}) and equation of motion (\ref{qV}). Variation of
(\ref{P-mu}) yields
\begin{eqnarray}
\delta P =\rho_i\delta \mu_i +\mu_i\delta \rho_i - \delta\epsilon
.
\end{eqnarray}
Using that
\begin{eqnarray}
\delta\epsilon = (\partial\epsilon [\rho_i
]/\partial\rho_i)\delta\rho_i =\mu_i\delta\rho_i,
\end{eqnarray}
we obtain
\begin{eqnarray}
\delta P =\rho_i\delta \mu_i,
\end{eqnarray}
demonstrating that $\delta P =0$, if $\delta\mu_i =0$.

Varying eq. (\ref{graveq}) and with the help of (\ref{qV})
expressing the variation of the term $\delta V (\vec{r})\Delta V
(\vec{r})/ 4\pi e^2$ (being obtained by partial integration) via
$(\rho_p -\rho_e )V(\vec{r})$, we arrive at the relation
\begin{eqnarray}\label{deltaro-gr}
\left[ \widetilde{C}_{0p}\rho_p -\widetilde{C}_{0n}\rho_n \right]
\delta \rho_p (\vec{r}) +\left[ \widetilde{C}_{0n}\rho_n -K_1
\rho_B \right] \delta \rho_B (\vec{r})-\rho_p\delta V (\vec{r})=0
.
\end{eqnarray}
>From eqs (\ref{deltaro}), (\ref{deltaro-gr}) we  recover the first
relation  (\ref{co}).

We could also vary constant values of chemical potentials
(\ref{chem-n}).  Then  we would obtain
\begin{eqnarray}
\widetilde{C}_{0n}\delta \rho_n (\vec{r})-K_1 \delta\rho_B
(\vec{r})=0,\,\,\,\, \widetilde{C}_{0p}\delta \rho_p (\vec{r})-K_1
\delta\rho_B (\vec{r}) -\delta V (\vec{r})=0.
\end{eqnarray}
Subtraction of these equations reproduces (\ref{deltaro}). Also
with the help of the replacement $\delta\rho_n =\delta\rho_B
-\delta\rho_p$ we recover (\ref{deltarop1}).

Finally, we have shown that both methods using $\Omega$ and $G$ in
appropriate variables reproduce the very same results.

\section{Estimation of non-linear correction terms}\label{example}

To get a feeling how small might be non-linear corrections to solutions
corresponding to
the linearized Poisson equation  (\ref{Pois-lin}), we,
following \cite{MPV77}, solve  eq. (\ref{qV}) analytically with
a positive charge density $\rho^{\rm ch}= \rho \theta
(R-r)$ and negative electron density $\rho_e =-\frac{V^3}{3\pi^2}$,
for simplicity fixing the gauge by the condition $V (r\rightarrow
\infty )\rightarrow 0$, for $r\rightarrow
 \infty$, $\rho =const \sim \rho_0$,
$\theta (R-r) =1$ for $r<R$ and zero for $r>R$. We will also
restrict ourselves by consideration of the case $R\gg \lambda_{\rm D}$,
$\lambda_{\rm D}^{-2}=4\pi e^2 \rho /\mu_{e,\rm Gibbs} $. Then, in
variables $x=(r-R)/\lambda_{\rm D}$, $V=-\mu_e \chi (x)$, eq. (\ref{qV})
acquires the form
\begin{eqnarray}\label{xi}
\chi^{''}=\chi^{3}-\theta (-x),
\end{eqnarray}
where reducing 3-dimensional Laplacian to the 1-dimensional
Laplacian we dropped small $O(\lambda_{\rm D} /R)$ term. This equation
has {\em{exact}} solutions
\begin{eqnarray}\label{xi-ex1}
\chi (x) =1-3\left[1+\frac{\mbox{sh}(a-x/\sqrt{3})}{\sqrt{2}}
\right]^{-1}, \,\, x<0
\end{eqnarray}
\begin{eqnarray}\label{xi-ex2}
\chi (x) =\frac{\sqrt{2}}{(x+b)}, \,\, x>0,
\end{eqnarray}
with constants $a$ and $b$ found from matching of the potentials with
the help of the boundary conditions: $\mbox{sh} a= 11\sqrt{2}$,
$b=\frac{4\sqrt{2}}{3}$.

Linearized eq. (\ref{xi}) expressed in terms of $\psi \simeq
1-\chi$ in the region $x<0$ renders
\begin{eqnarray}\label{xi1}
\psi^{''}-3\psi =0.
\end{eqnarray}
Solution is $\chi \simeq 1-C\mbox{exp}(x\sqrt{3})$. Using boundary
conditions one gets $C\simeq 0.24$. Comparison of this approximate
solution and the exact one (\ref{xi-ex1}) demonstrates coincidence
with deviations $\lsim 1.5$~\%, while the non-linear
correction $3\psi^2$ to the l.h.s. of (\ref{xi1}) is not small
compared to the linear term $3\psi$.

\section{Peculiarities of the Coulomb limit at finite
quark concentration $f^{\rm I}$} \label{Coul-pec}

In the main text we have demonstrated that our general results
coincide
with those previously obtained in the literature
for the limiting case of single droplet
($f\equiv f^{\rm I}\rightarrow 0$)
of a tiny size. We also have argued that the limit
of single small transverse size slab is rather peculiar and can't be
reproduced without inclusion of screening effects. Indeed, the
unscreened potential
of the single slab diverges at large distances from it. It is however not the
case for the periodic slab structures at finite $f$.
Let us now consider other peculiarities of
the case when $f$ is finite, for both geometries.

Assuming the constant (step function) charge densities
$\rho^{\rm I}_{\rm ch}\neq
\rho^{\rm II}_{\rm ch}$ ref. \cite{RPW83}
derived the following expression for the energy density  of the
droplet ($d=3$) and the slab ($d=1$):
\begin{eqnarray}\label{Peth}
&&\epsilon_{\rm C}^{(d)} =
2\pi e^2(\rho^{\rm II}_{\rm ch}-\rho^{\rm I}_{\rm ch}
)^2 R^2 f \Phi_d
(f),\\
 &&\Phi_d (f)=\left[ 2 (d-2)^{-1} (1-\frac{1}{2}d\, f^{(1-2/d)}
)+f\right](d+2)^{-1}.\nonumber
\end{eqnarray}

Using the global charge neutrality condition (for constant values of
$\rho^{\rm I,II}_{\rm ch}$):
\begin{eqnarray}\label{Peth-cond}
f\rho^{\rm I}_{\rm ch} +(1-f) \rho^{\rm II}_{\rm ch}=0 ,
\end{eqnarray}
we then find for the energies per droplet/slab volume:
\begin{eqnarray}\label{Peth-d3}
&&\widetilde{\epsilon}_{\rm C}^{(d=3)}
=\frac{4\pi e^2 (\rho_{\rm ch}^{\rm I})^2
R^2}{5 (f-1)^2}\left[ 1-\frac{3}{2} f^{1/3} +\frac{1}{2} f\right] ,
\end{eqnarray}
\begin{eqnarray}\label{Peth-d1}
&&\widetilde{\epsilon}_{\rm C}^{(d=1)}
=\frac{2\pi e^2 (\rho_{\rm ch}^{\rm I})^2
R^2}{3f}.
\end{eqnarray}
Adding to these expressions the corresponding surface energy density terms
$\beta_0\beta_1/\xi$ and $\beta_0\beta_1/3\xi$
respectively and minimizing the results with respect
to $\xi$ we arrive at the ratio
of the droplet to slab energy densities
\begin{eqnarray}
\frac{\widetilde{\epsilon}_{\rm C}^{(d=3)}(\xi_{\rm min}^{(d=3)})}
{\widetilde{\epsilon}_{\rm C}^{(d=1)}(\xi_{\rm min}^{(d=1)})}=3 (2/5)^{1/3}
\frac{(1-\frac{3}{2} f^{1/3} +
\frac{1}{2} f)^{1/3}f^{1/3}}{(1-f)^{2/3}},
\end{eqnarray}
whereas it follows that the ratio becomes to be larger than unit for $f>0.32$.
This means that
slabs become to be energetically preferable
for $f>0.32$ compared to spherical droplets,
cf. the boundary value of the concentration,
when rods transform to slabs,
as it follows from Fig. 17 of \cite{GP95}.
Note that again one used here mentioned above ``the
two part'' approach of Glendenning
to the problem, which allows for the minimization of
the energy densities at fixed $f$, see discussion in
sect. \ref{Comparison}.

For spherical droplets ($d=3$) and for the constant charge
densities we
obtain
\begin{eqnarray}\label{v1-sp}
&&V_{\rm I}^{'}=\frac{4\pi e^2 \rho^{\rm I}_{\rm ch} r}{3},\nonumber \\
&&V_{\rm II}^{'}=\frac{4\pi e^2 \rho^{\rm II}_{\rm ch} }{3}\left(
r-\frac{R_{\rm
W}^3}{r^2}\right) ,
\end{eqnarray}
and we immediately recover (\ref{Peth-d3}).

On the other hand, in the case $\xi \ll 1$,
$\widetilde{\alpha}_1 \xi \ll 1$
from our general expressions
(\ref{1-sol}), (\ref{psi})   we get
\begin{eqnarray}\label{v2-sp}
&&V_{\rm I}^{'}\simeq \frac{1}{3}V_0^{\rm I} (\kappa^{\rm I})^2 r,
\nonumber \\
&&V_{\rm II}^{'}\simeq  \frac{R}{3R_{\rm W}} V_0^{\rm
II}(\kappa^{\rm II})^2 \left( r-\frac{R_{\rm W}^3}{r^2}\right) .
\end{eqnarray}
Comparing  (\ref{v2-sp})
and (\ref{v1-sp}) we
express the charged densities through the
values of the constants $V_0^{\rm I}$ and $V_0^{\rm II}$ that gives the ratio
$$V_0^{\rm I}/V_0^{\rm
II}\simeq \alpha_0^2 f^{1/3} \rho^{\rm I}_{\rm ch}/\rho^{\rm II}_{\rm ch}.$$
Expanding $\widetilde{V}_0^{\rm I}$ and $\widetilde{V}_0^{\rm
II}$ in (\ref{const1-1cor11}), (\ref{const2-2cor}) in $\xi \ll 1$,
$\widetilde{\alpha}_1 \xi \ll 1$ and using that
$\widetilde{V}_0^{\rm I}=V_0^{\rm I}$, $V_0^{\rm
II}=-\widetilde{V}_0^{\rm II}\alpha_0 \xi /f^{1/3}$ we  reproduce the
global charge neutrality condition (\ref{Peth-cond}) in this case.
One can also find the same values of the charged densities using expressions
(\ref{qrhoch}), (\ref{r2}) in the limit $r\rightarrow 0$, e.g. we get
\begin{eqnarray}\label{rhcul3}
(\rho^{\rm I}_{\rm ch})^2 =\frac{\beta_0 \kappa_{\rm I}^2 (1-f)^2 \alpha_0^4}
{6\pi e^2 [f+\alpha_0^2 (1-f) ]^2}.
\end{eqnarray}
Comparison of (\ref{Peth-d3}) with its genetic partner expression
(\ref{emV-cor11 }) in the limit  $\xi \ll 1$,
$\widetilde{\alpha}_1 \xi \ll 1$ also  allows to recover the same expression
(\ref{rhcul3}).
Thus, we see that our general equations
reproduce the Coulomb limit (\ref{v1-sp})
and the
expression for the energy (\ref{Peth-d3}) for $d=3$, for
tiny-size  droplets.

As we see from our general solutions,
the value $\rho^{\rm I}_{\rm ch}$ is unambiguously determined
through the screening parameters $\kappa^{\rm I}$, $\kappa^{\rm II}$
and the chemical potentials, which determine the value of the
coefficient $\beta_0$, cf. (\ref{param1}).
Thus we have no room to vary $\rho^{\rm I}_{\rm ch}$ anymore.
In the calculations \cite{RPW83}
the value $\rho^{\rm I}$
(or $\rho^{\rm I}_{\rm ch}$, which is unambiguously expressed through
$\rho^{\rm I}$)
is considered as a free parameter, being recovered by the
minimization of the total energy density (including the Coulomb plus surface
energy densities), at the value of the concentration
$f$.
Since the value $\rho^{\rm I}_{\rm ch}$ depends on the geometry
of the structure the value $f$ also depends then on the geometry.
However, the latter dependence is very weak.
Thereby, instead of this more complicated procedure
with appropriate accuracy one may drop a small contribution
of the Coulomb and surface energies of the spatial structures to
the total energy, cf. \cite{G01}.
Then the quantities $\rho^{\rm I,II}_{\rm ch}$ and $f$
can be considered as approximately independent on the structure geometry,
whereas they essentially depend on the bulk properties of the matter.
In the
limit of tiny-size structures
$\rho^{\rm I,II}_{\rm ch}$ and $f$ are
given by the same expressions as we get from our general equations.

Let us consider now slabs. For constant
densities and $d=1$, we have
\begin{eqnarray}\label{v11}
&&V_{\rm I}^{'}=4\pi e^2 \rho^{\rm I}_{\rm ch} x,
\nonumber \\
&&V_{\rm II}^{'}= 4\pi e^2 \rho^{\rm II}_{\rm ch} (x\mp R_{\rm W}),
\end{eqnarray}
and we immediately recover (\ref{Peth-d1}). On the other hand,
in the limit $\xi \ll 1, \alpha_2 \xi \ll 1$
from our general equations (\ref{qsol-1}),
(\ref{qsol1-1}), (\ref{qsol1-2})  we obtain
\begin{eqnarray}\label{v22}
&&V_{\rm I}^{'}\simeq  V_0^{\rm I}(\kappa^{\rm I})^2 x,\nonumber \\
&&V_{\rm II}^{'}\simeq  V_0^{\rm II}(\kappa^{\rm II})^2 (x\mp
R_{\rm W}).
\end{eqnarray}
We suppressed here superscript (1)
symbolizing slabs.
Comparison of (\ref{v11}) and (\ref{v22}) yields
$$V_0^{\rm I}/V_0^{\rm
II}\simeq \alpha_0^2 \rho^{\rm I}_{\rm ch}/\rho^{\rm II}_{\rm ch}.$$
Using this
relation with the help of eqs (\ref{const1-1}), (\ref{const2-1})
we can easily reproduce the condition (\ref{Peth-cond}). With the help of
(\ref{eVo}), (\ref{eVo1}) we can also recover eq. (\ref{Peth-d1}).
Comparison of (\ref{Peth-d1}) and $\widetilde{\epsilon}_{V,\rm
I}^{(1)}+\widetilde{\epsilon}_{V,\rm
II}^{(1)}$
from (\ref{limC}) leads us to the same expression
(\ref{rhcul3}) for
$\rho^{\rm I}_{\rm ch}$, as in three dimensional case.
Thus
the Coulomb limit is totally recovered also in the case $d=1$
for finite values of $f$ ($\xi \ll 1$,
$\alpha_2 \xi \ll 1$). Other limit case, $\xi \ll 1$,
$\alpha_2 \xi \gg 1$, of a tiny fraction volume can't be reproduced by the
bare Coulomb solution (\ref{Peth-d1}), as we have mentioned,
leading  even to a
different dependence on
$\xi$.
In this case the correct asymptotic of the
$\delta\widetilde{\omega}^{(1)}_{\rm tot}$ is given by eq.
(\ref{Coulsl}) rather than by
eq. (\ref{Peth-d1}).

In our general consideration, as we argued in
sect. \ref{Comparison},
comparing the effective energies
$\delta\widetilde{\omega}_{\rm tot}^{(3)}(\xi_{\rm min}^{(3)})$ for droplets
and $\delta\widetilde{\omega}^{(1)}_{\rm tot}(\xi_{\rm min}^{(1)})$ for slabs
we
recover the Coulomb limit
only for tiny values of $\beta_1$. We illustrate it by Fig. 8,
where we present the values $\widetilde{\epsilon}_V^{\rm min}+
\widetilde{\epsilon}_{\rm S}^{\rm min}\equiv
\widetilde{\epsilon}_V^{(3)} (\xi_{\rm min}^{(3)} )+
\widetilde{\epsilon}_{\rm S}^{(3)} (\xi_{\rm min}^{(3)} )$,
as given by minimization in $\xi$
of the expressions (\ref{emV}), (\ref{ebV}),
(\ref{emV-cor1}), (\ref{ebV-cor1}) and (\ref{esV})
for droplets and $\widetilde{\epsilon}_V^{\rm min}+
\widetilde{\epsilon}_{\rm S}^{\rm min}\equiv
\widetilde{\epsilon}_V^{(1)} (\xi_{\rm min}^{(1)} )+
\widetilde{\epsilon}_{\rm S}^{(1)} (\xi_{\rm min}^{(1)} )$,
as given by minimization in $\xi$
of the expressions
(\ref{eVo}), (\ref{eVo1}) and
(\ref{esoc}) for slabs.

\newpage
\begin{figure}[h]
\begin{center}
  \epsfsize=0.9\textwidth
\centerline{\epsffile{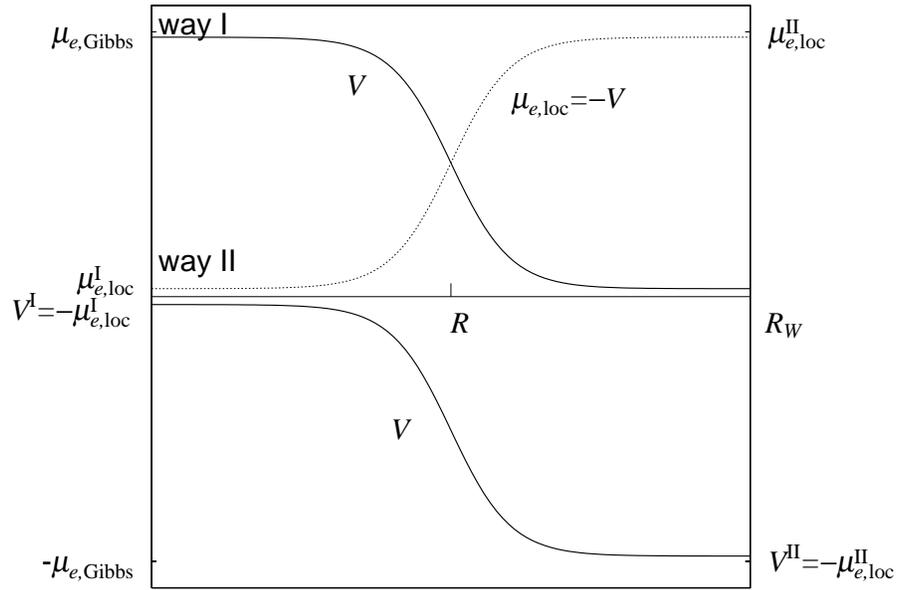}}
\end{center}
 \caption{Schematic view of the electric potential well
with two gauge choices, {\bf{way I}}:
$V =V(V^0=0)$,
and {\bf{way II}}: $V=V(V^0=-\mu_{e,{\rm Gibbs}});
V({\bf{way \,\,II}})=V({\bf{way \,\,I}})-\mu_{e,{\rm Gibbs}}$.
}
\end{figure}

\begin{figure}
[htb]
\vspace{-20mm}
  \epsfsize=0.82\textwidth
  \epsffile{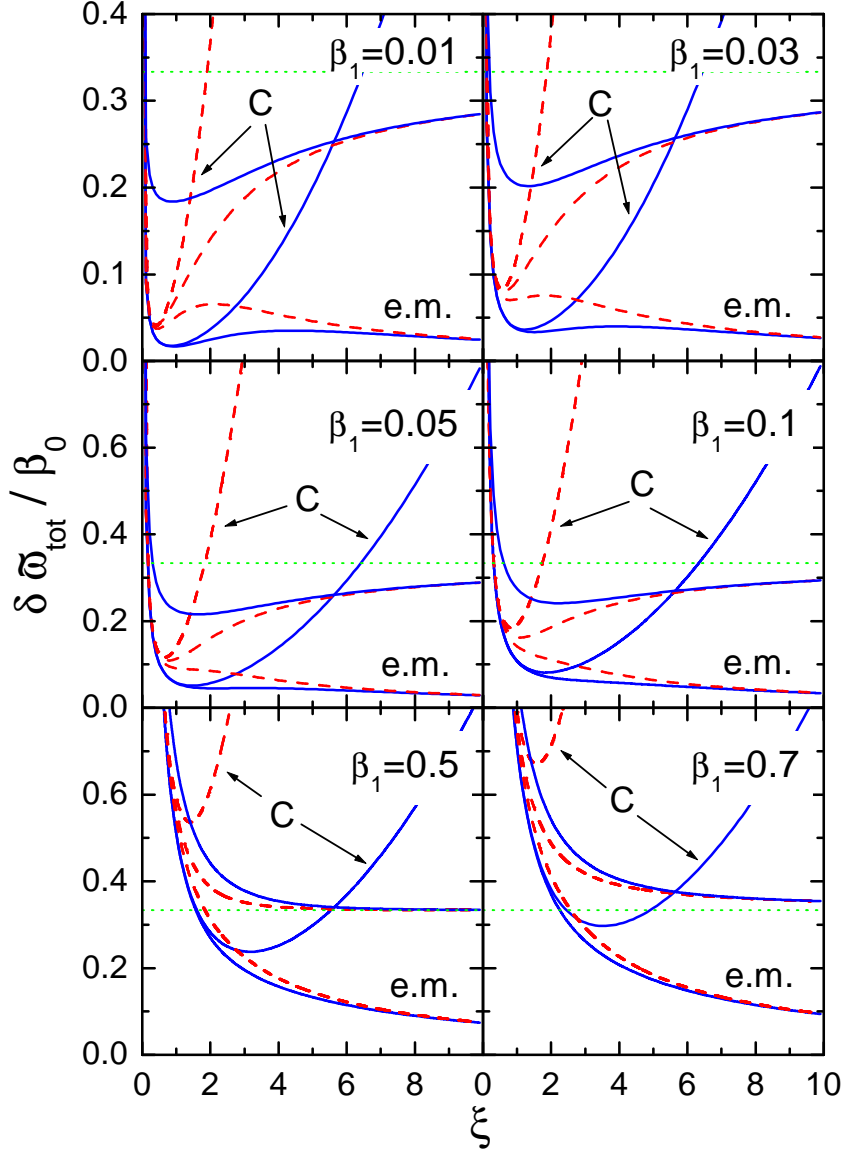 }
  \caption{Fraction dependence of the
effective energy per droplet volume
as a function of the scaled
radius $\xi=R/\lambda_{\rm D}^{\rm I}$, where $\beta_0$ is the parameter of the
energy
scale and $\beta_1$ is the parameter proportional to the
 surface tension, see (\ref{b0est}), (\ref{b1est}). Solid
lines are given for $f=0.5$ and dashed lines for $f=1/100$.
The Coulomb curves ``C'' are calculated with the help of
eqs (\ref{Peth-d3}), (\ref{rhcul3}),
demonstrating the Coulomb limit, and
``e.m.'' curves relate to  ``the electric field energy''
plus surface energy (without ``correlation'' contributions),
$\alpha_0=\lambda_{\rm D}^{\rm I}/\lambda_{\rm D}^{\rm II}$
is fixed as $\alpha_0=1$.}
\end{figure}

\newpage
\begin{figure}[htb]
 \vspace{2mm}
  \epsfsize=0.99\textwidth
  \epsffile{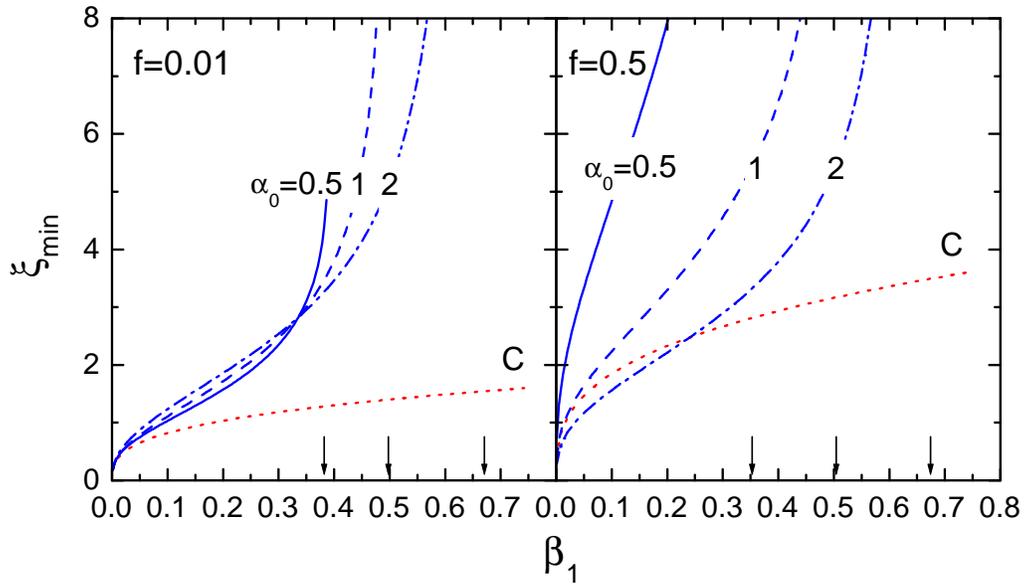}
  \caption{The radius $\xi_{\rm min}=
R_{\rm \min}/\lambda_{\rm D}^{\rm I}$ of the spherical
droplet corresponding to the minimum of
the effective energy,
$\delta \widetilde{\omega}_{\rm tot}$, in dimensional units.
The ``C'' curves are presented
for $\alpha_0 =1$, as in Fig.~2.}
\end{figure}

\newpage
\begin{figure}[htb]
 \vspace{2mm}
  \epsfsize=0.82\textwidth
  \epsffile{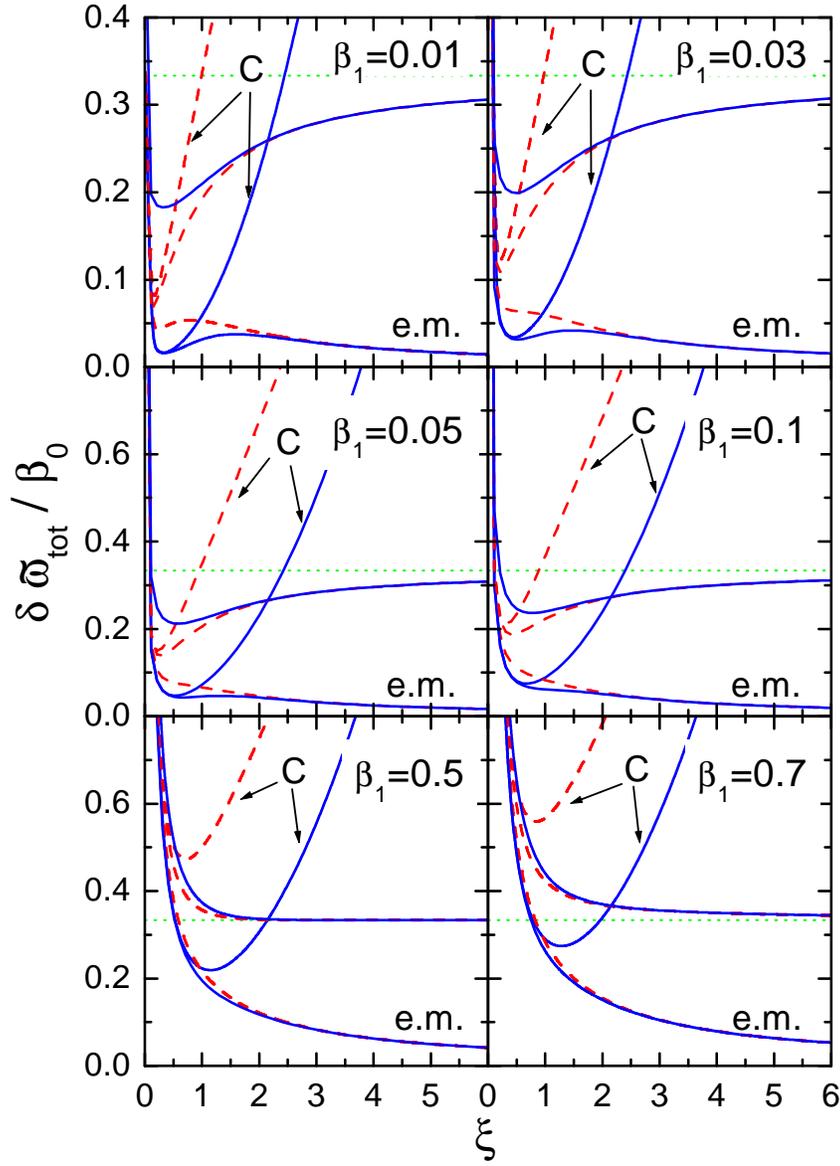}
  \caption{The same, as in Fig. 2, but for slabs.
The curves labeled by ``C'' are calculated with the help of
eq. (\ref{Coulsl}) for $f=0.01$ (dashed) and
(\ref{Peth-d1}), (\ref{rhcul3}) for $f=0.5$ (solid).
 }
\end{figure}

\newpage
\begin{figure}[htb]
 \vspace{-20mm}
  \epsfsize=0.99
\textwidth
  \epsffile{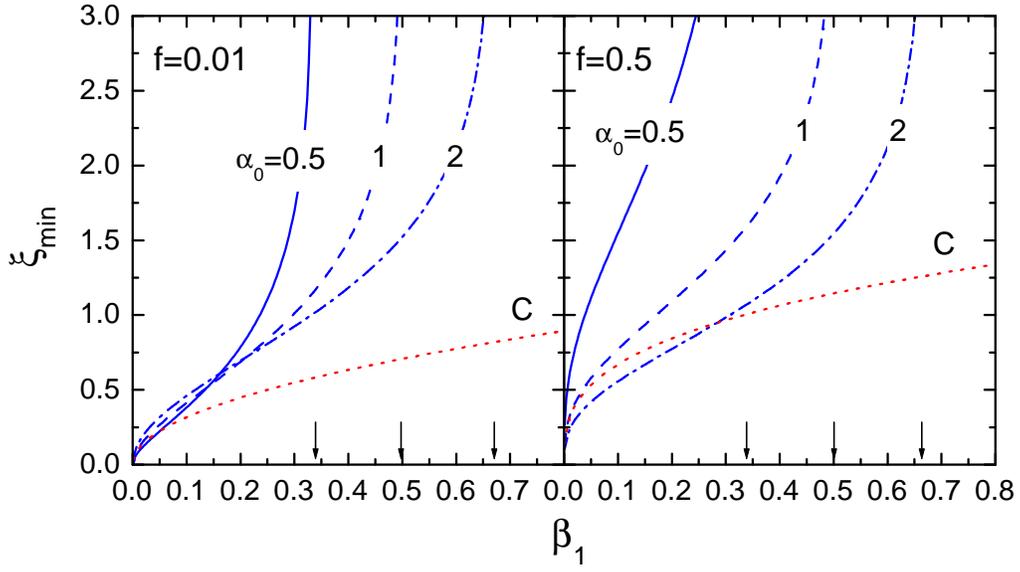}
  \caption{The transverse slab radius $\xi_{\rm min}$ corresponding to
the minimum of the effective energy,
$\delta\widetilde{\omega}_{\rm tot}^{(1)}$, in dimensionless units.
The ``C'' curves are the same, as in Fig. 4, shown for $\alpha_0 =1$. }
\end{figure}

\newpage
\begin{figure}[htb]
 \vspace{20mm}
  \epsfsize=
0.99\textwidth
  \epsffile{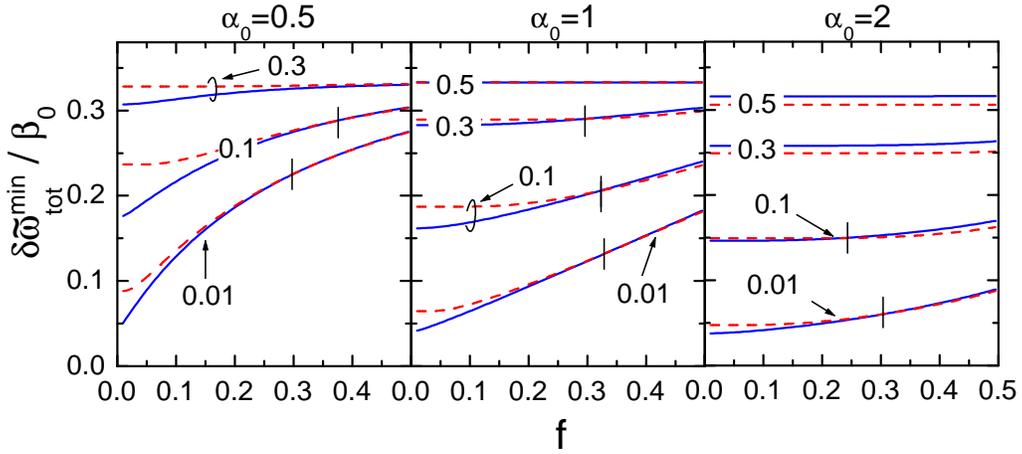}
  \caption{Comparison of the effective energies of  droplets (solid curves)
and  slabs (dashed curves). The numbers at the curves are the values of
$\beta_1$. Crossing points of the curves are indicated.}
\end{figure}

\newpage
\begin{figure}[htb]
 \vspace{2mm}
  \epsfsize=0.99\textwidth
  \epsffile{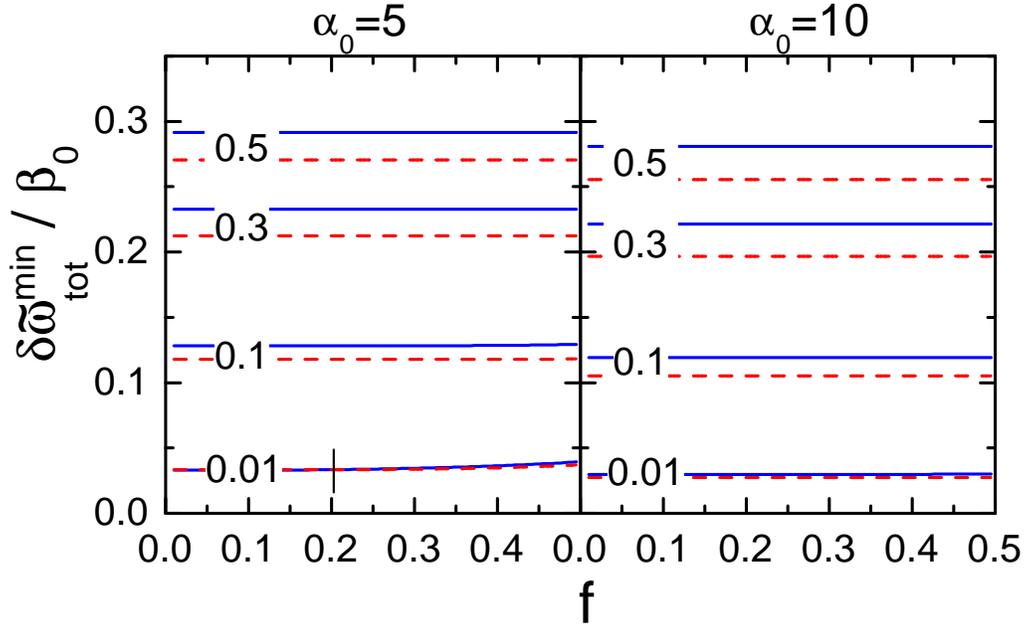}
  \caption{The same, as in Fig.~6, but for large values of $\alpha_0$.}
\end{figure}

\newpage
\begin{figure}[htb]
 \vspace{2mm}
  \epsfsize=0.99\textwidth
  \epsffile{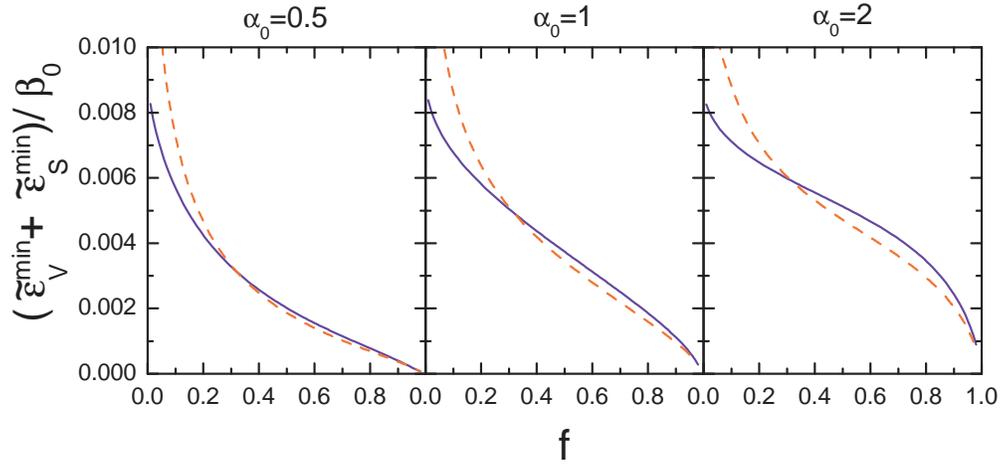}
  \caption{Comparison of the effective energies of  droplets (solid curves)
and  slabs (dashed curves) in the Coulomb limit
(for
$\beta_1 =10^{-3}$).}
\end{figure}
\end{document}